\documentclass[amsmath,amssymb,nofootinbib,prd]{revtex4}
\pdfoutput=1
\usepackage{graphicx}
\usepackage{epsfig}
\usepackage{rotate}
\usepackage{amsmath}
\usepackage{amssymb}
\usepackage{amsfonts}
\usepackage{enumerate}
\usepackage{afterpage}
\usepackage{color}

\newcommand{\ud}{\mathrm{d}}
\newcommand{\p}{\partial}
\newcommand{\cH}{\mathcal{H}}
\newcommand{\Perp}{\mathcal{P}}

\newcommand{\M}{\mathcal{M}}
\newcommand{\D}{\mathcal{D}}
\newcommand{\Q}{\mathcal{Q}}
\newcommand{\F}{\mathcal{ F}}

\def\be{\begin{equation}}
\def\ee{\end{equation}}
\def\bea{\begin{eqnarray}}
\def\eea{\end{eqnarray}}

\begin{document}

\title{Observed galaxy number counts on the lightcone up to second order: III. Magnification Bias}

\author{Daniele Bertacca}

\affiliation{Physics Department, University of the Western
Cape, Cape Town 7535, South Africa\\}

\begin{abstract}

We study up to second order the galaxy number over-density that depends on magnification in redshift space on cosmological scales for a concordance model. The result contains all general relativistic effects up to second order which arise from observing on the past light cone, including all redshift and lensing distortions, contributions from velocities, Sachs-Wolfe, integrated SW and time-delay terms.
We find several new terms and contributions that could be potentially important for an accurate calculation of the bias on estimates of non-Gaussianity and on precision parameter estimates.

\end{abstract}

\date{\today}

\maketitle

\section{Introduction}
Galaxy catalogues in observational cosmology  depend on the apparent flux from the source. Practically, this can be translated into the dependence of the observed number counts on the magnification that modifies the spatial distribution of the sources (for example of galaxies or quasars) (see, for example,  \cite{Turner:1984ch, Kaiser:1996tp,  Bartelmann:1999yn, Matsubara:2000pr, Gaztanaga:2002qk, Refregier:2003ct, Schneider:2005ka, Hui:2007cu, Hui:2007tm, LoVerde:2007ke, Schmidt:2008mb, Schmidt:2009ri, Schmidt:2009rh, Bartelmann:2010fz, Schmidt:2010ex, Liu:2013yna}).
 
 At the same time, on cosmological scales,  relativistic effects alter the observed number over-density giving the well-known corrections both from the usual redshift space distortions and  gravitational lensing convergence \cite{Matsubara:2000pr},  Doppler, Sachs-Wolfe, integrated SW and time-delay type terms. The full relativistic effects have been studied at first order in perturbation theory by~\cite{Yoo:2009au, Yoo:2010ni, Bonvin:2011bg, Challinor:2011bk, Jeong:2011as} and at second order by~\cite{Bertacca:2014dra, Bertacca:2014wga, Yoo:2014sfa, DiDio:2014lka}.
(A similar analysis at second order has been done on the cosmological distance-redshift relation in \cite{Umeh:2012pn}--\cite{Marozzi:2014kua}, or to the weak lensing in \cite{Dodelson:2005zj, Bernardeau:2009bm,Bernardeau:2011tc}.)
 
This paper completes our studies on derivation of  the formula for the observed galaxy number over-density up to second order on cosmological scales.  
Following the ``cosmic rulers" approach of  \cite{Jeong:2011as,Schmidt:2012ne} at first order and of \cite{Bertacca:2014wga} at second order, we generalize the previous  results  obtained in \cite{Bertacca:2014dra, Bertacca:2014wga} adding the magnification bias dependence (see also  \cite{Yoo:2014sfa}).

In the last part of the work, we look at the galaxy number over-density up to second order in redshift space on cosmological scales  for a concordance model, in the Poisson gauge. The result contains all general relativistic effects up to second order that arise from observing on the past light cone, including all redshift effects, lensing distortions from convergence and shear,  contributions from velocities, Sachs-Wolfe, integrated SW and time-delay terms.
We find several new terms and contributions which could be important for the accurate calculation on estimates of the primordial non-Gaussianity \cite{Bertacca:2014n}.

The paper is organised in the following way: in Section~\ref{CR} we briefly review and generalize the cosmic rulers to obtain the second-order perturbations of galaxy number counts in redshift space;
in Section~\ref{CR} we analyze the magnification both at first and second order, and, in Section \ref{ng} we present the galaxy number over-density  up to second order with all relativistic effects that arise from observing on the past lightcone and as a function of the magnification bias.
We perturb a flat Robertson-Walker universe in  the Poisson Gauge  for a $\Lambda$CDM model in Section~\ref{Poiss-pert}. Finally, Section~\ref{Sec:Conclusions} is devoted to conclusions.

Conventions: units $c=G = 1$;  signature is $(-, +, +, +)$; Greek indices run over $0, 1, 2, 3$, and Latin over $1, 2, 3$.

\section{Cosmic laboratory} \label{CR}

In this section we summarize briefly the comic rulers  defined in Ref. \cite{Bertacca:2014wga} (at first order see also  \cite{Jeong:2011as,Schmidt:2012ne}).
First of all, it is useful to define the parallel and perpendicular projection operators to the observed line-of-sight direction (see also \cite{Jeong:2011as,Schmidt:2012ne}). For any  spatial vectors and tensors:
\begin{eqnarray}
\label{Projection}
A_{\parallel} = n^{i} n^{j} A_{ij } \;, \quad
B_{\perp}^i =  \Perp^{ij} B_j = B^i -n^i B_{\parallel}\;,  
\end{eqnarray}
where $\Perp^i_{j}= \delta^{i}_j-n^in_j$. The directional derivatives are defined as
\begin{eqnarray}
\label{Projection2}
\p_\parallel = n^i  {\p \over   \p \bar x^i} \;,  \quad \quad \p^2_\parallel  = \p_\parallel \p_\parallel\;,\quad \quad  \p_{\perp i} =  \Perp^j_i \p_j=   {\p \over   \p \bar x^i} -  n_i \p_\parallel\;, ,\quad \quad   {\p n^j \over   \p \bar x^i}  &=& \frac{1}{\bar \chi}\Perp_i^j\;, 
\end{eqnarray}
and we have 
\begin{eqnarray}
\label{Projection3}
 {\p B^i \over   \p \bar x^j}  &=& n^i n_j \p_{\parallel} B_{\parallel} + n^i \p_{\perp j} B_{\parallel} +\p_{\perp j} B_{\perp}^i+ n_j \p_{\parallel} B_{\perp}^i +\frac{1}{\bar \chi} \Perp^i_j B_{\parallel}\;,\nonumber \\
{\rm and}& &\quad \quad  \nabla^2_\perp = \p_{\perp i}\p_\perp^i = \delta^{ij} {\p \over   \p \bar x^i}  {\p \over   \p \bar x^j} - \partial_\parallel^2 - \frac{2}{\bar \chi} \p_\parallel\;.
\end{eqnarray}

Redshift-space or {\it redshift frame} is the ``cosmic laboratory'' where we probe the observations. In redshift-space we use coordinates which effectively flatten our past lightcone so that the photon geodesic from an observed galaxy has the following conformal space-time coordinates  \cite{Jeong:2011as,Schmidt:2012ne, Bertacca:2014wga}:
\begin{equation}
\bar{x}^\mu=(\bar \eta,\; \bar {\bf x})=(\eta_0-\bar \chi, \; \bar \chi \, {\bf n}).
\end{equation}
Here $\bar \chi(z)$ is the comoving distance to the observed redshift in redshift-space, ${\bf n}$ is the observed direction to the  galaxy, i.e. $n^i=\bar x^i/\bar \chi=\delta^{ij} (\p \bar \chi/\p \bar x^j)$ . Using $\bar \chi$ as an affine parameter in the redshift frame, the total derivative along the past light cone is $\ud / \ud \bar \chi = - \p/ \p \bar \eta + n^i \p/\p \bar x^i$.
Defining the photon 4-momentum $p^\mu=\nu (a)k^\mu/a$, where $a$ is the scale factor, $\nu \propto 1/a $ is the frequency,  we perturb the comoving null geodesic vector $k^\mu$, at second order, in the following way 
 \begin{equation}
\label{kmu}
 k^\mu(\bar \chi) = \frac{\ud  x^\mu }{\ud \bar \chi}(\bar \chi)= \frac{\ud }{\ud \bar \chi}  \left(\bar{x}^\mu + \delta x^\mu\right)(\bar \chi) = \left(-1+\delta \nu^{(1)}+\frac{1}{2}\delta \nu^{(2)},\; n^i+\delta n^{i (1)}+\frac{1}{2}\delta n^{i (2)} \right)(\bar \chi)\;,
\end{equation}
 where in the redshift frame at zero order
\begin{equation}
\label{kmu-0}
 \bar{k}^\mu=\frac{\ud  \bar{x}^\mu }{\ud \bar \chi}=\left(-1, \; {\bf n} \right)\;.
\end{equation}
Defining $x^\mu (\chi)$ as the coordinates in the {\it physical frame}, where $\chi$ is the physical comoving distance of the source,  we can set up a mapping between redshift space and real space  (the ``physical frame")  in the following way
\begin{eqnarray}
\label{chi}
\chi = \bar \chi+ \delta \chi  \quad &{\rm where} &\quad \delta \chi= \delta \chi^{(1)} +\frac{1}{2}\delta \chi ^{(2)}\,, \\
\label{xph2}
x^\mu (\chi) = \bar{x}^\mu (\bar \chi)+ \Delta x^\mu (\bar \chi) \quad &{\rm where}& \quad \Delta x^\mu (\bar \chi)= \Delta x^{\mu (1)} (\bar \chi)+\frac{1}{2}\Delta x^{\mu (2)} (\bar \chi).
\end{eqnarray}

Taking into account the observed redshift $(1+z)\big|_{\chi} = (u_\mu p^\mu)\big|_{\chi}/(u_\mu p^\mu)|_o$, and assuming that $a_o = 1$ we can obtain explicitly the scale factor  \cite{Bertacca:2014wga}
\begin{eqnarray}
\label{Deltalna-1}
\Delta \ln a^{(1)}&=&- \delta \nu^{(1)} - E_{\hat{0}0}^{(1)} + E_{\hat{0}\|}^{(1)} \;, \\
\label{Deltalna-2}
 \Delta\ln a^{(2)}&=& - \delta\nu^{(2)} - E_{\hat{0}0}^{(2)}+ E_{\hat{0}\|}^{(2)} +  2 E_{\hat{0}\|}^{(1)} \left( \delta \nu^{(1)} + \delta n_{\|}^{(1)}\right) +2\left(\delta x^{0(1)} + \delta x_{\|}^{(1)}\right)  \p_\|\left( E_{\hat{0}\|}^{(1)} -  E_{\hat{0} 0}^{(1)}\right)   \nonumber \\
&&- \frac{2}{\cH} \left( E_{\hat{0}\|}^{(1)} -  E_{\hat{0} 0}^{(1)}\right) \left(\frac{\ud \, \Delta \ln a}{\ud \bar \chi}\right)^{(1)} +2 \left[ - \left( E_{\hat{0}\|}^{(1)} -  E_{\hat{0} 0}^{(1)}\right) + \frac{1}{\cH} \left(\frac{\ud \, \Delta \ln a}{\ud \bar \chi}\right)^{(1)} \right] \delta \nu^{(1)} \nonumber \\
&& -2 \delta x^{0 (1)} \left( \frac{\ud \delta \nu}{\ud \bar \chi}\right)^{(1)}  + 2 E_{\hat{0}\perp i}^{(1)} \delta n_\perp^{i (1)} +2 \left[\p_{\perp i} \left( E_{\hat{0}\|}^{(1)} -  E_{\hat{0} 0}^{(1)}\right) - \frac{1}{\bar \chi} E_{\hat{0}\perp i}^{(1)}  \right] \delta x_\perp^{i (1)} \;,
\end{eqnarray}
and the comoving distance
\begin{eqnarray}
\label{chi_1}
\delta \chi^{(1)} & = &  \delta x^{0 (1)}- \Delta x^{0 (1)} \;,\\ 
\label{chi_2}
\delta \chi^{(2)} & = &  \delta x^{0 (2)} -\frac{1}{\cH} \Delta \ln a^{(2)}+ \frac{(\cH' + \cH^2)}{\cH^3} \left( \Delta \ln a^{(1)} \right)^2 - \frac{2}{\cH}\delta \nu^{(1)} \Delta \ln a^{(1)} +2 \delta \nu^{(1)} \delta x^{0 (1)}\;,
\end{eqnarray}
where we have defined the  parallel and perpendicular parts of the tetrad in the comoving frame\footnote{Note that in general $E_{\hat \alpha}^{i}$ is not a  3-space tensor in the index $i$, so that  $E_{\hat 0}^\| \neq E_{\hat 0 \|}$ and $E_{\hat 0}^{\perp i} \neq \delta^{ij} E_{\hat 0 \perp i}$.
}:
\begin{eqnarray}
E_{\hat \alpha}^i =n^i  E_{\hat \alpha}^\| +E_{\hat \alpha}^{\perp i}\;,  \quad \quad {\rm where} \quad \quad &E_{\hat \alpha}^\| = n_i E_{\hat \alpha}^i &  \quad \quad {\rm and}  \quad \quad E_{\hat \alpha}^{\perp i} = \Perp^i_j E_{\hat \alpha}^j \;,\\
E_{\hat \alpha i} = n_i  E_{\hat \alpha \|} +E_{\hat \alpha \perp i}\;,  \quad \quad {\rm where}  \quad \quad &E_{\hat \alpha \|} = n^i E_{\hat \alpha i}&  \quad \quad {\rm and}  \quad \quad E_{\hat \alpha \perp i} = \Perp^j_i E_{\hat \alpha j} \;.
\end{eqnarray}
Here, up to second order, the scale factor is  
\begin{equation}
\label{a}
a(x^0(\chi))=\bar a \left(1 +\Delta \ln a^{(1)}+\frac{1}{2} \Delta \ln a^{(2)}\right)\;,
\end{equation}
where $\bar a = a(\bar x^0(\bar \chi))=1/(1+z)$, prime is $\p/\p \bar x^0 = \p/\p \eta$ and $\cH=\bar a'/\bar a$.
Then, we have, at first order,
\begin{eqnarray} 
\label{Deltax0-1}
\Delta x^{0(1)} & = & \frac{\Delta \ln a^{(1)}}{ \cH } \;, \\
\label{Dx_||-1}
\Delta x_{\parallel}^{(1)} &=&   \delta x^{0 (1)} + \delta x_{\parallel}^{(1)} - \Delta x^{0 (1)}  \;, \\
\label{Dx_perp-1}
\Delta x_{\perp}^{i(1)} &=& \delta x_{\perp}^{i(1)}\;,
\end{eqnarray}
and, at second order,
\begin{eqnarray} 
\label{Deltax0-2}
\Delta x^{0(2)}& = &\frac{1}{\cH} \Delta \ln a^{(2)}-  \frac{(\cH' + \cH^2)}{\cH^3} \left( \Delta \ln a^{(1)} \right)^2  \;,\\
\label{Dx_||-2}
\Delta x_{\parallel}^{(2)}&=& \delta x^{0 (2)} +  \delta x_{\parallel}^{(2)} -  \Delta x^{0 (2)}+ 2 \left(\delta \nu^{(1)}+ \delta n_{\parallel}^{(1)} \right)\left(  \delta x^{0 (1)}-  \frac{\Delta \ln a^{(1)}}{ \cH }\right)\;, \\
\label{Dx_perp-2}
\Delta x_{\perp}^{i(2)}&=& 2 \delta n_{\perp}^{i(1)}   \delta x^{0 (1)} -  \frac{2}{\cH} \delta n_{\perp}^{i(1)} \Delta \ln a^{(1)} +  \delta x_{\perp}^{i(2)}  \;.
\end{eqnarray}
Here let us also define $\Delta x_{\perp i}^{(n)}=\delta_{ij} \Delta x_{\perp}^{i(n)}$.

The next task is to study the physical number density  of galaxies $n_g$ as a function of the physical comoving coordinates $x^\mu$ and the magnification $\M$.
In particular, we consider the cumulative physical number density sample with a flux larger than a observed limit $\bar \F$ which can be translated  in terms of a the inferred threshold luminosity $\bar L(z)$ ($n_g=N$  in \cite{Challinor:2011bk, DiDio:2013bqa}). 
The  physical number  density  contained within a volume $\mathcal{\bar V}$ is given by
 \begin{eqnarray}
 \label{N}
\mathcal{N}=
&&\int_{ \mathcal{\bar V}}  \sqrt{-g(x^\alpha)}\: n_g(x^\alpha,  \M)\: \varepsilon_{\mu\nu\rho\sigma}
u^\mu(x^\alpha) \frac{\partial x^\nu}{\partial \bar x^1} \frac{\partial x^\rho}{\partial \bar x^2} \frac{\partial x^\sigma}{\partial \bar x^3}
 \ud^3\bar {\bf x}= \nonumber\\
 &&\int_{ \mathcal{\bar V}} \sqrt{-\hat g(x^\alpha)}\: a^3(x^0) n_g(x^\alpha,  \M)\; 
 \varepsilon_{\mu\nu\rho\sigma} \, E_{\hat 0}^\mu(x^\alpha) \frac{\partial x^\nu}{\partial \bar x^1} \frac{\partial x^\rho}{\partial \bar x^2}\frac{\partial x^\sigma}{\partial \bar x^3} \ud^3\bar {\bf x}\;,
\end{eqnarray}
where $\varepsilon_{\mu\nu\rho\sigma}$ is the Levi-Civita tensor, $\sqrt{- \hat g}=\sqrt{- g}/ a^4$, $\hat g^{\mu \nu}$ the comoving metric and 
 $u^\mu=E_{\hat 0}^\mu/a$ is the four velocity vector as a function of comoving location.
In the redshift frame is, by definition,
\begin{equation}
\label{N2}
\mathcal{N} =
\int_{ \mathcal{\bar V}} {\bar a^3(\bar x^0) \,  n_g\left(\bar x^0,\bar {\bf x}, \bar L \right)} \, \ud^3\bar {\bf x} \;.
\end{equation}

Using the cosmic rulers we get
\begin{eqnarray}
\label{deltaSqrtg}
\sqrt{-\hat g(x^\alpha)}=1+\Delta\sqrt{-\hat g (\bar x^\alpha)}^{\,(1)}+\frac{1}{2}\Delta\sqrt{-\hat  g (\bar x^\alpha)}^{\,(2)}\;,
\end{eqnarray}
where\footnote{Here we define $\hat g_\mu^{\mu (n)} =\hat  g^{\mu \nu (0)}\hat  g_{\mu\nu}^{(n)}$.}
\begin{eqnarray}
\label{deltaSqrtg-1}
\Delta\sqrt{-\hat g (\bar x^\alpha)}^{\,(1)}&=&\frac{1}{2} \hat g_\mu^{\mu (1)}(\bar x^\alpha) \;,\\
\label{deltaSqrtg-2}
\Delta \sqrt{-\hat g (\bar x^\alpha)}^{\,(2)}&=& 
\frac{1}{4} \hat g_\mu^{\mu (1)} (\bar x^\alpha) \; \hat g_\nu^{\nu (1)} (\bar x^\alpha)+\frac{1}{2} \hat g_\mu^{\mu (2)} (\bar x^\alpha)-\frac{1}{2}\hat g_\mu^{\nu (1)} (\bar x^\alpha) \; \hat g_\nu^{\mu (1)}  (\bar x^\alpha) + \left(\frac{\p  \hat g_\mu^{\mu}}{\p \bar x^\nu} \right)^{(1)} (\bar x^\alpha) \;  \Delta x^{\nu (1)},
\end{eqnarray}

From Eq. (\ref{a}), 
\begin{equation}
\label{a3}
a^3 = \bar a^3 \left[1 + 3 \; \Delta \ln a^{(1)} + 3 \left(\Delta \ln a^{(1)}\right)^2 + \frac{3}{2} \Delta \ln a^{(2)}\right]\;.
\end{equation}

Defining
\begin{eqnarray}
\label{V}
V(x^\alpha)=\varepsilon_{\mu\nu\rho\sigma} \, E_{\hat 0}^\mu(x^\alpha) \frac{\p x^\nu}{\p \bar x^1} \frac{\p x^\rho}{\p \bar x^2}\frac{\p x^\sigma}{\p \bar x^3}=1+\Delta V (\bar x^\alpha)^{\,(1)}+\frac{1}{2}\Delta  V(\bar x^\alpha)^{\,(2)}\;,
\end{eqnarray}
we find, at first and second order,
\begin{eqnarray}
\label{V1}
\Delta V (\bar x^\alpha)^{\,(1)}&=&E_{\hat 0}^{0(1)} +  E_{\hat 0}^{ \parallel(1)}+   \p_{\parallel} \Delta x_{\parallel}^{(1)}  + \frac{2}{\bar \chi} \Delta x_{\parallel}^{(1)}  - 2 \kappa^{(1)}\;, \\
\label{V2}
\Delta V (\bar x^\alpha)^{\,(2)}&=&E_{\hat 0}^{0(2)} +  E_{\hat 0}^{ \parallel(2)}+   \p_{\parallel} \Delta x_{\parallel}^{(2)}  + \frac{2}{\bar \chi} \Delta x_{\parallel}^{(2)}  - 2 \kappa^{(2)}
+ 2 \left(\kappa^{(1)}\right)^2 -2 \big|\gamma^{(1)}\big|^2+  \vartheta_{ij}^{(1)} \vartheta^{ij(1)}+ \frac{2}{\bar \chi^2} \left( \Delta x_{\parallel}^{(1)} \right)^2  \nonumber \\
&+&    \frac{4}{\bar \chi}  \Delta x_{\parallel}^{(1)} \p_{\parallel} \Delta x_{\parallel}^{(1)}-\frac{4}{\bar \chi}  \Delta x_{\parallel}^{(1)}\kappa^{(1)}  - 4 \kappa^{(1)}  \p_{\parallel} \Delta x_{\parallel}^{(1)}+ \frac{2}{\bar \chi}  \Delta x_{\perp i}^{(1)} \p_{\parallel}   \Delta x_{\perp}^{i (1)} - 2\left( \p_{\parallel}   \Delta x_{\perp}^{i (1)}  \right)  \left( \p_{\perp i}   \Delta x_{\parallel}^{(1)}  \right) \nonumber \\
&+& \frac{2}{\cH}{\left( E_{\hat 0}^{0 (1)}+E_{\hat 0 }^{\| (1)}\right)}' \Delta \ln a^{(1)} +   2\p_{\parallel} \left( E_{\hat 0}^{0 (1)}+E_{\hat 0 }^{\| (1)}\right)\Delta x_{\parallel}^{(1)} + 2 \p_{\perp i}\left( E_{\hat 0}^{0 (1)}+E_{\hat 0 }^{\| (1)}\right)\Delta x_{\perp}^{i (1)}  \nonumber \\
 & +&  2 E_{\hat 0}^{0 (1)}  \p_{\parallel} \Delta x_{\parallel}^{(1)}  -2E_{\hat 0}^{\|(1)} \p_{\parallel}  \Delta x^{0 (1)} - 4 \left( E_{\hat 0}^{0 (1)}+E_{\hat 0 }^{\| (1)}\right) \kappa^{(1)}   -2E_{\hat 0 }^{\perp i (1)} \p_{\perp i}  \left(\Delta x^{0 (1)}+\Delta x_{\parallel}^{(1)}\right)  \nonumber \\
 & +&  \frac{4}{\bar \chi} \left( E_{\hat 0}^{0 (1)}+E_{\hat 0 }^{\| (1)}\right)  \Delta x_{\parallel}^{(1)} \;,
\end{eqnarray}

where 
 \begin{eqnarray}
 \label{kappa-n}
\kappa^{(n)}=- \frac{1}{2}  \p_{\perp i} \Delta x_{\perp}^{i (n)} 
\end{eqnarray}
is  the coordinate weak lensing convergence term at order $n$,
\begin{equation}
\label{shear}
\gamma_{ij}^{(1)}=- \p_{\perp (i}   \Delta x_{\perp j)}^{(1)}-\Perp_{ij}\kappa^{(1)} 
\end{equation}
 is the coordinate weak lensing shear term\footnote{Here $ A_{[i} B_{j]} = \left(A_{i} B_{j}- B_{i}A_{j}\right)/2$ and $A_{(i} B_{j)} = \left(A_{i} B_{j}+ B_{i}A_{j}\right)/2$.}, $\gamma_{ij}^{(1)}\gamma^{ij (1)}=2|\gamma^{(1)}|^2$ and $\vartheta_{ij}^{(1)}=-\p_{\perp [i}   \Delta x_{\perp j]}^{(1)}$.
From Eq.\ (\ref{xph2}), we note that 
 \begin{eqnarray} 
 \label{partialparallep}
\p_{\parallel}  \Delta x^{\mu (n)} (\bar \chi, {\bf n})=   \p_{\bar \chi}  \Delta x^{\mu (n)} \;,
\end{eqnarray}
where $\p_{\bar \chi} $ is applied to all terms that are functions of $\bar x^0=\bar \eta(\bar \chi)$ and/ or $\bar x^i=\bar x^i (\bar \chi)$.

The next subsection will be devoted to obtain at second order the magnification and  Sec. \ref{ng} we will analyze in detail $n_g$ in order to obtain the observed galaxy over-density.

\subsection{Magnification}\label{MB}

The magnification is defined in the following way

\begin{equation}
\label{M}
\M=\left(\frac{\D_A}{\bar \D_A}\right)^{-2}=\left(\frac{\D_L}{\bar \D_L}\right)^{-2}\;,
\end{equation}
 where $\D_L$ is the luminosity distance,  $\D_A$ is the angular distance which are related by
 \begin{equation}
\D_L=(1+z)^2\D_A, \quad \quad {\rm and} \quad\quad \bar  \D_A = \bar a(\bar x^0(\bar \chi)) \bar \chi\;.
\end{equation}
 Here $\bar \D_L$ and  $\bar  \D_A $ are luminosity and angular distance at zero order.
 
Using the cosmic rulers prescription, let us define $(\D_A/\bar \D_A)^{2}$ in the following way  \cite{Jeong:2011as}
\begin{equation}
\label{M-1}
\M^{-1}=\left(\frac{\D_A}{\bar \D_A}\right)^{2}= \frac{\sqrt{-g(x^\alpha)}}{ {\bar a(\bar x^0(\bar \chi))}^2}\: \varepsilon_{\mu\nu\rho\sigma} u^\mu v^\nu  \frac{\partial x^\rho}{\partial \bar x^i} \frac{\partial x^\sigma}{\partial \bar x^j} \alpha^i \beta^j\;,
\end{equation}
where $\{n^i,\alpha^j,\beta^k\}$ is a three dimensional orthonormal basis\footnote{e.g. $n^i=\epsilon^{ijk} \alpha_j \beta_k $, $\alpha^i \alpha_i=\beta^i \beta_i=1$ etc.} and $v^\nu$ is the unit spatial part of the null vector $p^\mu$, orthogonal to $u^\mu$ and directed away from the observer, i.e. 
\begin{equation}
\label{vmu_1}   
v^\mu=\frac{p^\mu}{p^\rho u_\rho}+u^\mu=\frac{k^\mu}{k^\rho u_\rho}+u^\mu\;,
\end{equation}
where we have used $p^\mu = \nu k^\mu/a$. Now, making the change of the variable from $\chi$ to $\bar \chi$, we obtain
\begin{equation} 
k^\mu(\chi)=\frac{\ud x^\mu(\chi)}{\ud  \chi}=\left(\frac{\ud \bar \chi}{\ud \chi}\right) \frac{\ud x^\mu(\chi)}{\ud \bar \chi}\;
\end{equation}
and
\begin{equation}
\label{vmu_2}   
v^\mu=\frac{ \left(\ud x^\mu(\chi)/\ud \bar \chi \right)}{\left(\ud x^\rho (\chi)/\ud \bar \chi \right) u_\rho}+u^\mu\;.
\end{equation}
Here $\ud x^\rho (\chi)/\ud \bar \chi$ is equivalent to $\p_{\bar \chi} x^\rho (\chi(\bar \chi), {\bf n})$ \cite{Bertacca:2014wga, Jeong:2011as}.
Taking in to account that $u^\mu = E_{\hat 0}^\mu/a$, $u_\mu=a E_{\hat 0 \mu}$ and, using Eq.(\ref{vmu_2}),  Eq. (\ref{M-1}) turns out
\begin{eqnarray}
\label{M-1_pert-2}
\M^{-1}&=&\frac{\sqrt{-g(x^\alpha)}}{\left(\ud x^\rho (\chi)/\ud \bar \chi \; u_\rho \right) {\bar a(\bar x^0(\bar \chi))}^2}\: \varepsilon_{\mu\nu\rho\sigma} u^\mu \frac{\ud x^\nu(\chi)}{\ud \bar \chi}  \frac{\partial x^\rho}{\partial \bar x^j} \frac{\partial x^\sigma}{\partial \bar x^k} \alpha^j \beta^k \nonumber \\
&=&\frac{\sqrt{-\hat g(x^\alpha)}}{\left(\ud x^\rho (\chi)/\ud \bar \chi \; E_{\hat 0 \rho} \right)} \left(\frac{a(x^0)}{ {\bar a(\bar x^0)}}\right)^2 \varepsilon_{\mu\nu\rho\sigma} E_{\hat 0}^\mu \frac{\p x^\nu}{\p \bar x^i}  \frac{\partial x^\rho}{\partial \bar x^i} \frac{\partial x^\sigma}{\partial \bar x^j} n^i \alpha^j \beta^k\;.
\end{eqnarray}
We note immediately that Eq. (\ref{M-1_pert-2}) generalizes for any gauges the relation obtained in \cite{Jeong:2011as}. Using the cosmic rulers, 
we will expand the relation to second order  this relation in order to obtain $\M$ and $\D_A$ (or $\D_L$). Let us point out that,  at second order, $\D_L$ is also already obtained with different methods in   \cite{Umeh:2012pn}--\cite{Clarkson:2014pda} and \cite{Yoo:2014sfa}.

Using the relations obtained in Sec. \ref{CR}, it easy to proof  that, up to second order, 
\begin{equation}
 \varepsilon_{\mu\nu\rho\sigma} E_{\hat 0}^\mu \frac{\p x^\nu}{\p \bar x^i}  \frac{\partial x^\rho}{\partial \bar x^i} \frac{\partial x^\sigma}{\partial \bar x^j} n^i \alpha^j \beta^k=1+\Delta V (\bar x^\alpha)^{\,(1)}+\frac{1}{2}\Delta  V(\bar x^\alpha)^{\,(2)}
\end{equation}
and
\begin{equation}
\left(\frac{\ud x^\rho (\chi)}{\ud \bar \chi}E_{\hat 0 \rho} \right)=1+ \Delta\left(\frac{\ud x^\rho (\chi)}{\ud \bar \chi}E_{\hat 0 \rho} \right)^{(1)}+\frac{1}{2}\Delta\left(\frac{\ud x^\rho (\chi)}{\ud \bar \chi}E_{\hat 0 \rho} \right)^{(2)}\;,
\end{equation}
where
\begin{eqnarray}
\Delta\left(\frac{\ud x^\rho (\chi)}{\ud \bar \chi}E_{\hat 0 \rho} \right)^{(1)}&=&  - \p_{\bar \chi} \Delta x^{0(1)} - E_{\hat{0}0}^{(1)} + E_{\hat{0}\|}^{(1)}\;, \\
\Delta\left(\frac{\ud x^\rho (\chi)}{\ud \bar \chi}E_{\hat 0 \rho} \right)^{(2)} &=&  -  \p_{\bar \chi}\Delta x^{0(2)} - E_{\hat{0}0}^{(2)} + E_{\hat{0}\|}^{(2)}  - \frac{2}{\cH}{\left( E_{{\hat 0} 0}^{ (1)} -E_{{\hat 0} \|}^{ (1)}\right)}' \Delta \ln a^{(1)} -   2\p_{\parallel} \left(  E_{{\hat 0} 0}^{ (1)} -E_{{\hat 0} \|}^{ (1)}\right)\Delta x_{\|}^{(1)} \nonumber \\
&& - 2 \p_{\perp i}\left(  E_{{\hat 0}0}^{ (1)} -E_{{\hat 0} \|}^{ (1)}\right)\Delta x_{\perp}^{i (1)}   +2 E_{{\hat 0}0}^{ (1)}  \p_{\bar \chi}\Delta x^{0(1)} +2E_{\hat 0 \, \|}^{(1)}  \p_{\bar \chi}\Delta x_{\|}^{(1)} +2E_{\hat 0 \, \perp i}^{(1)}  \p_{\bar \chi}\Delta x_{\perp}^{i(1)} \nonumber \\
&& -\frac{2}{\bar \chi}E_{\hat 0 \, \perp i}^{(1)}\Delta x_{\perp}^{i(1)} \;.
\end{eqnarray}

Then, Eq.\ (\ref{M-1_pert-2}) yields
\begin{equation}
\label{M-1-pert}
\M^{-1}= 1+\Delta\left(\M^{-1}\right)^{(1)}+\frac{1}{2}\Delta\left(\M^{-1}\right)^{(2)}\;,
\end{equation}
where
\begin{eqnarray}
\label{M-1_1}
\Delta\left(\M^{-1}\right)^{(1)} &=& \frac{1}{2} \hat g_\mu^{\mu (1)} + E_{\hat 0}^{0(1)} +  E_{\hat 0}^{ \parallel(1)}+  E_{{\hat 0} 0}^{ (1)} -E_{{\hat 0} \|}^{ (1)}+ 2 \Delta \ln a^{(1)}  +\p_{\bar \chi}\left(\Delta x^{0(1)} +\Delta x_{\parallel}^{(1)}  \right) + \frac{2}{\bar \chi} \Delta x_{\parallel}^{(1)}  - 2 \kappa^{(1)}\;, \nonumber \\ \\
\label{M-1_2}
\Delta\left(\M^{-1}\right)^{(2)} &=& \frac{1}{2} \hat g_\mu^{\mu (2)} + E_{\hat 0}^{0(2)} +  E_{\hat 0}^{ \parallel(2)}+  E_{{\hat 0} 0}^{(2)} -E_{{\hat 0} \|}^{(2)}+ 2 \Delta \ln a^{(2)}  +\p_{\bar \chi}\left(\Delta x^{0(2)} +\Delta x_{\|}^{(2)}  \right) + \frac{2}{\bar \chi} \Delta x_{\parallel}^{(2)}  - 2 \kappa^{(2)} , \nonumber \\
&&+ \frac{1}{4} \hat g_\mu^{\mu (1)} \; \hat g_\nu^{\nu (1)} -\frac{1}{2}\hat g_\mu^{\nu (1)}  \; \hat g_\nu^{\mu (1)}  +2   \left(\kappa^{(1)}\right)^2 -4 \kappa^{(1)}  \p_{\bar \chi} \Delta x_{\parallel}^{(1)} -  \frac{4}{\bar \chi}  \Delta x_{\parallel}^{(1)}  \kappa^{(1)} - 4 \left( E_{\hat 0}^{0 (1)}+E_{\hat 0 }^{\| (1)}\right) \kappa^{(1)} \nonumber \\
&& + 2 \left( E_{\hat 0}^{0 (1)}+E_{\hat 0 }^{\| (1)}+ E_{{\hat 0} 0}^{ (1)} -E_{{\hat 0} \|}^{ (1)}+ \frac{2}{\bar \chi}  \Delta x_{\parallel}^{(1)} \right) \p_{\bar \chi} \Delta x_{\|}^{(1)} -2\left( E_{\hat 0 0}^{(1)}+  E_{\hat 0 }^{\| (1)}\right) \p_{\bar \chi}\left(\Delta x^{0(1)} +\Delta x_{\parallel}^{(1)}  \right)  \nonumber \\
&& + \frac{2}{\bar \chi^2} \left( \Delta x_{\|}^{(1)} \right)^2  + 2 \left[  E_{\hat 0}^{0(1)} +  E_{\hat 0}^{ \parallel(1)}+  E_{{\hat 0} 0}^{ (1)} -E_{{\hat 0} \|}^{ (1)}  +\p_{\bar \chi}\left(\Delta x^{0(1)} +\Delta x_{\parallel}^{(1)}  \right) + \frac{2}{\bar \chi} \Delta x_{\parallel}^{(1)}  - 2 \kappa^{(1)}\right]\nonumber \\
&&\times  \left(\frac{1}{2} \hat g_\mu^{\mu (1)} +E_{{\hat 0} 0}^{ (1)} -E_{{\hat 0} \|}^{ (1)}+ \p_{\bar \chi}\Delta x^{0(1)}\right) + 2\p_{\parallel} \left( \frac{1}{2} \hat g_\mu^{\mu (1)}+ E_{\hat 0}^{0 (1)}+E_{\hat 0 }^{\| (1)}+E_{{\hat 0} 0}^{ (1)} -E_{{\hat 0} \|}^{ (1)}\right)\Delta x_{\parallel}^{(1)}\nonumber \\
&&  -2 \big|\gamma^{(1)}\big|^2+  \vartheta_{ij}^{(1) }\vartheta^{ij(1)}+2\left(\frac{1}{\bar \chi}  \Delta x_{\perp i}^{(1)} -\p_{\perp i}   \Delta x_{\parallel}^{(1)} \right) \p_{\bar \chi}   \Delta x_{\perp}^{i (1)} +\frac{4}{\bar \chi} \left( E_{\hat 0}^{0 (1)}+E_{\hat 0 }^{\| (1)}\right)  \Delta x_{\|}^{(1)} \nonumber \\
&&+ 2 \p_{\perp i}\left( \frac{1}{2} \hat g_\mu^{\mu (1)}+ E_{\hat 0}^{0 (1)}+E_{\hat 0 }^{\| (1)} + E_{{\hat 0} 0}^{ (1)} -E_{{\hat 0} \|}^{ (1)}\right)\Delta x_{\perp}^{i (1)} 
+2 E_{{\hat 0} \perp i}^{ (1)} \left(\frac{1}{\bar \chi}\Delta x_{\perp}^{i (1)} - \p_{\bar \chi} \Delta x_{\perp}^{i (1)} \right)  \nonumber \\
&& -2 E_{\hat 0 }^{\perp i (1)} \p_{\perp i}\left(\Delta x^{0(1)} +\Delta x_{\parallel}^{(1)}  \right)  + \frac{2}{\cH}{\left(  \frac{1}{2} \hat g_\mu^{\mu (1)}+ E_{\hat 0}^{0 (1)}+E_{\hat 0 }^{\| (1)}+ E_{{\hat 0} 0}^{ (1)} -E_{{\hat 0} \|}^{ (1)}\right)}' \Delta \ln a^{(1)}  \nonumber \\
&& + 4 \left[\frac{1}{2} \hat g_\mu^{\mu (1)} + E_{\hat 0}^{0(1)} +  E_{\hat 0}^{ \parallel(1)}+  E_{{\hat 0} 0}^{ (1)} -E_{{\hat 0} \|}^{ (1)}  +\p_{\bar \chi}\left(\Delta x^{0(1)} +\Delta x_{\parallel}^{(1)}  \right) + \frac{2}{\bar \chi} \Delta x_{\parallel}^{(1)}  - 2 \kappa^{(1)}\right]\Delta \ln a^{(1)}  \nonumber \\
&& + 2\left(\Delta \ln a^{(1)} \right)^2\;.
\end{eqnarray}

Consequently the magnification turns out
\begin{equation}
\M= 1+\Delta \M^{(1)}+\frac{1}{2}\Delta \M^{(2)}\;,
\end{equation}
where
\begin{eqnarray}
\label{M_1}
\Delta \M^{(1)} &=&-\Delta\left(\M^{-1}\right)^{(1)} =- \frac{1}{2} \hat g_\mu^{\mu (1)} - E_{\hat 0}^{0(1)} -  E_{\hat 0}^{ \parallel(1)}-  E_{{\hat 0} 0}^{ (1)} +E_{{\hat 0} \|}^{ (1)}- 2 \Delta \ln a^{(1)}  -\p_{\bar \chi}\left(\Delta x^{0(1)} +\Delta x_{\|}^{(1)}  \right) - \frac{2}{\bar \chi} \Delta x_{\|}^{(1)}   \nonumber \\
&&+ 2 \kappa^{(1)} \;, \\
\label{M_2}
\Delta \M^{(2)} &=&-\Delta\left(\M^{-1}\right)^{(2)} +2\left[\Delta\left(\M^{-1}\right)^{(1)}\right]^2 =- \frac{1}{2} \hat g_\mu^{\mu (2)} - E_{\hat 0}^{0(2)} -  E_{\hat 0}^{ \|(2)}-  E_{{\hat 0} 0}^{ (2)} +E_{{\hat 0} \|}^{ (2)}- 2 \Delta \ln a^{(2)}    - \frac{2}{\bar \chi} \Delta x_{\|}^{(2)}  \nonumber \\
&& -\p_{\bar \chi}\left(\Delta x^{0(2)} +\Delta x_{\|}^{(2)}  \right) + 2 \kappa^{(2)} +\frac{1}{4} \left(  \hat g_\mu^{\mu (1)}\right)^2 +\frac{1}{2}\hat g_\mu^{\nu (1)}  \; \hat g_\nu^{\mu (1)} +\frac{6}{\bar \chi^2} \left( \Delta x_{\|}^{(1)} \right)^2  +2 \big|\gamma^{(1)}\big|^2  -  \vartheta_{ij}^{(1) }\vartheta^{ij(1)} \nonumber \\
&& +2 \left[  E_{\hat 0}^{0(1)} +  E_{\hat 0}^{ \parallel(1)}  + \frac{2}{\bar \chi} \Delta x_{\parallel}^{(1)}\right]\left[ E_{\hat 0}^{0(1)} +  E_{\hat 0}^{ \parallel(1)}  +  E_{{\hat 0} 0}^{ (1)} -E_{{\hat 0} \|}^{ (1)} +\p_{\bar \chi}\left(\Delta x^{0(1)} +\Delta x_{\parallel}^{(1)}  \right) \right]  \nonumber \\
&& +2  \p_{\bar \chi}\left(\Delta x^{0(1)} +\Delta x_{\parallel}^{(1)}  \right) \p_{\bar \chi}\Delta x_{\parallel}^{(1)}  + \hat g_\mu^{\mu (1)}\left[ E_{\hat 0}^{0(1)} +  E_{\hat 0}^{ \parallel(1)} +  E_{{\hat 0} 0}^{ (1)} -E_{{\hat 0} \|}^{ (1)}+\p_{\bar \chi}\left(\Delta x^{0(1)} +\Delta x_{\parallel}^{(1)}  \right)  + \frac{2}{\bar \chi} \Delta x_{\parallel}^{(1)}\right] \nonumber \\
 &&+ 2\left( E_{\hat 0 0}^{(1)}+  E_{\hat 0 }^{\| (1)}\right) \p_{\bar \chi}\left(\Delta x^{0(1)} +\Delta x_{\parallel}^{(1)}  \right) -2\p_{\parallel} \left( \frac{1}{2} \hat g_\mu^{\mu (1)}+ E_{\hat 0}^{0 (1)}+E_{\hat 0 }^{\| (1)}+E_{{\hat 0} 0}^{ (1)} -E_{{\hat 0} \|}^{ (1)}\right)\Delta x_{\parallel}^{(1)}\nonumber \\
&& - 4 \kappa^{(1)}\left[ \frac{1}{2} \hat g_\mu^{\mu (1)} + E_{\hat 0}^{0(1)} +  E_{\hat 0}^{ \parallel(1)} +  E_{{\hat 0} 0}^{ (1)} -E_{{\hat 0} \|}^{ (1)} +\p_{\bar \chi}\left(\Delta x^{0(1)} +\Delta x_{\parallel}^{(1)}  \right) \right]  -  \frac{12}{\bar \chi} \kappa^{(1)}  \Delta x_{\parallel}^{(1)} +6 \left(\kappa^{(1)}\right)^2\nonumber \\
 &&  - 2\left(\frac{1}{\bar \chi}  \Delta x_{\perp i}^{(1)} -\p_{\perp i}   \Delta x_{\parallel}^{(1)} \right) \p_{\bar \chi}   \Delta x_{\perp}^{i (1)}  - 2 \p_{\perp i}\left( \frac{1}{2} \hat g_\mu^{\mu (1)}+ E_{\hat 0}^{0 (1)}+E_{\hat 0 }^{\| (1)} + E_{{\hat 0} 0}^{ (1)} -E_{{\hat 0} \|}^{ (1)}\right)\Delta x_{\perp}^{i (1)} \nonumber \\
 && -2 E_{{\hat 0} \perp i}^{ (1)} \left(\frac{1}{\bar \chi}\Delta x_{\perp}^{i (1)} - \p_{\bar \chi} \Delta x_{\perp}^{i (1)} \right) +2 E_{\hat 0 }^{\perp i (1)} \p_{\perp i}\left(\Delta x^{0(1)} +\Delta x_{\parallel}^{(1)}  \right) \nonumber \\
 && - \frac{2}{\cH}{\left(  \frac{1}{2} \hat g_\mu^{\mu (1)}+ E_{\hat 0}^{0 (1)}+E_{\hat 0 }^{\| (1)}+ E_{{\hat 0} 0}^{ (1)} -E_{{\hat 0} \|}^{ (1)}\right)}' \Delta \ln a^{(1)}   +6\left(\Delta \ln a^{(1)} \right)^2 \nonumber \\
&& + 4 \left[\frac{1}{2} \hat g_\mu^{\mu (1)} + E_{\hat 0}^{0(1)} +  E_{\hat 0}^{ \parallel(1)}+  E_{{\hat 0} 0}^{ (1)} -E_{{\hat 0} \|}^{ (1)}  +\p_{\bar \chi}\left(\Delta x^{0(1)} +\Delta x_{\parallel}^{(1)}  \right) + \frac{2}{\bar \chi} \Delta x_{\parallel}^{(1)}  - 2 \kappa^{(1)}\right]\Delta \ln a^{(1)}\;.
\end{eqnarray}

In similar way, we can obtain the luminosity distance
\begin{equation}
\frac{\D_L}{\bar \D_L}= 1+\frac{\D_L^{(1)}}{\bar \D_L}+\frac{1}{2}\frac{\D_L^{(2)}}{\bar \D_L}\;,
\end{equation}
where
\begin{eqnarray}
\label{DL_1}
\frac{\D_L^{(1)}}{\bar \D_L} &=&\frac{\D_A^{(1)}}{\bar \D_A}=\frac{1}{2}\Delta\left(\M^{-1}\right)^{(1)} =-\frac{1}{2}\Delta\M^{(1)}= \frac{1}{4} \hat g_\mu^{\mu (1)} +\frac{1}{2} E_{\hat 0}^{0(1)} + \frac{1}{2} E_{\hat 0}^{ \parallel(1)}+  \frac{1}{2}E_{{\hat 0} 0}^{ (1)} -\frac{1}{2} E_{{\hat 0} \|}^{ (1)}+ \Delta \ln a^{(1)}    \nonumber \\
&&+\frac{1}{2}\p_{\bar \chi}\left(\Delta x^{0(1)} +\Delta x_{\parallel}^{(1)}  \right)+ \frac{1}{\bar \chi} \Delta x_{\parallel}^{(1)}  - \kappa^{(1)} 
\end{eqnarray}
and
\begin{eqnarray}
\label{DL_2}
\frac{\D_L^{(2)}}{\bar \D_L} = \frac{\D_A^{(2)}}{\bar \D_A}&=&\frac{1}{2}\Delta\left(\M^{-1}\right)^{(2)} -\frac{1}{4}\left[\Delta\left(\M^{-1}\right)^{(1)}\right]^2= - \frac{1}{2}\Delta \M^{(2)} +\frac{3}{4}\left(\Delta \M^{(1)}\right)^2= \frac{1}{4} \hat g_\mu^{\mu (2)} + \frac{1}{2} E_{\hat 0}^{0(2)} +\frac{1}{2}  E_{\hat 0}^{ \|(2)}  \nonumber \\
&& +\frac{1}{2}  E_{{\hat 0} 0}^{(2)} - \frac{1}{2} E_{{\hat 0} \|}^{(2)}+  \Delta \ln a^{(2)}  +\frac{1}{2} \p_{\bar \chi}\left(\Delta x^{0(2)} +\Delta x_{\|}^{(2)}  \right) + \frac{1}{\bar \chi} \Delta x_{\parallel}^{(2)}  -  \kappa^{(2)} + \frac{1}{16} \hat g_\mu^{\mu (1)} \; \hat g_\nu^{\nu (1)} \nonumber \\
&& -\frac{1}{4}\hat g_\mu^{\nu (1)}  \; \hat g_\nu^{\mu (1)}  -  \kappa^{(1)} \left[E_{\hat 0}^{0 (1)}+E_{\hat 0 }^{\| (1)}+E_{{\hat 0} 0}^{ (1)} -E_{{\hat 0} \|}^{ (1)}+\p_{\bar \chi}\left(\Delta x^{0(1)} +\Delta x_{\parallel}^{(1)}  \right)\right]   -\frac{1}{2} \hat g_\mu^{\mu (1)} \kappa^{(1)} \nonumber \\
&& +\frac{1}{4} \hat g_\mu^{\mu (1)} \left[  E_{\hat 0}^{0(1)} +  E_{\hat 0}^{ \parallel(1)}+  E_{{\hat 0} 0}^{ (1)} -E_{{\hat 0} \|}^{ (1)} +\p_{\bar \chi}\left(\Delta x^{0(1)} +\Delta x_{\parallel}^{(1)}  \right) + \frac{2}{\bar \chi} \Delta x_{\parallel}^{(1)} \right]  \nonumber \\
&&    - \p_{\bar \chi}\Delta x_{\parallel}^{(1)}\p_{\bar \chi}\left(\Delta x^{0(1)} +\Delta x_{\parallel}^{(1)}  \right)   -\left( E_{\hat 0 0}^{(1)}+  E_{\hat 0 }^{\| (1)}\right) \p_{\bar \chi}\left(\Delta x^{0(1)} +\Delta x_{\parallel}^{(1)}  \right)   - \big|\gamma^{(1)}\big|^2+ \frac{1}{2} \vartheta_{ij}^{(1) }\vartheta^{ij(1)}  \nonumber \\
&& +\frac{1}{\bar \chi} \Delta x_{\parallel}^{(1)} \left[  E_{\hat 0}^{0 (1)}+E_{\hat 0 }^{\| (1)}+ E_{{\hat 0} 0}^{ (1)} -E_{{\hat 0} \|}^{ (1)}+\p_{\bar \chi}\left(\Delta x^{0(1)} +\Delta x_{\parallel}^{(1)}  \right)   \right] +  \frac{1}{4}\bigg[-\left(E_{\hat 0}^{0(1)} +  E_{\hat 0}^{ \parallel(1)}\right)  \nonumber \\
&&+3 \left(E_{{\hat 0} 0}^{ (1)} -E_{{\hat 0} \|}^{ (1)}\right) + 3\p_{\bar \chi}\left(\Delta x^{0(1)} +\Delta x_{\parallel}^{(1)}  \right)\bigg] \bigg[E_{\hat 0}^{0(1)} +  E_{\hat 0}^{ \parallel(1)}+  E_{{\hat 0} 0}^{ (1)} -E_{{\hat 0} \|}^{ (1)}+\p_{\bar \chi}\left(\Delta x^{0(1)} +\Delta x_{\parallel}^{(1)}  \right)\bigg] \nonumber \\
&&  + \p_{\parallel} \left( \frac{1}{2} \hat g_\mu^{\mu (1)}+ E_{\hat 0}^{0 (1)}+E_{\hat 0 }^{\| (1)} +E_{{\hat 0} 0}^{ (1)} -E_{{\hat 0} \|}^{ (1)}\right)\Delta x_{\parallel}^{(1)}  +\left(\frac{1}{\bar \chi}  \Delta x_{\perp i}^{(1)} -\p_{\perp i}   \Delta x_{\parallel}^{(1)} \right) \p_{\bar \chi}   \Delta x_{\perp}^{i (1)}  \nonumber \\
&&+  \p_{\perp i}\left( \frac{1}{2} \hat g_\mu^{\mu (1)}+ E_{\hat 0}^{0 (1)}+E_{\hat 0 }^{\| (1)} + E_{{\hat 0} 0}^{ (1)} -E_{{\hat 0} \|}^{ (1)}\right)\Delta x_{\perp}^{i (1)} + E_{{\hat 0} \perp i}^{ (1)} \left(\frac{1}{\bar \chi}\Delta x_{\perp}^{i (1)} - \p_{\bar \chi} \Delta x_{\perp}^{i (1)} \right)  \nonumber \\
&& - E_{\hat 0 }^{\perp i (1)} \p_{\perp i}\left(\Delta x^{0(1)} +\Delta x_{\parallel}^{(1)}  \right)  + \frac{1}{\cH}{\left(  \frac{1}{2} \hat g_\mu^{\mu (1)}+ E_{\hat 0}^{0 (1)}+E_{\hat 0 }^{\| (1)}+ E_{{\hat 0} 0}^{ (1)} -E_{{\hat 0} \|}^{ (1)}\right)}' \Delta \ln a^{(1)} \nonumber \\
&& +  \left[\frac{1}{2} \hat g_\mu^{\mu (1)} + E_{\hat 0}^{0(1)} +  E_{\hat 0}^{ \parallel(1)}+  E_{{\hat 0} 0}^{ (1)} -E_{{\hat 0} \|}^{ (1)}  +\p_{\bar \chi}\left(\Delta x^{0(1)} +\Delta x_{\parallel}^{(1)}  \right) + \frac{2}{\bar \chi} \Delta x_{\parallel}^{(1)}  - 2 \kappa^{(1)}\right]\Delta \ln a^{(1)}\;.  \nonumber \\
\end{eqnarray}

\section{The observed over-density with magnification bias}
\label{ng}

Using the results obtained above, in this section we analyze in detail $n_g$ and, finally, we present the galaxy number over-density  up to second order with all relativistic effects that arise from observing on the past lightcone and as function of the magnification bias (see also \cite{Yoo:2014sfa}).

Expanding, $n_g(x^\mu,\M)$, we find
\begin{eqnarray}
\label{expng}
n_g(x^\alpha,\M)&=&n_g(\bar x^0+ \Delta x^0, \bar\M+\Delta \M)^{(0)}+n_g(\bar x^\alpha+ \Delta x^\alpha ,  \bar\M+\Delta \M)^{(1)}+\frac{1}{2}n_g(\bar x^\alpha ,  \bar\M)^{(2)}\nonumber\\
&=& \bar n_g+\frac{\p \bar n_g}{\p \M} \Delta \M^{(1)}+\frac{1}{2} \frac{\p \bar n_g}{\p \M}\Delta \M^{(2)}+\frac{1}{2} \frac{\p^2 \bar n_g}{{\p \M}^2} \left( \Delta \M^{(1)}\right)^2+\frac{\p \bar n_g}{\p \bar x^0} \, \Delta  x^{0(1)}+\frac{\p^2 \bar n_g}{\p \M \p \bar x^0} \, \Delta x^{0(1)} \Delta \M^{(1)}\nonumber \\
&&+\frac{1}{2}\frac{\p \bar n_g}{\p \bar x^0} \, \Delta \bar x^{0(2)}+\frac{1}{2} \frac{\p^2 \bar n_g}{{\p \bar x^{0}}^2} \left(  \Delta x^{0(1)}\right)^2+n_g^{(1)}+\frac{\p n_g^{(1)}}{\p \M} \Delta \M^{(1)}+\frac{\p  n_g^{(1)}}{\p \bar x^0} \, \Delta  x^{0(1)}+\frac{\p  n_g^{(1)}}{\p \bar x^i} \, \Delta  x^{i(1)} +\frac{1}{2}n_g^{(2)}\;, \nonumber\\
\end{eqnarray}
where
$\bar n_g={n_g(\bar x^0, \bar L)}^{(0)}$ is the background number density of sources with luminosity exceeding $\bar L$ and, from the second line, $n_g^{(1)}={n_g(\bar x^\alpha, \bar L)}^{(1)}$ and $n_g^{(2)}={n_g(\bar x^\alpha, \bar L)}^{(2)}$.
Now, we define  
\begin{equation}
\delta_g^{(1)}=\delta_g(\bar x^\alpha, \bar L)^{(1)}=\frac{{n_g(\bar x^\alpha, \bar L)}^{(1)}}{{n_g(\bar x^0, \bar L)}^{(0)}} \quad \quad {\rm and} \quad \quad \delta_g^{(2)}=\delta_g(\bar x^\alpha, \bar L)^{(2)}=\frac{{n_g(\bar x^\alpha, \bar L)}^{(2)}}{{n_g(\bar x^0, \bar L)}^{(0)}}\;.
\end{equation}
Knowing that $\bar a=a(\bar x^0)$,
\begin{equation}
\frac{\p \bar n_g}{\p \bar x^0} (\bar x^0, \bar L)=\frac{\p \bar n_g}{\p \bar x^0}\bigg|_{\bar L}=\cH\frac{\p \bar n_g}{\p \ln \bar a}\bigg|_{\bar L}= \cH  \frac{\p\bar n_g}{\p \ln \bar a} (\bar a, \bar L)\;,
\end{equation}
$\Delta \M^{(1)} = -\Delta (\M^{-1})^{(1)} $ and $\Delta \M^{(2)} =-\Delta (\M^{-1})^{(2)} +2 [\Delta(\M^{-1})^{(1)}]^2$, we find
\begin{equation}
\label{Deltang}
n_g(x^\alpha,\M)=\bar n_g(\bar x^0, \bar L) +\Delta n_g(\bar x^\alpha, \bar L)^{(1)}+\frac{1}{2}\Delta n_g(\bar x^\alpha, \bar L)^{(2)}
\end{equation}
where
\begin{eqnarray}
\frac{\Delta n_g(\bar x^\alpha, \bar L)^{(1)}}{\bar n_g}&=& \delta_g^{(1)} + \frac{\p \ln \bar n_g}{\p \ln \bar a} \Delta \ln a^{(1)} - \Q \, \Delta \left(\M^{-1}\right)^{(1)} \;,\\
\frac{\Delta n_g(\bar x^\alpha, \bar L)^{(2)}}{\bar n_g}&=& \delta_g^{(2)} + \frac{\p \ln \bar n_g}{\p \ln \bar a} \Delta \ln a^{(2)} - \Q \, \Delta \left(\M^{-1}\right)^{(2)} + \left(2\Q+\Q^2-\frac{\p \Q}{\p \bar L}\right)  \left[ \Delta\left(\M^{-1}\right)^{(1)}\right]^2\nonumber\\
&& + 2 \p_{\parallel}\delta_g^{(1)} \Delta x_{\parallel}^{(1)} +  2 \p_{\perp i} \delta_g^{(1)}  \Delta x_{\perp }^{i(1)} -2\Q \, \delta_g^{(1)} \Delta\left(\M^{-1}\right)^{(1)}  -2\Q^{(1)} \Delta\left(\M^{-1}\right)^{(1)} \nonumber\\
&&+2 \bigg[-\left(\Q\frac{\p \ln \bar n_g}{\p \ln \bar a}+  \frac{\p \Q}{\p \ln \bar a}\right) \Delta\left(\M^{-1}\right)^{(1)} + \frac{\p \ln \bar n_g}{\p \ln \bar a} \delta_g^{(1)} \bigg] \Delta \ln a^{(1)}  +  \frac{2}{\cH}  \delta_g^{(1)}{'} \Delta \ln a^{(1)}   \nonumber \\
&&+ \left[-\frac{\p \ln \bar n_g}{\p \ln \bar a} + \left(\frac{\p \ln \bar n_g}{\p \ln \bar a}\right)^2+\frac{\p^2 \ln \bar n_g}{{\p \ln \bar a}^2} \right]  \left( \Delta \ln a^{(1)}\right)^2\;.
\end{eqnarray}
Here  we have defined the background and the first order magnification bias
\begin{equation}
\Q(\bar x^0, \bar L)=\frac{ \p \ln \bar n_g}{\p \M}\bigg|_{\bar a} =-\frac{\p \ln \bar n_g}{\p \ln \bar L}\bigg|_{\bar a} \quad \quad {\rm and} \quad \quad \Q^{(1)}(\bar x^\alpha, \bar L)=\frac{ \p  \delta_g^{(1)} }{\p \M}\Bigg|_{\bar a} =-\frac{\p  \delta_g^{(1)}}{\p \ln \bar L}\Bigg|_{\bar a}\;.
\end{equation}
Let us point out that usually  $\left({\p \Q}/{\p \ln \bar a}\right)$ and $\Q^{(1)}$ are not generally considered in the literature (for example see \cite{Yoo:2014sfa}). 

Then, from Eqs.\ (\ref{N}), (\ref{N2}), (\ref{deltaSqrtg}), (\ref{a3}), (\ref{V}), (\ref{M-1}) and (\ref{Deltang}), we  obtain the observed fractional number over-density
\begin{equation}
\label{Delta_g}
\Delta_g=\frac{ n_g (\bar x^0, \bar{\bf x}, \bar L) - \bar n_g(\bar x^0, \bar L)}{ \bar n_g(\bar x^0, \bar L)} = \Delta_g^{(1)} + \frac{1}{2}\Delta_g^{(2)}\;,
\end{equation}
where\footnote{Here we have used the following relation
\begin{eqnarray}
\label{pchiDx0+Dx||}
\p_{\bar \chi}\left(\Delta x^{0(1)} +\Delta x_{\|}^{(1)}  \right)=\frac{\ud}{\ud \bar \chi}\left[\delta x^{0(1)}(\bar \chi) +\delta x_{\|}^{(1)} (\bar \chi) \right]=\delta\nu^{(1)}+\delta n_{\|}^{(1)}\;.
\end{eqnarray}}
\begin{eqnarray}
\label{Deltag-1}
\Delta_g^{(1)} &=& \frac{\Delta n_g^{(1)}}{\bar n_g}+ 3 \Delta \ln a^{(1)} + \Delta \sqrt{-g}^{(1)}+\Delta V^{(1)}=\delta_g^{(1)} + \frac{\left(1- \Q \right)}{2} \hat g_\mu^{\mu (1)} +\left( b_e-2 \Q\right) \, \Delta \ln a^{(1)}  +   \p_{\parallel} \Delta x_{\parallel}^{(1)}  \nonumber \\
&&+ \frac{2\left(1- \Q \right)}{\bar \chi} \Delta x_{\parallel}^{(1)}  - 2 \left(1- \Q \right) \kappa^{(1)} +  \left(1- \Q \right) \left(E_{\hat 0}^{0(1)} +  E_{\hat 0}^{ \parallel(1)}\right) - \Q \left(E_{{\hat 0} 0}^{ (1)} -E_{{\hat 0} \|}^{ (1)}+\delta\nu^{(1)}+\delta n_{\|}^{(1)} \right)\;,  
\end{eqnarray}
and 
\begin{eqnarray}
\label{Deltag-2}
&&\Delta_g^{(2)} = \frac{\Delta n_g^{(2)}}{\bar n_g}+ 3 \Delta \ln a^{(2)} + \Delta \sqrt{-g}^{(2)}+\Delta V^{(2)} + 6 \left( \frac{\Delta n_g^{(1)}}{\bar n_g}+ \Delta \sqrt{-g}^{(1)}+\Delta V^{(1)}\right) \, \Delta \ln a^{(1)}+ 6 \left( \Delta \ln a^{(1)}\right)^2\nonumber \\
&&+ 2  \frac{\Delta n_g^{(1)}}{\bar n_g}  \Delta V^{(1)} + 2 \Delta \sqrt{-g}^{(1)} \Delta V^{(1)}+ 2  \frac{\Delta n_g^{(1)}}{\bar n_g}  \Delta \sqrt{-g}^{(1)}  \nonumber \\
&& = \delta^{(2)}+ b_e \Delta \ln a^{(2)} + \Delta \sqrt{-g}^{(2)}+\Delta V^{(2)} - \Q \, \Delta \left(\M^{-1}\right)^{(2)}+ 2\bigg[ b_e\delta^{(1)}+\frac{1}{\cH} {\delta^{(1)}}'+  b_e\Delta V^{(1)} + b_e \Delta \sqrt{-g}^{(1)}  \nonumber \\
&&-  \left(Qb_e+\frac{\p \Q}{\p \ln \bar a}\right)  \Delta\left(\M^{-1}\right)^{(1)}\bigg] \, \Delta \ln a^{(1)}  + \left(-b_e+b_e^2 + \frac{\p b_e}{\p \ln \bar a}\right) \left( \Delta \ln a^{(1)}\right)^2 -2\Q^{(1)}  \Delta \left(\M^{-1}\right)^{(1)} \nonumber \\
&& + \left(2\Q+\Q^2-\frac{\p \Q}{\p \ln \bar L}\right)\left[\Delta \left(\M^{-1}\right)^{(1)}\right]^2 -2 \Q \delta^{(1)} \Delta \left(\M^{-1}\right) +2 \p_\|\delta^{(1)} \Delta x_\|^{(1)} +  2 \p_{\perp i} \delta_g^{(1)}  \Delta x_{\perp }^{i(1)}  \nonumber \\
&&-2\Q\Delta \left(\M^{-1}\right)^{(1)}  \Delta V^{(1)}+2  \delta^{(1)} \Delta V^{(1)} - 2 \Q \Delta \sqrt{-g}^{(1)} \Delta \left(\M^{-1}\right)^{(1)} + 2 \Delta \sqrt{-g}^{(1)} \Delta V^{(1)}  + 2  \delta^{(1)} \Delta \sqrt{-g}^{(1)} \label{Deltag-2_2} \nonumber \\
&& =\delta_g^{(2)} + \frac{\left(1- \Q \right)}{2} \hat g_\mu^{\mu (2)} +\left( b_e-2 \Q\right) \, \Delta \ln a^{(2)}  +   \p_{\parallel} \Delta x_{\parallel}^{(2)}    + \frac{2\left(1- \Q \right)}{\bar \chi} \Delta x_{\parallel}^{(2)}  - 2 \left(1- \Q \right) \kappa^{(2)}  \nonumber \\
&&+  \left(1- \Q \right) \left(E_{\hat 0}^{0(2)} +  E_{\hat 0}^{ \parallel(2)}\right) - \Q \left(E_{{\hat 0} 0}^{ (2)} -E_{{\hat 0} \|}^{ (2)}\right)- \Q \,\p_{\bar \chi}\left(\Delta x^{0(2)} +\Delta x_{\|}^{(2)}  \right)   +\frac{2}{\cH}  \delta_g^{(1)}{'} \Delta \ln a^{(1)}  + 2 \p_{\parallel}\delta_g^{(1)} \Delta x_{\parallel}^{(1)}  \nonumber \\
&&+  2 \p_{\perp}^{i} \delta_g^{(1)}  \Delta x_{\perp i}^{(1)}  +2 \delta_g^{(1)} \p_{\bar \chi} \Delta x_{\parallel}^{(1)}  +\left(1-\Q +\Q^2- \frac{\p \Q}{\p \ln \bar L}\right) \bigg[\frac{1}{4} \, \hat g_\mu^{\mu (1)} \; \hat g_\nu^{\nu (1)} + \frac{4}{\bar \chi} \left( E_{\hat 0}^{0 (1)}+E_{\hat 0 }^{\| (1)}\right)  \Delta x_{\|}^{(1)} \nonumber \\
&& - 4 \left( E_{\hat 0}^{0 (1)}+E_{\hat 0 }^{\| (1)}\right)  \kappa^{(1)}  +  \hat g_\mu^{\mu (1)}\left( E_{\hat 0}^{0 (1)}+E_{\hat 0 }^{\| (1)}\right) +  \frac{2}{\bar \chi} \Delta x_{\parallel}^{(1)}  \hat g_\mu^{\mu (1)} -2  \hat g_\mu^{\mu (1)} \kappa^{(1)} \bigg] +  \left(1- \Q \right)\bigg[ -\frac{1}{2}\hat g_\mu^{\nu (1)}  \; \hat g_\nu^{\mu (1)}  \nonumber \\
&&+    \frac{1}{\cH} \hat g_\mu^{\mu (1)}{'} \Delta \ln a^{(1)} + \left(\p_{\parallel} \hat g_\mu^{\mu (1)}\right) \Delta x_{\parallel}^{(1)} +   \left(\p_{\perp i}\hat g_\mu^{\mu (1)}\right) \Delta x_{\perp}^{i (1)}  -4 \kappa^{(1)}  \p_{\bar \chi} \Delta x_{\parallel}^{(1)} +   \frac{4}{\bar \chi}  \Delta x_{\parallel}^{(1)} \p_{\bar \chi} \Delta x_{\parallel}^{(1)} -2 \big|\gamma^{(1)}\big|^2  \nonumber \\
&&  +  \vartheta_{ij}^{(1) }\vartheta^{ij(1)} + \frac{2}{\bar \chi}  \Delta x_{\perp i}^{(1)} \left( \p_{\bar \chi}  \Delta x_{\perp}^{i (1)}  \right)- 2\left( \p_{\bar \chi}   \Delta x_{\perp}^{i (1)}  \right)  \left( \p_{\perp i}   \Delta x_{\parallel}^{(1)}  \right)+ \frac{2}{\cH}{\left( E_{\hat 0}^{0 (1)}+E_{\hat 0 }^{\| (1)}\right)}' \Delta \ln a^{(1)} \nonumber \\
&& +   2\p_{\parallel} \left( E_{\hat 0}^{0 (1)}+E_{\hat 0 }^{\| (1)}\right)\Delta x_{\parallel}^{(1)}  + 2 \p_{\perp i}\left( E_{\hat 0}^{0 (1)}+E_{\hat 0 }^{\| (1)}\right)\Delta x_{\perp}^{i (1)}-2E_{\hat 0 }^{\| (1)}  \left(\delta\nu^{(1)}+\delta n_{\|}^{(1)} \right) + 2\left( E_{\hat 0}^{0 (1)}+E_{\hat 0 }^{\| (1)}\right)  \p_{\bar \chi} \Delta x_{\parallel}^{(1)} \nonumber \\
&& +  \hat g_\mu^{\mu (1)}  \p_{\bar \chi} \Delta x_{\|}^{(1)} +  \hat g_\mu^{\mu (1)}  \delta_g^{(1)}-2E_{\hat 0 }^{\perp i (1)} \p_{\perp i} \left(\Delta x^{0 (1)}+\Delta x_{\|}^{(1)}\right) + 2  \left( E_{\hat 0}^{0 (1)}+E_{\hat 0 }^{\| (1)}\right) \delta_g^{(1)} +  \frac{4}{\bar \chi} \Delta x_{\parallel}^{(1)}  \delta_g^{(1)}-4 \delta_g^{(1)}\kappa^{(1)}\bigg]  \nonumber \\
&&  + \bigg(1-\Q +2\Q^2  -2\frac{\p \Q}{\p \ln \bar L}\bigg)   \left[2 \left(\kappa^{(1)}\right)^2+  \frac{2}{\bar \chi^2} \left(\Delta x_{\|}^{(1)}\right)^2 -\frac{4}{\bar \chi} \kappa^{(1)} \Delta x_{\|}^{(1)}\right] \nonumber \\
&&+2\left( \Q^2- \frac{\p \Q}{\p \ln \bar L}\right)\left(E_{{\hat 0} 0}^{ (1)} -E_{{\hat 0} \|}^{ (1)}\right) \left(\delta\nu^{(1)}+\delta n_{\|}^{(1)} \right)+\left(\Q^2- \frac{\p \Q}{\p \ln \bar L}\right)\bigg[2 \left( E_{\hat 0}^{0 (1)}+E_{\hat 0 }^{\| (1)}\right)\left(\delta\nu^{(1)}+\delta n_{\|}^{(1)} \right) \nonumber \\
&& +2 \left( E_{\hat 0}^{0 (1)}+E_{\hat 0 }^{\| (1)}\right) \left(E_{{\hat 0} 0}^{ (1)} -E_{{\hat 0} \|}^{ (1)}\right)  -4 \left(E_{{\hat 0} 0}^{ (1)} -E_{{\hat 0} \|}^{ (1)}\right) \kappa^{(1)}-4\left(\delta\nu^{(1)}+\delta n_{\|}^{(1)} \right)\kappa^{(1)} +   \frac{4}{\bar \chi}  \Delta x_{\|}^{(1)} \left(E_{{\hat 0} 0}^{ (1)} -E_{{\hat 0} \|}^{ (1)}\right) \nonumber \\
&&+   \frac{4}{\bar \chi}  \Delta x_{\|}^{(1)}\left(\delta\nu^{(1)}+\delta n_{\|}^{(1)} \right)  +\left(\delta\nu^{(1)}+\delta n_{\|}^{(1)} \right)^2 + \left(E_{{\hat 0} 0}^{ (1)} -E_{{\hat 0} \|}^{ (1)}\right)^2 +\left( E_{\hat 0}^{0 (1)}+E_{\hat 0 }^{\| (1)}\right)^2+\hat g_\mu^{\mu (1)}  \left(E_{{\hat 0} 0}^{ (1)} -E_{{\hat 0} \|}^{ (1)}\right) \nonumber \\
&& + \hat g_\mu^{\mu (1)} \left(\delta\nu^{(1)}+\delta n_{\|}^{(1)} \right) \bigg] +\Q \bigg[ -\frac{2}{\cH}{\left( E_{{\hat 0} 0}^{ (1)} -E_{{\hat 0} \|}^{ (1)}\right)}' \Delta \ln a^{(1)} -2  \p_{\|} \left(E_{{\hat 0} 0}^{ (1)} -E_{{\hat 0} \|}^{ (1)}\right)\Delta x_{\|}^{(1)} - 2   \left(E_{{\hat 0} 0}^{ (1)} -E_{{\hat 0} \|}^{ (1)}\right) \p_{\bar \chi}\Delta x_{\|}^{(1)}\nonumber \\
&& -2  \p_{\perp i}\left( E_{{\hat 0} 0}^{ (1)} -E_{{\hat 0} \|}^{ (1)}\right)\Delta x_{\perp}^{i (1)} +2E_{{\hat 0} \perp i}^{ (1)} \left(-\frac{1}{\bar \chi}\Delta x_{\perp}^{i (1)}+\p_{\bar \chi} \Delta x_{\perp}^{i (1)} \right)+2 E_{{\hat 0} 0}^{ (1)} \left(\delta\nu^{(1)}+\delta n_{\|}^{(1)} \right) -2\left( E_{{\hat 0} 0}^{ (1)} -E_{{\hat 0} \|}^{ (1)}\right)  \delta_g^{(1)}\nonumber \\
&&-2 \left(\delta\nu^{(1)}+\delta n_{\|}^{(1)} \right) \delta_g^{(1)} \bigg]+\Bigg\{2\left[b_e\left(1-\Q\right) +2 \Q^2 -2\frac{\p \Q}{\p \ln \bar L}-\frac{\p \Q}{\p \ln \bar a} \right]\left(\frac{1}{2} \hat g_\mu^{\mu (1)} + E_{\hat 0}^{0(1)} +  E_{\hat 0}^{ \|(1)} + \frac{2}{\bar \chi} \Delta x_{\parallel}^{(1)}  - 2 \kappa^{(1)}\right) \nonumber \\
&& +2\left[\Q \left(2-b_e\right) +2 \Q^2 -2\frac{\p \Q}{\p \ln \bar L}-\frac{\p \Q}{\p \ln \bar a} \right]\left(E_{{\hat 0} 0}^{ (1)} -E_{{\hat 0} \|}^{ (1)}+\delta\nu^{(1)}+\delta n_{\|}^{(1)} \right)+2\left(b_e-2\Q \right)\left( \delta_g^{(1)}+\p_{\bar \chi} \Delta x_{\|}^{(1)} \right)\nonumber \\
&& -4\Q^{(1)}\Bigg\}\, \Delta \ln a^{(1)} +\left(-b_e+b_e^2+ \frac{\p b_e}{\p \ln \bar a}+6\Q-4 \Q b_e+4 \Q^2 -4\frac{\p \Q}{\p \ln \bar L}-4\frac{\p \Q}{\p \ln \bar a} \right) \left(\Delta \ln a^{(1)}\right)^2\nonumber \\
&&-2\Q^{(1)} \left(\frac{1}{2} \hat g_\mu^{\mu (1)} + E_{\hat 0}^{0(1)} +  E_{\hat 0}^{ \parallel(1)}+  E_{{\hat 0} 0}^{ (1)} -E_{{\hat 0} \|}^{ (1)}  +\delta\nu^{(1)}+\delta n_{\|}^{(1)}+ \frac{2}{\bar \chi} \Delta x_{\parallel}^{(1)}  - 2 \kappa^{(1)}\right)\;.
\end{eqnarray}

Here
\begin{eqnarray}
\label{be}
b_e =\frac{\p \ln \bar n_g (\bar a , \bar L)}{\p \ln \bar a}+3
\end{eqnarray}
is the evolution bias  term related to the comoving number density.
Before concluding this section let us add the following comments:
\begin{itemize}
\item[-] In this case, $b_e$ in Eq.\ (\ref{be}) is  defined with partial derivates w.r.t.\ $\bar a$. 
Generally, at linear order, $b_e$ is defined with the total derivates instead of the partial one because the redshift distribution of $n_g$  is usually defined in terms of a fixed thersold $\bar L$. 
At second order, to correctly obtain  all the terms we can not use this approach.

Instead, if we apply the total derivates we find the following  relation
\begin{eqnarray}
\label{beQ}
b_{e\Q} = \frac{\ud \ln \left[\bar a^3 \bar n_g (\bar a , \bar L(\bar a))\right]}{\ud \ln \bar a}=b_e-\frac{\p \ln \bar L}{\p \ln \bar a}\,\Q=b_e+2\left(1+\frac{1}{\bar \chi \cH}\right)\Q\;,
\end{eqnarray}
where for last step we have used\footnote{For simplicity, in Eq.\ (\ref{beQ}), we have assumed that $\bar \F$ does not depend explicitly on time and/or if we are considering the bolometric relation between the flux density and the (bolometric) Luminosity.} $\bar L=4 \pi \bar \F \, \bar \D_L^2(\bar a) =4 \pi\bar \F \, \bar \chi^2/\bar a^2 $.

\item[-] Obviously, the perturbative expansion made in Eq.\ (\ref{expng}) is correct if and only if
\begin{equation}
\frac{\p}{\p \M} \left(\frac{\p \bar n_g}{\p \bar x^0}\right)=\frac{\p}{\p \bar x^0} \left(\frac{\p \bar n_g}{\p \M}\right) \quad \quad {\rm or,~ equivalently,} \quad \quad  \frac{\p b_e}{\p \ln \bar L}= -\frac{\p \Q}{\p \ln \bar a}\;.
\end{equation}
We think that this consistency equation is very important not only for a theoretical point of view but also for  feedback in the data analysis of survey catalogs.

\item[-] For simplicity,  defining $\delta_g^{(1)}$ in Poisson gauge (see next section),  assuming no velocity bias between galaxy and matter,   we can rewrite the magnification bias at first order in the following way
\begin{equation}
 \Q^{(1)}(\bar x^\alpha, \bar L)=-\frac{\p  \delta_g^{(1)}}{\p \ln \bar L} =-\frac{\p   \delta_{g \, {\rm CO}}^{(1)}}{\p \ln \bar L} + \cH \frac{\p b_e}{\p  \ln \bar L} v  =-\frac{\p   b_1(\bar x^0,\bar L)}{\p \ln \bar L} \,\delta_{m \, {\rm CO}}^{(1)}  + \cH \frac{\p b_e}{\p  \ln \bar L} v
\;,
\end{equation}
where  $\delta_{g \, {\rm CO}}^{(1)}$ and  $\delta_{m \, {\rm CO}}^{(1)}$  are galaxy and cold dark matter over-density in the comoving-time orthogonal  (CO) gauge\footnote{Let us point out that   the comoving-time orthogonal  gauge becomes the usual  comoving-synchronous  gauge when the perturbations are dominated by pressure-free matter, for example in the $\Lambda$CDM model.} (see also Ref.\  \cite{Bertacca:2014wga}) respectively.
As a result  the terms in Eq.\  (\ref{Deltag-2}) proportional to $\Q^{(1)}$ can not be neglected because the bias $b_1$ depends on the redshift and the luminosity $\bar L$.

\item[-] Taking into account only the following terms of Eq.\ (\ref{Deltag-2})
\begin{equation}
\label{New-limit}
  - 2 \left(1- \Q \right) \kappa^{(2)} -4 \left(1- \Q \right) \delta_g^{(1)}\kappa^{(1)} +2 \bigg(1-\Q +2\Q^2  -2\frac{\p \Q}{\p \ln \bar L}\bigg) \left(\kappa^{(1)}\right)^2 -2 \left(1- \Q \right) \big|\gamma^{(1)}\big|^2 +  \vartheta_{ij}^{(1) }\vartheta^{ij(1)} 
\end{equation}
we can compare our result with the second order contribution obtained in Eq.\ (14) of   \cite{Schmidt:2008mb}  via the standard approach
\begin{equation}
\label{New-case}
2 c_1 \delta_g ~ \kappa+c_2 \kappa^2+c_1 \gamma^2\;,
\end{equation}
where $c_1=-2(1-\Q)$ and $c_2=2-6\Q +4\Q^2$. Approximating for simplicity that $\gamma^{(1)}\sim\gamma$ and $\kappa^{(1)}\sim\kappa$, we note immediately that $\kappa^{(2)}$ and $ \vartheta_{ij}^{(1) }\vartheta^{ij(1)} $ are usually omitted from standard analyses. Moreover, comparing $\kappa^2$ terms, we see that the coefficient of $\Q$ are different  [in Eq.\ (\ref{New-limit}) it is $-2$ and in Eq.\ (\ref{New-case}) it is $-6$]. 
This discrepancy is related to the term \[-2\Q\Delta \left(\M^{-1}\right)^{(1)}\Delta V^{(1)}\] that we find in the intermediate step of  Eq.\ (\ref{Deltag-2}).
Finally, there is no ${\p \Q/\p \ln \bar L}$ part in Eq.\ (\ref{New-case}).
\item[-] Using the relations obtained in Sec.\ III of \cite{Bertacca:2014wga}, we can obtain,  in a complete general way, the magnification,   the luminosity distance  and the observed over-density  in a general gauge both at first and second order.
\end{itemize}

\section{Perturbation terms in the Poisson Gauge  for a concordance model}\label{Poiss-pert}

We present the observed galaxy number over-density up to second order in redshift space on cosmological scales  for a $\Lambda$CDM model (without the magnification, see \cite{Bertacca:2014dra}).
The standard assumption at first order is that galaxy velocity equals CDM velocity on large scales, and we are assuming that it is reasonable to extend this assumption to second order,  since we are dealing only with large scales (i.e. well above the nonlinear scale).
Moreover, we assume a concordance background and at first order we neglect anisotropic stress, vector and tensor perturbations.  

In the Poisson gauge, the metric and peculiar velocity are
 \begin{eqnarray} 
 \label{Poiss-metric}
 \ud s^2 &=& a(\eta)^2\left\{-\left(1 + 2\Phi +\Phi^{(2)}\right)\ud\eta^2+2\omega_{i}^{(2)}\ud\eta \, \ud x^i+\left[\delta_{ij} \left(1 -2\Phi -\Psi^{(2)}\right)+\frac{1}{2}\hat h_{ij}^{(2)}\right]\ud x^i\ud x^j\right\} ,\\
v^{i  }&=& \p^i v +\frac{1}{2}v^{i (2)},~~ v^{i (2)}= \p^i v^{(2)}+ \hat v^{i (2)}, 
\end{eqnarray}
where we omit the superscript (1) on familiar quantities such as  $\Phi$ and $\p^iv$. At second order, the first-order scalars generate vector perturbations $\omega_{i}^{(2)}, \hat v^{i (2)}$ and a tensor perturbation $\hat h_{ij}^{(2)}$.

Now, taking into account the relations obtained in appendix \ref{A} or in Sec. IV  of Ref.\ \cite{Bertacca:2014wga}, for Eqs.\ (\ref{Deltalna-1}),  (\ref{chi_1}), (\ref{Deltax0-1}), (\ref{Dx_||-1}), (\ref{Dx_perp-1}), (\ref{kappa-n}), (\ref{shear})  we find at first order

\begin{eqnarray}
\label{Poiss-Deltalna-1}
\Delta \ln a^{(1)}  &=&\left (\Phi _o-v _{\| \, o}\right) - \Phi + \p_\| v + 2I^{(1)}  \;, \\
\delta \chi^{(1)}  &=& -\left(\bar \chi+\frac{1}{\cH}\right)\left(\Phi _o-v _{\| \, o}\right)+  \frac{1}{\cH}\left(\Phi - \p_\| v \right) -T^{(1)}   -2\left(\bar \chi+\frac{1}{\cH}\right) I^{(1)} - 2\int_0^{\bar \chi} \ud \tilde \chi \tilde \chi  \Phi {'} 
 \;, \\
\label{Poiss-Dx^0-1}
\Delta x^{0 (1) }&=&\frac{1}{\cH} \Delta \ln a^{(1)} \;, \\
\label{Poiss-Dx||-1}
\Delta x _\|^{(1)}  &=& - T^{(1)}   - \frac{1}{\cH}\Delta \ln a^{(1)}  \;,  \\
 \Delta x_{\perp}^{i  (1)}&=& - \bar \chi \, v^{i  }_{\perp \, o }+ 2  \bar \chi S_\perp^{i (1)} - \bar \chi \p_\perp^i T^{(1)}  \;,\\
 \kappa^{(1)} &=&- v_{\| \, o}  + \int_0^{\bar \chi} \ud \tilde \chi  \left(\bar \chi-\tilde \chi\right) \frac{\tilde \chi}{ \bar \chi} \,   \tilde \nabla^2_\perp \Phi\;, \\
\gamma_{ij}^{(1)} &=& - \Perp_{ij} v^{(1)}_{\|\, o}-n_{(j}  v_{\perp i) \, o}^{(1)}   + 2 \int_0^{\bar \chi} \ud \tilde \chi \left[\left(\bar \chi-\tilde \chi\right) \frac{\tilde \chi}{ \bar \chi}  \tilde \p_{\perp (i}  \tilde \p_{\perp j)} \Phi \right]-\Perp_{ij} \kappa^{(1)}\;.\label{gamp}
\end{eqnarray}

Here\footnote{Note that $\tilde \p_i= \p /\p \tilde x^i$.} \cite{Bertacca:2014dra}
\begin{eqnarray}
\label{Poiss-iota}
I^{(1)}   =  - \int_0^{\bar \chi} \ud \tilde \chi \, \Phi {'}  \;,\quad \quad \quad S^{i (1)} = -  \int_0^{\bar \chi} \ud \tilde \chi \left( \tilde\p^i \Phi   -\frac{1}{\tilde \chi} n^i\Phi  \right)\;,
\label{Poiss-varsigma}
\end{eqnarray}
where $I^{(1)}$ is the integrated Sachs-Wolfe (ISW) effect at first order,
\begin{eqnarray}
S_{\perp}^{i (1)} =\Perp^i_j S_{\perp}^{j (1)}= - \int_0^{\bar \chi} \ud \tilde \chi \, \tilde\p^i_\perp \Phi   \;, \quad \quad \quad S_{\|}^{(1)}  = n_i S^{i (1)} =\Phi _o  -  \Phi + I^{(1)}  + \int_0^{\bar \chi} \ud \tilde \chi \frac{\Phi}{\tilde \chi} \;. 
\end{eqnarray}
and
\begin{eqnarray}
\label{Poiss-s-1}
 T^{(1)} =- 2 \int_0^{\bar \chi} \ud \tilde \chi \Phi 
\end{eqnarray}
is a radial displacement at first order and corresponds to the usual (Shapiro) time delay (STD) term \cite{Challinor:2011bk}.
Another useful relation at first order is the following
\begin{equation}
\p_\| \Delta x _\|^{(1)} = 2 \Phi   -  \p_\| v - \frac{1}{\cH}  \p_\|^2 v  +\frac{1}{\cH} \Phi {'} - \frac{\cH'}{\cH^2}\Delta \ln a^{(1)}  \;.
\end{equation}

Instead, at second order, Eqs.\  (\ref{Deltalna-2}),  (\ref{Deltax0-2}), (\ref{Dx_||-2}), (\ref{Dx_perp-2}), (\ref{kappa-n}) turn out (see also  \cite{Bertacca:2014dra})
\begin{eqnarray}
\label{Poiss-Deltalna-2}
 && \Delta\ln a^{(2)} =- \Phi^{(2)}+ \p_\| v^{(2)}+ \hat v^{(2)}_\| +  3  {\Phi}^2  -  \left( \p_\| v \right)^2+ \p_{\perp i} v \,  \p^i_{\perp} v  -2  \p_\| v \,  \Phi  - \frac{2}{\cH}\left( \Phi  -  \p_\| v \right) \left(  \Phi {'} - \p_\|^2 v  \right)  - 4 \bigg(  \Phi   +  \frac{1}{\cH}  \p_\|^2 v  \nonumber \\
&&   - \frac{1}{\cH}  \Phi {'} \bigg) I^{(1)}   +2  \;   \left(2\Phi{'} +\p_\| \Phi -  \p_\|^2 v \right) T^{(1)}  +   4 \bar \chi \p_{\perp i}\left(-\Phi  +   \p_\| v  \right)   S_{\perp}^{i (1)}   +2 \bigg[\bar \chi \p_{\perp i}\left(\Phi  -   \p_\| v  \right)   + \p_{\perp i} v \bigg]  \p_{\perp}^i T^{(1)}+ 8\Phi  \kappa^{(1)}   \nonumber \\
&&    + 2I^{(2)} + 8  \left(I^{(1)}\right)^2    +4 \int_0^{\bar \chi}  \ud \tilde{\chi} \Bigg[   \Phi{''}  T^{(1)}  + 2 \Phi \Phi{'}  +  2 \Phi{'}  I^{(1)}    + 2 \Phi    \tilde \p_{\perp j}S_{\perp}^{j(1)}    - 2  \tilde \chi  \tilde \p_{\perp i} \Phi{'}    S_\perp^{i(1)}   - 2 \bigg( \frac{\ud \Phi}{\ud \tilde \chi}     -  \frac{1}{\tilde\chi}  \Phi  \bigg) \kappa^{(1)}  \nonumber\\
    &&       + \tilde \chi  \tilde \p_{\perp i} \Phi{'}   \p_\perp^i T^{(1)} \Bigg]  + \Phi^{(2)}_{\, o}- v^{(2)}_{\| \, o} - \Phi _o^2 + 8 \Phi _{\, o} v _{\| \, o} +  v _{k\, o} v^{k  }_o   + 2\left(\Phi _o-v _{\| \, o}\right) \bigg(  -  \Phi - \frac{1}{\cH}  \p_\|^2 v  + \frac{1}{\cH}  \Phi {'}  + 2 I^{(1)}  \bigg) \nonumber \\
&&  + 8 v_{\| \, o}    \int_0^{\bar \chi}   \frac{\ud \tilde{\chi}}{\tilde \chi}  \Phi +2   \bar \chi   v^{i (1)}_{\perp \, o }  \bigg(  \p_{\perp i} \Phi   -   \p_{\perp i}   \p_\| v    - 2 \p_{\perp i}  I^{(1)}    \bigg)\;,
\end{eqnarray}

\begin{eqnarray}
\label{Poiss-Dx0-2_3}
&&\Delta x^{0(2)}=   -\frac{1}{\cH} \Phi^{(2)} + \frac{1}{\cH} \p_\| v^{(2)}+ \frac{1}{\cH} \hat v^{(2)}_\| +  \frac{2}{\cH} I^{(2)}-\left(\frac{\cH'}{\cH^3}-\frac{2}{\cH}\right)  {\Phi}^2   - \left( \frac{\cH'}{\cH^3} + \frac{2}{\cH}\right)\left( \p_\| v \right)^2+  \frac{1}{\cH} \p_{\perp i} v   \p^i_{\perp} v +2  \frac{\cH'}{\cH^3}\Phi   \p_\| v  \nonumber \\
&&  +  \frac{2}{\cH^2} \left( \Phi  -  \p_\| v \right) \left(  - \Phi {'} + \p_\|^2 v  \right)
 + 4 \bigg\{ \frac{\cH'}{\cH^3} \Phi -\left( \frac{\cH'}{\cH^3}+ \frac{1}{\cH}\right)  \p_\| v   -  \frac{1}{\cH^2}  \left(- \Phi {'} + \p_\|^2 v  \right) - \left( \frac{\cH'}{\cH^3}- \frac{1}{\cH}\right) I^{(1)}  \bigg\} I^{(1)} \nonumber \\
&&  + 4\frac{\bar \chi}{\cH} \p_{\perp i}\left(-\Phi  +   \p_\| v  \right)  S_{\perp}^{i (1)}  +  \frac{2}{\cH} \left(2\Phi{'} +\p_\| \Phi -  \p_\|^2 v \right) T^{(1)}  +\frac{8} {\cH} \Phi \kappa^{(1)}    + \frac{2}{\cH} \left[\bar \chi \p_{\perp i}\left( \Phi  -   \p_\| v  \right) +  \p_{\perp i} v \right] \p^i_\perp T^{(1)} \nonumber \\
      && + \frac{4} {\cH} \int_0^{\bar \chi}  \ud \tilde{\chi} \Bigg[   \Phi{''}  T^{(1)}   + 2 \Phi \Phi{'}+  2 \Phi{'}  I^{(1)}    + 2 \Phi    \tilde \p_{\perp j}S_{\perp}^{j(1)}     - 2 \bigg( \frac{\ud}{\ud \tilde \chi} \Phi    -  \frac{1}{\tilde\chi}  \Phi  \bigg) \kappa^{(1)}   - 2  \tilde \chi  \tilde \p_{\perp i} \Phi{'}   S_\perp^{i(1)} + \tilde \chi   \tilde \p_{\perp i} \Phi{'}   \p_\perp^i T^{(1)}\Bigg]  \nonumber\\
&&+ \frac{1}{\cH} \Phi^{(2)}_o-\frac{1}{\cH} v^{(2)}_{\| \, o} -\left(\frac{\cH'}{\cH^3}+ \frac{2}{\cH}\right) \Phi_{o}^2+2\left( \frac{\cH'}{\cH^3}+\frac{5}{\cH}\right)\Phi _o v _{\| \, o}- \frac{\cH'}{\cH^3} v _{\| o}^2 + \frac{1}{\cH}v _{\perp i \, o}v^{i  }_{\perp \, o} + \frac{8} {\cH} v_{\| \, o}    \int_0^{\bar \chi}   \frac{\ud \tilde{\chi}}{\tilde \chi}  \Phi \nonumber \\
&&+ 2 \left(\Phi _o-v _{\| \, o}\right)   \bigg[\frac{\cH'}{\cH^3} \Phi-\left( \frac{\cH'}{\cH^3}+ \frac{1}{\cH}\right)  \p_\| v  - \frac{1}{\cH^2} \p_\|^2 v   + \frac{1}{\cH^2} \Phi {'}   -  2 \frac{\cH'}{\cH^3} I^{(1)}  \bigg]    +2 \frac{\bar \chi}{\cH} v_{\perp \, o}^{i }  \left( \p_{\perp i} \Phi  -  \p_{\perp i}   \p_\| v        - 2 \p_{\perp i}  I^{(1)}   \right) \;,  \nonumber \\ 
\end{eqnarray}

\begin{eqnarray} 
\label{Poiss-Dx_||-2_2}
&&\Delta x_{\parallel}^{(2)}=   -\frac{1}{ \cH} \Delta \ln a^{(2)}-  T^{(2)}  - \frac{4}{\cH}   \Phi \;  \Delta \ln a^{(1)} + \left(\frac{\cH' }{\cH^3} +\frac{1}{\cH} \right)\left( \Delta \ln a^{(1)}  \right)^2  -4\Phi   T^{(1)}   -4\bar \chi S_{\perp }^{i (1)}S_{\perp }^{j (1)} \delta_{ij}   \nonumber \\
&&  +4 \int_0^{\bar \chi}  \ud \tilde{\chi} \bigg( -  { \Phi}^2 -  \Phi{'} T^{(1)}  - 2\Phi \kappa^{(1)}  - \tilde \chi \tilde \p_{\perp i}\Phi  \p_\perp^i T^{(1)}   \bigg)  +8 \int_0^{\bar \chi}  \ud \tilde{\chi} ~ (\bar \chi - \tilde \chi) \bigg[    -  \Phi   \tilde \p_{\perp m}S_{\perp}^{m(1)}     
 +  \left( \frac{\ud}{\ud \tilde \chi} \Phi   -  \frac{1}{\tilde\chi} \Phi  \right) \kappa^{(1)}  \nonumber \bigg]\nonumber \\
&&- \bar \chi \left( 8 \Phi _o v _{\| \, o}   +   v _{\perp k \, o} v^{k  }_{\perp \, o}\right) -4  v_{\| \, o}\left( T^{(1)}+ 2\bar \chi  \int_0^{\bar \chi}   \frac{\ud \tilde{\chi}}{\tilde \chi} \Phi \right) \;,
\end{eqnarray}

\begin{eqnarray} 
\label{Poiss-Dx_perp-2_2}
&&\Delta x_{\perp}^{i(2)}=  \int_0^{\bar \chi} \ud \tilde \chi \bigg(  2\omega^{i(2)}_{\perp} - n^j \hat h_{jk}^{(2)} \Perp^{ki}  - 8\tilde \chi  \tilde \p^i_{\perp }  \Phi  ~ I^{(1)} - 8  \tilde \p^i_{\perp }  \Phi \int_0^{\tilde \chi} \ud \tilde{\tilde \chi} \tilde{\tilde \chi}  \Phi {'}  +4 \Phi \tilde \p_\perp^i T^{(1)} \bigg) \nonumber \\
&& + \int_0^{\bar \chi} \ud \tilde \chi \left(\bar \chi-\tilde \chi\right) \bigg\{-\bigg[ \tilde \p^i_\perp \left( \Phi^{(2)} + 2 \omega^{(2)}_{\| } + \Psi^{(2)} - \frac{1}{2} \hat h^{(2)}_{\| } \right)  + \frac{1}{\tilde \chi} \left(-2\, \omega^{i(2)}_{\perp} +  n^k \hat h_{kj}^{(2)} \Perp^{ij}  \right)\bigg]   \nonumber \\
&&  + 8 \Phi  \tilde \p^i_\perp \Phi-8   \tilde \p_{\perp }^i  \Phi  I^{(1)}  - 4 \tilde \p_{\perp }^i    \Phi{'} ~ T^{(1)}  
           +  \frac{4}{\tilde \chi}  \Phi  \p^i_\perp T^{(1)} + 4 \bigg(\tilde \chi \Perp^{im} \tilde \p_{\perp j}  \tilde  \p_{\perp m}   \Phi +  \Perp^i_j \Phi{'}  \bigg) \left(- 2   S_\perp^{j(1)} +  \p_\perp^j T^{(1)}\right)  \bigg\} \nonumber \\
&&- 4 \left(  T^{(1)}  +  \frac{1}{\cH}  \Delta \ln a^{(1)}  + 2   \bar \chi I^{(1)}  + 2  \int_0^{\bar \chi} \ud \tilde \chi \tilde \chi  \Phi {'} \right)  S_{\perp}^{i (1)} + \bar \chi \bigg[ -2\omega^{i(2)}_{\perp \, o}-  v^{i(2)}_{\perp \, o} + \frac{1}{2} n^j \hat h_{ j k\, o}^{(2)} \Perp^{ki}
+ 2 \Phi _o v^{i  }_{\perp \, o } -  v _{\| \, o}v^{i }_{\perp \, o}  \bigg]  \nonumber \\
&&  + 2 v^{j  }_{\perp \, o } \bigg[  \Perp_j^i T^{(1)} +  \frac{1}{\cH} \Perp_j^i \Delta \ln a^{(1)}  +\bar \chi^2  \Perp^{im} \p_{\perp j}   \p_{\perp m}   \left( T^{(1)}+2 \bar \chi  \int_0^{\bar \chi}   \frac{\ud \tilde{\chi}}{\tilde \chi} \Phi \right)     \bigg] 
\end{eqnarray}
and
\begin{eqnarray}
 \label{Poiss-kappa-2-0}
 &&\kappa^{(2)}= \frac{1}{2}  \int_0^{\bar \chi} \ud \tilde \chi  \left(\bar \chi-\tilde \chi\right) \frac{\tilde \chi}{ \bar \chi}   \tilde \nabla^2_\perp \left( \Phi^{(2)} + 2 \omega^{(2)}_{\| }+\Psi^{(2)} - \frac{1}{2} \hat h^{(2)}_{\| } \right)       + \frac{1}{2}  \int_0^{\bar \chi} \ud \tilde \chi \bigg(-2  \tilde \p_\perp^i \omega_i^{ (2)} + \frac{4}{\tilde\chi} \omega_\|^{ (2)} + \Perp^{ij} n^k  \tilde \p_i  \hat h_{jk}^{ (2)}   \nonumber \\
 && - \frac{3}{\tilde \chi}  \hat h_\|^{ (2)}\bigg)      -2\bigg( 2  I^{(1)} +\frac{2}{ \bar \chi}\int_0^{\bar \chi} \ud \tilde \chi \tilde \chi \Phi {'}  + \frac{1}{ \bar \chi} T^{(1)}   +  \frac{1}{\bar \chi \cH}\Delta \ln a  \bigg) \left(\kappa^{(1)}  - \frac{\bar \chi}{2}\nabla^2_\perp T^{(1)}\right)      -2S_{\perp}^{i }   \bigg[ -  \p_{\perp i } T^{(1)} -  \frac{1}{\cH} \p_{\perp i}  \Delta \ln a^{(1)} \nonumber \\
 &&    -2 \left(\bar \chi\p_{\perp i}  I^{(1)} + \p_{\perp i}  \int_0^{\bar \chi} \ud \tilde \chi   \tilde \chi  \Phi{'}  \right)\bigg]         + 2 \int_0^{\bar \chi} \ud \tilde \chi \, \frac{\tilde \chi}{ \bar \chi}  \bigg[   \frac{4}{\tilde \chi}   \Phi  S_{\|}^{(1)}  - 2 \Phi   \tilde \p_{\perp m} S^{m (1)} + 2\tilde \chi   \tilde \nabla^2_{\perp} \Phi I^{(1)}  + 2  \tilde \nabla^2_{\perp} \Phi \int_0^{\tilde \chi} \ud \tilde{\tilde \chi} \tilde{\tilde \chi}  \Phi {'}    \nonumber\\
&& + 2  \tilde \chi \p_{\perp i}  \Phi   \tilde \p_\perp^i I^{(1)}  + 2 \p_{\perp i}  \Phi   \tilde \p^i_{\perp} \int_0^{\tilde \chi} \ud \tilde{\tilde \chi} \tilde{\tilde \chi}  \Phi {'}    -  \tilde \p_{\perp i} \Phi  \p_\perp^i T^{(1)}  - \frac{2}{\tilde \chi}\Phi  \kappa^{(1)}  \Bigg]  + 2\int_0^{\bar \chi} \ud \tilde \chi \left(\bar \chi-\tilde \chi\right)\frac{\tilde \chi}{ \bar \chi}  \bigg[ - 2\tilde \p_{\perp i} \Phi \tilde\p^i_\perp  \Phi \nonumber\\
&&  + 2 \tilde\p^i_\perp  \Phi \tilde \p_{\perp i} I^{(1)}  - 2 \Phi  \tilde \nabla^2_\perp \Phi      - \tilde \nabla^2_{\perp} \Phi  ~ T^{(1)}    -  \tilde \p_{\perp }^i    \Phi{'}  \tilde \p_{\perp i} T^{(1)} + 2 I^{(1)}  \tilde \nabla^2_\perp \Phi  +  \frac{2}{\tilde\chi}  \bigg( -     \frac{1}{\tilde\chi}\Phi +  \frac{\ud}{\ud \tilde \chi} \Phi\bigg)  \kappa^{(1)}  +  \frac{2}{\tilde\chi} \tilde \p_{\perp i}  \Phi ~ S_{\perp}^{i(1)}  \nonumber\\
&&   - \frac{2}{\tilde\chi}   \Phi ~    \tilde \p_{\perp m}S_{\perp}^{m(1)}  -   \frac{2}{\tilde \chi}  \tilde \p_{\perp i} \Phi\p_\perp^i T^{(1)}    + \bigg(  \tilde \p_{\perp i} \tilde \nabla^2_{\perp} \Phi + \frac{1}{\tilde\chi} \tilde \p_{\perp i}  \Phi{'}    \bigg)\left( 2  \tilde \chi S_\perp^{i(1)} - \tilde \chi \p_\perp^i T^{(1)}\right)   -   \bigg( \tilde \p^{(j}_{\perp} \tilde \p^{m)}_{\perp} \Phi + \frac{1}{\tilde \chi}  \Perp^{jm} \Phi{'}  \nonumber\\
&&  + \frac{1}{\tilde \chi^2}\Perp^{jm}  \Phi  \bigg) \left(\gamma_{mj}^{(1)}+  \Perp_{mj} \kappa^{(1)} \right) -    \frac{1}{\tilde \chi} n^{[ j} \p^{m]}_\perp \Phi ~ \theta_{mj}^{(1)}  \Bigg]   - 2 \omega_{\| \, o}^{(2)}- v_{\| \, o}^{(2)}+\frac{3}{4}\hat h_{\| \, o}^{ (2)}    + \frac{1}{2} v _{\perp i \, o } v^{i  }_{\perp \, o} + 2\Phi_o  v_{\| o} - 2v _{\| o}^2 \nonumber\\
 &&     - 4   v_{\| \, o}   \ \int_0^{\bar \chi}   \frac{\ud \tilde{\chi}}{\tilde \chi} \Phi   -   v^i_{\perp \, o } \bigg[  - 2  S_{\perp i}^{(1)} +  2\p_{\perp i} T^{(1)}  + \bar \chi^2   \p_{\perp i}  \nabla^2_{\perp}\left( T^{(1)}+2 \bar \chi  \int_0^{\bar \chi}   \frac{\ud \tilde{\chi}}{\tilde \chi} \Phi \right)     +  \frac{1}{\cH} \p_{\perp i}  \Delta \ln a^{(1)}  \bigg] \nonumber\\
 &&= \frac{1}{2}  \int_0^{\bar \chi} \ud \tilde \chi  \left(\bar \chi-\tilde \chi\right) \frac{\tilde \chi}{ \bar \chi}   \tilde \nabla^2_\perp \left( \Phi^{(2)} + 2 \omega^{(2)}_{\| }+\Psi^{(2)} - \frac{1}{2} \hat h^{(2)}_{\| } \right)      + \frac{1}{2}  \int_0^{\bar \chi} \ud \tilde \chi \bigg(-2  \tilde \p_\perp^i \omega_i^{ (2)} + \frac{4}{\tilde\chi} \omega_\|^{ (2)} + \Perp^{ij} n^k  \tilde \p_i  \hat h_{jk}^{ (2)}  - \frac{3}{\tilde \chi}  \hat h_\|^{ (2)}\bigg)    \nonumber \\
  &&    - \frac{2}{ \bar \chi}\bigg( 2 \bar\chi I^{(1)} +2\int_0^{\bar \chi} \ud \tilde \chi \tilde \chi \Phi {'}  +T^{(1)}   +  \frac{1}{\cH}\Delta \ln a  \bigg)   \left(\kappa^{(1)}  - \frac{\bar \chi}{2}\nabla^2_\perp T^{(1)}\right)   +2S_{\perp}^{i }   \bigg(  \p_{\perp i } T^{(1)} +  \frac{1}{\cH} \p_{\perp i}  \Delta \ln a^{(1)} 
  + 2\bar \chi\p_{\perp i}  I^{(1)}      \nonumber \\
    &&  + 2\p_{\perp i}  \int_0^{\bar \chi} \ud \tilde \chi   \tilde \chi  \Phi{'}    \bigg) +2 \int_0^{\bar \chi} \ud \tilde{\chi}  \frac{ \tilde \chi }{\bar \chi} \bigg[  + 2\tilde \chi   \tilde \nabla^2_{\perp} \Phi I^{(1)}  + 2  \tilde \nabla^2_{\perp} \Phi \int_0^{\tilde \chi} \ud \tilde{\tilde \chi} \tilde{\tilde \chi}  \Phi {'}   + 2  \tilde \chi \p_{\perp i}  \Phi   \tilde \p_\perp^i I^{(1)} + 2 \p_{\perp i}  \Phi   \tilde \p_{\perp i} \int_0^{\tilde \chi} \ud \tilde{\tilde \chi} \tilde{\tilde \chi}  \Phi {'}   + \frac{4}{\tilde \chi}   \Phi  S_{\|}^{(1)}   \nonumber \\ 
&& - 2 \Phi   \tilde \p_{\perp m} S^{m (1)}   -  \tilde \p_{\perp i} \Phi  \p_\perp^i T^{(1)}  - \frac{2}{\tilde \chi}\Phi  \kappa^{(1)}\bigg]  + 2\int_0^{\bar \chi} \ud \tilde \chi \left(\bar \chi-\tilde \chi\right)\frac{\tilde \chi}{ \bar \chi}  \Bigg[ - 2\tilde\p^i_\perp  \Phi   \tilde \p_{\perp i} \Phi + 2 \tilde\p^i_\perp  \Phi  \tilde \p_{\perp i} I^{(1)}   - 2 \Phi  \tilde \nabla^2_\perp \Phi + 2  \tilde \nabla^2_\perp \Phi I^{(1)}    \nonumber\\
&&   - \tilde \nabla^2_{\perp} \Phi  ~ T^{(1)}    -   \tilde \p_{\perp }^i    \Phi{'}  \tilde \p_{\perp i} T^{(1)}   +  \frac{2}{\tilde\chi}  \bigg( -     \frac{1}{\tilde\chi}\Phi +  \frac{\ud}{\ud \tilde \chi} \Phi\bigg)  \kappa^{(1)}  
 +  \frac{1}{\tilde\chi} \tilde \p_{\perp i}  \Phi ~ S_{\perp}^{i(1)}    - \frac{3}{2 \tilde \chi}  \tilde \p_{\perp i} \Phi   \p_\perp^i T^{(1)}    + \tilde \chi \bigg(  \tilde \p_{\perp i} \tilde \nabla^2_{\perp} \Phi + \frac{1}{\tilde\chi} \tilde \p_{\perp i}  \Phi{'}    \bigg) \nonumber\\
&& \times \left( 2   S_\perp^{i(1)} -  \p_\perp^i T^{(1)}\right)+  \tilde\chi \tilde \p^{(j}_{\perp} \tilde \p^{m)}_{\perp} \Phi \left( 2  \p_{\perp(m}  S_{\perp j)}^{(1)} -   \p_{\perp (m}   \p_{\perp j)}  T^{(1)}\right) +  2\Phi{'} \tilde \p_{\perp m} S_{\perp}^{m(1)} - \left( \Phi{'}   + \frac{1}{\tilde \chi} \Phi  \right) \tilde \nabla_\perp^2 T^{(1)} \Bigg] \nonumber\\
 &&    - 2 \omega_{\| \, o}^{(2)}- v_{\| \, o}^{(2)}+\frac{3}{4}\hat h_{\| \, o}^{ (2)} + 2 \Phi_{\, o}   v_{\| \, o}   + \frac{1}{2} v _{\perp i \, o } v^{i  }_{\perp \, o}  + v_{\| o}^2      + \frac{2}{\bar \chi}   v_{\| \, o}   \bigg( +\bar \chi \kappa^{(1)}      +T^{(1)}    - 2 \bar\chi I^{(1)} - 2\int_0^{\bar \chi} \ud \tilde \chi \tilde \chi \Phi {'}   \bigg)  \nonumber \\
  &&  +   v^i_{\perp \, o } \bigg[   2  S_\perp^{i(1)} - 2\p_\perp^i T^{(1)}  - \bar \chi^2   \p_{\perp i}  \nabla^2_{\perp}\left( T^{(1)}+2 \bar \chi  \int_0^{\bar \chi}   \frac{\ud \tilde{\chi}}{\tilde \chi} \Phi \right)    -  \frac{1}{\cH} \p_{\perp i}  \Delta \ln a^{(1)}  \bigg]\;.
  \end{eqnarray}

%
%
%
%
 Here
\begin{equation}
\label{Poiss-s-1}
 T^{(2)} = - \int_0^{\bar \chi} \ud \tilde \chi \left(\Phi^{(2)} +2 \omega^{(2)}_{\| }+ \Psi^{(2)}- \frac{1}{2}h^{(2)}_{\| }\right) \;,
 \end{equation}
and 
\begin{equation}
I^{(2)} =  -\frac{1}{2} \int_0^{\bar \chi} \ud \tilde \chi \left(\Phi^{(2)}{'} +2  \omega^{(2)}_{\| }{'} + \Psi^{(2)}{'}- \frac{1}{2} \hat h^{(2)}_{\| }{'} \right)\;.
\end{equation}
Using Eqs.  (\ref{partialparallep}) and (\ref{ConsEq}) for $\Delta x^{(2)}_\| $, we find
   \begin{eqnarray} 
\label{Poiss-dDx_||-2_4}
 && \p_\| \Delta x_{\parallel}^{(2)} =+ \Phi^{(2)} +  \Psi^{(2)} -\frac{1}{2}\hat h_{\|}^{(2)}+  \frac{1}{ \cH} \Psi^{(2)}{'}-  \frac{1}{2 \cH} \hat h^{(2)}_{\| }{'} -\frac{1}{ \cH }\p_\|^2 v^{(2)} -\frac{1}{ \cH }   \p_\|\hat v^{(2)}_\|  - \p_\| v^{(2)}- \hat v^{(2)}_\|  -   \frac{\cH'}{\cH^2} \Delta \ln a^{(2)}  \nonumber \\
&& + \frac{2}{\cH^2}\left(\p_\|^2 v  \right)^2 +4   {\Phi}^2+2\left( \p_\| v  \right)^2  + \frac{2}{\cH^2}\left( \Phi {'}  \right)^2 - 2 \Phi \p_\| v     + \frac{2}{\cH^2} \p_\| v  \p_\|^2 \Phi   + \frac{6}{ \cH }\Phi  \Phi {'}+2\frac{\cH'}{\cH^3}\Phi \Phi {'} +\frac{4}{ \cH }\p_\| v  \p_\| \Phi    \nonumber \\ 
&& - \frac{8}{\cH} \Phi \p_\|^2 v   - \frac{2}{\cH^2} \Phi \p_\|^3 v   -\frac{2}{\cH}\Phi \p_\| \Phi + \frac{2}{\cH^2}\Phi \frac{\ud \,}{\ud \bar \chi}\Phi {'} - \frac{2}{\cH^2}\p_\| v \frac{\ud \,}{\ud \bar \chi}\Phi {'}  -2\frac{\cH'}{\cH^3}\Phi  \p_\|^2 v    +\frac{4}{\cH} \p_\| v \p_\|^2 v    +2\frac{\cH'}{\cH^3} \p_\| v \p_\|^2 v - \frac{2}{\cH^2}\Phi  \p_\|^2 \Phi    \nonumber \\ 
&&-2\frac{\cH'}{\cH^3} \p_\| v \Phi {'} -  \frac{4}{\cH^2}\p_\|^2 v   \Phi {'}  +\frac{2}{\cH}\p_{\perp i} v \p^i_\perp \Phi   -\frac{4}{ \cH } \p_{\perp i} v   \p_{\perp}^i \p_\|  v  +\frac{4}{ \bar \chi \cH } \p_{\perp i} v    \p_{\perp}^i v    - 2\p_{\perp i} v   \p_{\perp}^i v  +  \frac{2}{\cH^2}\p_\| v \p_\|^3v  \nonumber \\  
&&    +2\left[ \frac{1}{\cH} \p_\|^3 v + \p_\|^2 v -\frac{1}{\cH} \p_\| \Phi {'} -2\p_\| \Phi \right] T^{(1)}  + 4\bar \chi \bigg(  \frac{1}{\cH} \p_{\perp i}\Phi {'}  - \frac{1}{\cH} \p_{\perp i} \p_\|^2 v -\p_{\perp i} \p_\| v + 2 \p_{\perp i}  \Phi \bigg) S_\perp^{i(1)}   -4 S_{\perp }^{i (1)}S_{\perp }^{j (1)} \delta_{ij}  \nonumber \\ 
&&  + 4  \bigg( \frac{\cH' }{\cH^3}\p_\|^2 v -\frac{\cH' }{\cH^3}\Phi {'}   +\frac{1}{\cH} \Phi {'}   +  \frac{1}{\cH} \p_\|^2 v  +  \frac{1}{\cH^2} \p_\|^2 \Phi  + \frac{1}{\cH^2} \p_\|^3 v  + \frac{1}{\cH}   \p_\| \Phi  - \frac{1}{\cH^2}\frac{\ud \,}{\ud \bar \chi}\Phi {'}    \bigg) I^{(1)}    +2 \bigg[-\left(1-\frac{2}{\bar \chi \cH}\right) \p_{\perp i} v  \nonumber \\ 
&&+ \bar \chi \left(1-\frac{2}{\bar \chi \cH}\right)\p_{\perp i} \p_\| v  - 2 \bar \chi  \p_{\perp i} \Phi - \frac{\bar \chi }{\cH}   \p_{\perp i}\Phi {'} + \frac{\bar \chi}{\cH} \p_{\perp i} \p_\|^2 v   \bigg]   \p_\perp^i T^{(1)}  -8 \Phi  \kappa^{(1)} +2 \bigg[-2\frac{\cH' }{\cH^2} \Phi   - \frac{2}{\cH} \frac{\ud}{\ud \bar \chi}\Phi \nonumber \\
 &&- \left(\frac{\cH' }{\cH^3} +\frac{1}{\cH} \right)  \bigg(-\p_\|^2 v -\cH  \p_\| v  + \Phi {'} \bigg) \bigg] \Delta \ln a^{(1)}  +\left[-\frac{\cH'' }{\cH^3} +3\left( \frac{\cH' }{\cH^2} \right)^2 + \frac{\cH' }{\cH^2} \right] \left( \Delta \ln a^{(1)}  \right)^2 
\nonumber \\
  &&  +8 \int_0^{\bar \chi}  \ud \tilde{\chi} \Bigg[    -   \Phi \tilde \p_{\perp m}S_{\perp}^{m(1)}   +  \bigg( \frac{\ud}{\ud \bar \chi} \Phi    -  \frac{1}{\tilde\chi} \Phi  \bigg) \kappa^{(1)}  \Bigg]    -8\Phi _o v _{\|  o}     -  v _{\perp k \, o} v^{k  }_{\perp \, o}\Phi   - 8  v_{\| \, o} \int_0^{\bar \chi}   \frac{\ud \tilde{\chi}}{\tilde \chi} \Phi       \nonumber \\
&&+ 2\left(\Phi _o-v _{\| \, o}\right) \bigg[\left(\frac{\cH' }{\cH^3}+ \frac{1}{\cH}\right) \p_\|^2 v  +\left(-\frac{\cH' }{\cH^3}+ \frac{1}{\cH}\right)  \Phi {'}   + \frac{1}{\cH^2} \p_\|^2 \Phi  + \frac{1}{\cH^2}\p_\|^3 v + \frac{1}{\cH}  \p_\| \Phi - \frac{1}{\cH^2}\frac{\ud \,}{\ud \bar \chi}\Phi {'}  \bigg]  \nonumber \\
&&
-2v_{\perp \, o}^{i }\bigg[    -\frac{\bar \chi}{\cH}\p_{\perp i}  \left(-\Phi {'} +\p_\|^2v +\cH \p_\| v  \right) - 2  S_{\perp}^{j(1)}\delta_{ij}+2  \bar \chi  \p_{\perp i} \Phi   
     \bigg]\;.
\end{eqnarray}

At this point, using these results, we can focus on the main results of this paper. For $\M^{-1}$, from Eqs. (\ref{M-1_1}) and (\ref{M-1_2}), we find
\begin{eqnarray}
\Delta\left(\M^{-1}\right)^{(1)} =-2 \Phi  + 2 \left(1  - \frac{1}{\cH \bar \chi} \right)\Delta \ln a^{(1)} -\frac{2}{\bar \chi} T^{(1)}   - 2 \kappa^{(1)}  \;,
\end{eqnarray}
and
\begin{eqnarray}
&&\Delta\left(\M^{-1}\right)^{(2)} =-2  \Psi^{(2)}  - \frac{1}{2}\hat h_{\|}^{(2)} + 2\left( 1  -\frac{1}{\bar \chi \cH}\right) \Delta \ln a^{(2)}     - \frac{2}{\bar \chi} T^{(2)} - 2 \kappa^{(2)}    +4 \p_{\|}\Phi  T^{(1)}  +\frac{2}{\bar \chi^2} \left(   T^{(1)} \right)^2   +  \frac{4}{\bar \chi}  \kappa^{(1)}  T^{(1)}\nonumber \\
&& +2   \left(\kappa^{(1)}\right)^2   -2 \big|\gamma^{(1)}\big|^2+  \vartheta_{ij}^{(1) }\vartheta^{ij(1)}   - 8  \bar \chi \p_{\perp i}\Phi S_\perp^{i (1)} + 4 \bar \chi \p_{\perp i}\Phi \p_\perp^i T^{(1)}  + \frac{4}{\cH}S_{\perp}^{i (1)} \p_{\perp i}\Delta \ln a^{(1)}   -4 S_{\perp }^{i (1)}S_{\perp }^{j (1)} \delta_{ij} \nonumber \\
 &&   + 2 \left( 1 + \frac{1}{\bar \chi} \frac{\cH' }{\cH^3} +  \frac{1}{\bar \chi^2 \cH^2 }  -  \frac{3}{\bar \chi \cH} \right)\left(\Delta \ln a^{(1)} \right)^2     + 8 \int_0^{\bar \chi}  \ud \tilde{\chi} \bigg[   -  \Phi  \tilde \p_{\perp m}S_{\perp}^{m(1)}  +  \left( \frac{\ud \Phi}{\ud \tilde \chi}   -  \frac{1}{\tilde\chi} \Phi  \right) \kappa^{(1)}  \bigg]   \nonumber \\
&&  + \frac{8}{\bar \chi} \int_0^{\bar \chi}  \ud \tilde{\chi} \Bigg\{  \bigg[ -  { \Phi}^2 -  \Phi{'} T^{(1)} -2 \Phi  \kappa^{(1)}    - \tilde \chi \tilde \p_{\perp i}\Phi  \p_\perp^i T^{(1)}   \bigg]     +2(\bar \chi - \tilde \chi) \bigg[     -  \Phi   \tilde \p_{\perp m}S_{\perp}^{m(1)}     
 +  \left( \frac{\ud  \Phi }{\ud \tilde \chi}  -  \frac{1}{\tilde\chi} \Phi  \right) \kappa^{(1)}  \nonumber \bigg]\Bigg\}  \nonumber \\
&&+4 \left[ - 2\Phi   + \frac{1}{\cH}  \frac{\ud \Phi}{\ud \bar \chi}  - \frac{2}{\bar \chi} T^{(1)}     +   \frac{1}{\bar \chi^2 \cH}  T^{(1)}    +  \frac{1}{\bar \chi \cH}  \kappa^{(1)}  - 2 \kappa^{(1)} \right]\Delta \ln a^{(1)}   -24 \Phi_o  v_{\|  o}   -   v _{\perp k \, o} v^{k  }_{\perp \, o}  \nonumber \\
&&   -8  v_{\| \, o}\left( \frac{1}{\bar \chi} T^{(1)}+ 3  \int_0^{\bar \chi}   \frac{\ud \tilde{\chi}}{\tilde \chi} \Phi \right)  +2 \, v_{\perp  i\, o}  \left[ 2 \bar \chi  \p^i_{\perp }\Phi  +2 S^{i }_{\perp} - \frac{1}{\cH} \p^i_{\perp }\Delta \ln a^{(1)} \right]\;,
\end{eqnarray}
   \begin{eqnarray}
&&        \nonumber \\
  &&
 \;. \nonumber 
\end{eqnarray}

where
\begin{eqnarray}
\vartheta_{ij}^{(1)}\vartheta^{ij(1)} &=&  \frac{1}{2}v_{\perp i o}^{(1)}v_{\perp  o}^{i(1)} + \frac{2}{ \bar \chi}v_{\perp i \, o}^{(1)} \int_0^{\bar \chi} \ud \tilde \chi\bigg[ \left(\bar \chi-\tilde \chi\right)  \tilde \p^i_{\perp } \Phi\bigg]+\frac{2}{\bar \chi^2}  \int_0^{\bar \chi} \ud \tilde \chi\bigg[ \left(\bar \chi-\tilde \chi\right)  \tilde \p_{\perp i}  \Phi \bigg]  \int_0^{\bar \chi} \ud \tilde \chi\bigg[ \left(\bar \chi-\tilde \chi\right)  \tilde \p^i_{\perp}  \Phi \bigg] \nonumber \\
&=&  \frac{1}{2}v_{\perp i o}^{(1)}v_{\perp  o}^{i(1)} -v_{\perp i o}^{(1)}\left(2S_{\perp }^{i (1)}-  \p_\perp^i T^{(1)} \right)+  \frac{1}{2} \delta_{ij} \left(2S_{\perp }^{i (1)}-  \p_\perp^i T^{(1)} \right)\left(2S_{\perp }^{j (1)} -  \p_\perp^j T^{(1)} \right) \nonumber \\
&=&  \frac{1}{2\bar \chi^2}\Delta x_{\perp i}^{ (1)} \Delta x_{\perp}^{i  (1)}
\;. \label{varp}
 \end{eqnarray}

From  Eqs. (\ref{M_1}) and (\ref{M_2}), the magnification turns out
\begin{eqnarray}
\Delta \M^{(1)} = 2 \Phi  - 2 \left(1  - \frac{1}{\cH \bar \chi} \right)\Delta \ln a^{(1)} +\frac{2}{\bar \chi} T^{(1)}  + 2 \kappa^{(1)} \;, 
\end{eqnarray}
 \begin{eqnarray}
&&\Delta \M^{(2)} =  2  \Psi^{(2)}  + \frac{1}{2}\hat h_{\|}^{(2)} - 2 \Delta \ln a^{(2)}  +\frac{2}{\bar \chi \cH} \Delta \ln a^{(2)}     + \frac{2}{\bar \chi} T^{(2)} + 2 \kappa^{(2)} +  8\Phi^2    +  \frac{16}{\bar \chi}  \Phi T^{(1)}   - 4 \p_{\|}\Phi  T^{(1)}    +\frac{6}{\bar \chi^2} \left(   T^{(1)} \right)^2    \nonumber \\
&& +16\Phi \kappa^{(1)} +  \frac{12}{\bar \chi}  \kappa^{(1)}  T^{(1)} +6   \left(\kappa^{(1)}\right)^2   +2 \big|\gamma^{(1)}\big|^2 -  \vartheta_{ij}^{(1) }\vartheta^{ij(1)}   + 8  \bar \chi \p_{\perp i}\Phi S_\perp^{i (1)} - 4 \bar \chi \p_{\perp i}\Phi \p_\perp^i T^{(1)}  - \frac{4}{\cH}S_{\perp}^{i (1)} \p_{\perp i}\Delta \ln a^{(1)} \nonumber \\
&&  + 4 S_{\perp }^{i (1)}S_{\perp }^{j (1)} \delta_{ij}   + 2 \left(3  + \frac{3}{\cH^2 \bar \chi^2} -5  \frac{1}{\cH \bar \chi}  - \frac{1}{\bar \chi} \frac{\cH' }{\cH^3} \right)\left(\Delta \ln a^{(1)} \right)^2  +4 \bigg[ -2\left(1 - \frac{2}{\cH \bar \chi} \right)  \Phi - \frac{1}{\cH}  \frac{\ud \Phi}{\ud \bar \chi}+\frac{1}{\bar \chi}\left(-2 +  \frac{3}{\bar \chi \cH} \right)T^{(1)}  \nonumber \\
&& +\left(  - 2   + \frac{3}{ \bar \chi \cH} \right)\kappa^{(1)} \bigg]\Delta \ln a^{(1)}  - 8 \int_0^{\bar \chi}  \ud \tilde{\chi} \bigg[     -  \Phi  \tilde \p_{\perp m}S_{\perp}^{m(1)}  +  \left( \frac{\ud \Phi}{\ud \tilde \chi}   -  \frac{1}{\tilde\chi} \Phi  \right) \kappa^{(1)}  \bigg] \nonumber \\
  && 
  - \frac{8}{\bar \chi} \int_0^{\bar \chi}  \ud \tilde{\chi} \Bigg\{  -\bigg[ { \Phi}^2 +  \Phi{'} T^{(1)}   +  2 \Phi \kappa^{(1)}  + \tilde \chi \tilde \p_{\perp i}\Phi  \p_\perp^i T^{(1)}   \bigg]     +2(\bar \chi - \tilde \chi) \bigg[   -  \Phi   \tilde \p_{\perp m}S_{\perp}^{m(1)}     
 +  \left( \frac{\ud}{\ud \tilde \chi} \Phi   -  \frac{1}{\tilde\chi} \Phi  \right) \kappa^{(1)}  \nonumber \bigg]\Bigg\}  \nonumber \\
&& + 24\Phi _o v _{\| \, o} +   v _{\perp k \, o} v^{k  }_{\perp \, o}  + 8  v_{\| \, o}\left( \frac{1}{\bar \chi} T^{(1)}+ 3  \int_0^{\bar \chi}   \frac{\ud \tilde{\chi}}{\tilde \chi} \Phi \right)  +2 \, v_{\perp  i\, o}  \left[ - 2 \bar \chi  \p^i_{\perp }\Phi  - 2 S^{i }_{\perp} + \frac{1}{\cH} \p^i_{\perp }\Delta \ln a^{(1)} \right]  
\end{eqnarray}
and, from Eqs. (\ref{DL_1}) and (\ref{DL_2}),  the luminosity (or the angular) distance is 
\begin{eqnarray}
\frac{\D_L^{(1)}}{\bar \D_L} =\frac{\D_A^{(1)}}{\bar \D_A} = - \Phi  +  \left(1  - \frac{1}{\cH \bar \chi} \right)\Delta \ln a^{(1)} -\frac{1}{\bar \chi} T^{(1)}   -  \kappa^{(1)}
\end{eqnarray}
\begin{eqnarray}
&&\frac{\D_L^{(2)}}{\bar \D_L} =\frac{\D_A^{(2)}}{\bar \D_A}= - \Psi^{(2)}  - \frac{1}{4}\hat h_{\|}^{(2)} +  \left(1  - \frac{1}{\cH \bar \chi} \right)\Delta \ln a^{(2)}  -  \frac{1}{\bar \chi} T^{(2)} -  \kappa^{(2)}  - \Phi^2-  \frac{2}{\bar \chi}  \Phi T^{(1)} + 2 \p_{\|}\Phi  T^{(1)}       - 2\Phi \kappa^{(1)} \nonumber \\
&&        -  \big|\gamma^{(1)}\big|^2+ \frac{1}{2} \vartheta_{ij}^{(1) }\vartheta^{ij(1)}   - 4  \bar \chi \p_{\perp i}\Phi S_\perp^{i (1)} + 2 \bar \chi \p_{\perp i}\Phi \p_\perp^i T^{(1)} + \frac{2}{\cH}S_{\perp}^{i (1)} \p_{\perp i}\Delta \ln a^{(1)}  -2   S_{\perp }^{i (1)}S_{\perp }^{j (1)} \delta_{ij}  +  \left( \frac{1}{\bar \chi} \frac{\cH' }{\cH^3}  -  \frac{1}{\bar \chi \cH} \right) \nonumber \\
&& \times \left(\Delta \ln a^{(1)} \right)^2 + 2 \left[  -  \Phi - \frac{1}{\cH \bar \chi}  \Phi  + \frac{1}{\cH}  \frac{\ud \Phi}{\ud \bar \chi}  - \frac{1}{\bar \chi} T^{(1)}        -  \kappa^{(1)} \right]\Delta \ln a^{(1)}     + \frac{4}{\bar \chi} \int_0^{\bar \chi}  \ud \tilde{\chi} \Bigg\{  \bigg[-  { \Phi}^2   -  \Phi{'} T^{(1)} -2 \Phi  \kappa^{(1)} \nonumber \\
&& - \tilde \chi \tilde \p_{\perp i}\Phi  \p_\perp^i T^{(1)}   \bigg]     +2(\bar \chi - \tilde \chi) \bigg[   -  \Phi   \tilde \p_{\perp m}S_{\perp}^{m(1)}     +  \left( \frac{\ud}{\ud \tilde \chi} \Phi   -  \frac{1}{\tilde\chi} \Phi  \right) \kappa^{(1)}  \nonumber \bigg]\Bigg\}  + 4 \int_0^{\bar \chi}  \ud \tilde{\chi} \bigg[     -  \Phi  \tilde \p_{\perp m}S_{\perp}^{m(1)}    +  \left( \frac{\ud \Phi}{\ud \tilde \chi}   -  \frac{1}{\tilde\chi} \Phi  \right) \kappa^{(1)}  \bigg]  \nonumber \\
 &&       - 12 \Phi _o v _{\| \, o} -   \frac{1}{2} v _{\perp k \, o} v^{k  }_{\perp \, o}   -4  v_{\| \, o}\left( \frac{1}{\bar \chi} T^{(1)}+ 3  \int_0^{\bar \chi}   \frac{\ud \tilde{\chi}}{\tilde \chi} \Phi \right) + v_{\perp  i\, o}  \left[ 2\bar \chi  \p^i_{\perp }\Phi  +2 S^{i }_{\perp}  - \frac{1}{\cH} \p^i_{\perp }\Delta \ln a^{(1)} \right]\;.
\end{eqnarray}
Finally,  for the observed over-density, we find\footnote{When one chooses a suitable gauge, like the Poisson gauge that we use in this section, and setting initial conditions correctly there are no gauge modes because the observed over-density $\Delta_g$ is gauge invariant.}
 \begin{eqnarray}
\label{Deltag-1}
\Delta_g^{(1)} &=&\delta_g^{(1)} +\left[ b_e  - \frac{\cH'}{\cH^2} - 2 \Q  - 2 \frac{\left(1- \Q \right)}{\bar \chi \cH} \right]  \Delta \ln a^{(1)}  +\left(-1+2 \Q \right)  \Phi - \frac{1}{\cH}  \p_\|^2 v  + \frac{1}{\cH} \Phi {'} -2 \frac{\left(1- \Q \right)}{\bar \chi}  T^{(1)}  \nonumber\\
&& - 2 \left(1- \Q \right) \kappa^{(1)} \;,   
\end{eqnarray}
\begin{eqnarray}
\label{Deltag-2}
&&\Delta_g^{(2)} = \delta_g^{(2)}  +\left[ b_e-2 \Q  -   \frac{\cH'}{\cH^2} - \left(1 - \Q\right) \frac{2}{\bar \chi \cH} \right] \, \Delta \ln a^{(2)}  - \left(1 - \Q\right) \left( 2  \Psi^{(2)} +\frac{1}{2}\hat h_{\|}^{(2)} \right)    - \left(1- \Q \right) \frac{2}{\bar \chi} T^{(2)}- 2 \left(1- \Q \right) \kappa^{(2)} \nonumber \\  
&&  + \Phi^{(2)}+  \frac{1}{ \cH} \Psi^{(2)}{'}-  \frac{1}{2 \cH} \hat h^{(2)}_{\| }{'} -\frac{1}{ \cH }\p_\|^2 v^{(2)} -\frac{1}{ \cH }   \p_\|\hat v^{(2)}_\| +2\left(-1+2\Q\right)\Phi \delta_g^{(1)} - \frac{2}{\cH} \delta_g^{(1)} \p_\|^2 v  +\frac{2}{\cH}\delta_g^{(1)}  \Phi {'}+ \frac{2}{\cH}\left(2\Q+\frac{\cH'}{\cH^2}\right)\Phi \Phi {'}\nonumber \\ 
&&+ \left(-5 + 4\Q  +4 \Q^2 - 4 \frac{\p \Q}{\p \ln \bar L} \right){\Phi}^2 + \left( \p_\| v  \right)^2   - \frac{2}{\cH}\left(1+ 2\Q  +\frac{\cH'}{\cH^2}\right)\Phi \p_\|^2 v + \frac{2}{\cH^2}\left( \Phi {'}  \right)^2 + \frac{2}{\cH^2}\left(\p_\|^2 v  \right)^2  -\frac{2}{\cH}\Phi \p_\| \Phi   \nonumber \\
 &&  + \frac{2}{\cH^2} \p_\| v  \p_\|^2 \Phi  +\frac{4}{ \cH }\p_\| v  \p_\| \Phi   - \frac{2}{\cH^2} \Phi \p_\|^3 v    + \frac{2}{\cH^2}\Phi \frac{\ud \Phi {'} }{\ud \bar \chi}- \frac{2}{\cH^2} \p_\| v \frac{\ud \Phi {'} }{\ud \bar \chi}  +\frac{2}{\cH} \left(1+\frac{\cH'}{\cH^2} \right) \p_\| v \p_\|^2 v   - \frac{2}{\cH^2}\Phi  \p_\|^2 \Phi \nonumber \\
 && +\frac{2}{\cH} \left(1 -\frac{\cH'}{\cH^2} \right)  \p_\| v  \Phi {'}  -  \frac{4}{\cH^2}\p_\|^2 v   \Phi {'}  +  \frac{2}{\cH^2}\p_\| v \p_\|^3v +\frac{2}{\cH}\p_{\perp i} v \p^i_\perp \Phi   -\frac{4}{ \cH } \p_{\perp i} v   \p_{\perp}^i \p_\|  v   +\left(-1  +\frac{4}{ \bar \chi \cH } \right) \p_{\perp i} v    \p_{\perp}^i v  \nonumber \\
  &&  + \Bigg\{ \bigg[  -2 b_e - 4 \Q   + 4 b_e \Q  - 8 \Q^2  + 8 \frac{\p \Q}{\p \ln \bar L} +4 \frac{\p \Q}{\p \ln \bar a}   
  + 2  \frac{\cH'}{\cH^2} \left(1 - 2\Q\right)   + \frac{4}{\bar \chi \cH}\bigg(-1+\Q +2\Q^2  - 2\frac{\p \Q}{\p \ln \bar L} \bigg) \bigg] \Phi 
 \nonumber \\    
&&  +2 \left[ b_e - 2\Q   -  \frac{\cH'}{\cH^2}  - \frac{2}{\bar \chi \cH} \left(1 - \Q\right)  \right] \delta_g^{(1)}   - \frac{2}{\cH}  \frac{\ud  \delta_g^{(1)} }{\ud \bar \chi}+  \frac{2}{\cH}  \left[ -   b_e   + 2 \Q     +  \frac{\cH' }{\cH^2}    +  \frac{2}{\bar \chi \cH}  \left(1 - \Q\right)\right]  \p_\|^2 v  \nonumber \\    
 &&   + \frac{2}{\cH}\left[  -2  +  b_e  - \frac{\cH' }{\cH^2}  -   \frac{2}{\bar \chi \cH} \left(1 - \Q\right)\right]  \Phi {'}    - \frac{4}{\cH} \Q \p_\|  \Phi  + 4\bigg[  - \left(b_e-b_e\Q+2 \Q^2 -2\frac{\p \Q}{\p \ln \bar L}-\frac{\p \Q}{\p \ln \bar a} \right) +  \frac{\cH'}{\cH^2} \left(1 - \Q\right)  \nonumber \\    
&&  + \frac{1}{\bar \chi \cH}  \left(1-\Q +2\Q^2  -2\frac{\p \Q}{\p \ln \bar L}\right) \bigg] \left(\frac{1}{\bar \chi}T^{(1)}+\kappa^{(1)} \right)  \Bigg\}\,\Delta \ln a^{(1)}+ \Bigg\{-b_e+b_e^2+ \frac{\p b_e}{\p \ln \bar a}+6\Q-4 \Q b_e+4 \Q^2 \nonumber \\
&& -4\frac{\p \Q}{\p \ln \bar L}-4\frac{\p \Q}{\p \ln \bar a} + \left(1-2 b_e + 4\Q \right) \frac{\cH' }{\cH^2} -\frac{\cH'' }{\cH^3} +3\left( \frac{\cH' }{\cH^2} \right)^2 + \frac{6}{\bar \chi} \frac{\cH' }{\cH^3} \left(1 - \Q\right)   +  \frac{2}{\bar \chi^2 \cH^2} \bigg(1-\Q +2\Q^2  -2\frac{\p \Q}{\p \ln \bar L}\bigg) \nonumber \\
&& + \frac{2}{\bar \chi \cH} \left[ 1 - 2b_e - \Q  + 2b_e \Q  - 4 \Q^2 +4\frac{\p \Q}{\p \ln \bar L} +2 \frac{\p \Q}{\p \ln \bar a} \right]    \Bigg\} \left(\Delta \ln a^{(1)}\right)^2  + 4  \bigg[  +\frac{1}{\cH} \left(1 -\frac{\cH' }{\cH^2}\right)\Phi {'} + \frac{1}{\cH}   \p_\| \Phi\nonumber \\
 &&   +  \frac{1}{\cH} \left(1 +  \frac{\cH' }{\cH^2} \right)\p_\|^2 v    +  \frac{1}{\cH^2} \p_\|^2 \Phi  + \frac{1}{\cH^2} \p_\|^3 v   - \frac{1}{\cH^2}\frac{\ud \Phi {'} }{\ud \bar \chi}  \bigg] I^{(1)}  + \bigg[ - \frac{4}{\bar \chi}  \left(1- \Q \right) \delta_g^{(1)} - 2   \p_{\|}\delta_g^{(1)}   - \frac{4}{\bar \chi \cH} \left(1  - \Q \right)  \Phi {'} \nonumber \\ 
 &&+ \frac{4}{\bar \chi} \left( -1 + \Q+2 \Q^2- 2\frac{\p \Q}{\p \ln \bar L} \right)  \Phi +2\left(1-2 \Q\right) \p_{\|}\Phi  + \frac{4}{\bar \chi \cH} \left(1- \Q \right) \p_\|^2 v   +\frac{2}{\cH} \p_\|^3 v  -\frac{2}{\cH} \p_\| \Phi {'}    \bigg] T^{(1)}  +\bigg(1-\Q +2\Q^2   \nonumber \\ 
 &&  -2\frac{\p \Q}{\p \ln \bar L}\bigg) \left[  \frac{2}{\bar \chi^2} \left(  T^{(1)} \right)^2 + \frac{4}{\bar \chi} T^{(1)}   \kappa^{(1)}  \right] +4 \bigg[ - \left(1-\Q -2\Q^2 + 2\frac{\p \Q}{\p \ln \bar L}\right) \Phi  + \frac{1}{\cH} \left(1- \Q \right) \p_\|^2 v   -\frac{1}{\cH}\left(1- \Q \right) \Phi {'} \nonumber \\
&&      -  \left(1- \Q \right) \delta_g^{(1)} \bigg]\kappa^{(1)} +  \left(1- \Q \right)  \vartheta_{ij}^{(1) }\vartheta^{ij(1)}  + 2\left(1-\Q +2\Q^2- 2\frac{\p \Q}{\p \ln \bar L}\right) \left(\kappa^{(1)}\right)^2   -2  \left(1- \Q \right)\big|\gamma^{(1)}\big|^2  + 4 \bigg[  \frac{\bar \chi}{\cH}\bigg(  \p_{\perp i}\Phi {'} \nonumber \\
 &&  -  \p_{\perp i} \p_\|^2 v  \bigg)   +   \bar \chi  \p_{\perp i} \delta_g^{(1)}   +  \bar \chi  \p_{\perp i} \Phi  -2\bar \chi \left(1  -  \Q \right)\p_{\perp i} \Phi    +  \frac{1}{\cH}\left(1- \Q \right)   \p_{\perp i}  \Delta \ln a^{(1)} \bigg] S_{\perp}^{i (1)}  -4 (1-\Q)S_{\perp }^{i (1)}S_{\perp }^{j (1)} \delta_{ij} \nonumber \\
 && +2 \bigg[  \frac{2}{\bar \chi \cH} \p_{\perp i} v   -  \frac{\bar \chi}{\cH} \p_{\perp i} \Phi {'} + \frac{\bar \chi}{\cH} \p_{\perp i} \p_\|^2 v   -\frac{2}{\cH}\p_{\perp i}\p_\| v  \   -   \bar \chi  \p_{\perp i} \delta_g^{(1)}     - \bar \chi  \p_{\perp i} \Phi       +  2 \bar \chi   \left(1- \Q \right) \p_{\perp i} \Phi     \bigg]  \p_\perp^i T^{(1)}  \nonumber \\
&&     +4\Q^{(1)} \left[ \Phi  - \left( 1 - \frac{1}{\bar \chi \cH} \right) \Delta \ln a^{(1)}  +\frac{1}{\bar \chi} T^{(1)} +  \kappa^{(1)}\right]   + 8(1-\Q)\Bigg\{  \int_0^{\bar \chi}  \ud \tilde{\chi}\bigg[       -  \Phi  \tilde \p_{\perp m}S_{\perp}^{m(1)}  +  \left( \frac{\ud \Phi}{\ud \tilde \chi}   -  \frac{1}{\tilde\chi} \Phi  \right) \kappa^{(1)}  \bigg]   \nonumber \\
 && - \frac{1}{\bar \chi}\int_0^{\bar \chi}  \ud \tilde{\chi} \bigg(    { \Phi}^2  +  \Phi{'} T^{(1)} +2 \Phi \kappa^{(1)}   + \tilde \chi \tilde \p_{\perp i}\Phi  \p_\perp^i T^{(1)}   \bigg) +  \frac{1}{\bar \chi} \int_0^{\bar \chi}  \ud \tilde{\chi} ~ (\bar \chi - \tilde \chi) \bigg[  -  2 \Phi   \tilde \p_{\perp m}S_{\perp}^{m(1)}     
 + 2  \left( \frac{\ud  \Phi}{\ud \tilde \chi}   -  \frac{1}{\tilde\chi} \Phi  \right) \kappa^{(1)}  \nonumber \bigg] \Bigg\}   \nonumber \\
           &&- \left(1 - \Q\right) \left( 24\Phi _o v _{\| \, o}  +   v _{\perp k \, o} v^{k  }_{\perp \, o} \right)    - 8 (1-\Q)  v_{\| \, o}  \left(  \frac{1}{\bar \chi} T^{(1)}+ 3 \int_0^{\bar \chi}   \frac{\ud \tilde{\chi}}{\tilde \chi} \Phi \right)    + 2\left(\Phi _o-v _{\| \, o}\right) \bigg[\left(\frac{\cH' }{\cH^3}+ \frac{1}{\cH}\right) \p_\|^2 v   \nonumber \\
           &&  +\left(-\frac{\cH' }{\cH^3}+ \frac{1}{\cH}\right)  \Phi {'}      + \frac{1}{\cH^2} \p_\|^2 \Phi  + \frac{1}{\cH^2}\p_\|^3 v + \frac{1}{\cH}  \p_\| \Phi - \frac{1}{\cH^2}\frac{\ud \Phi {'} }{\ud \bar \chi} \bigg]   +v^{i  }_{\perp \, o }\bigg[  -2   \bar \chi  \p_{\perp i} \Phi  + 2 \frac{\bar \chi}{\cH} \p_{\perp i}  \left(-\Phi {'} +\p_\|^2v  \right)    - 2 \bar \chi    \p_{\perp i} \delta_g^{(1)}   \bigg]  \nonumber \\
 &&     + \left(1 - \Q\right) v_{\perp  i\, o} \bigg[   4  S^{i }_{\perp}  - \frac{2}{\cH}  \p^i_{\perp}  \Delta \ln a^{(1)} + 4 \bar \chi  \p^i_{\perp} \Phi  \bigg]   \;.
   \end{eqnarray}

Here found several new terms and contributions that we cannot a priori neglect. In particular: {\it i)} comparing our result with \cite{Yoo:2014sfa}, we have new magnification terms  proportional to  $({\p \Q}/{\p \ln \bar a})$ and $ \Q^{(1)}(\bar x^\alpha, \bar L)=- (\p b_1/\p \ln \bar L)\delta_{m \, {\rm CO}}^{(1)}$ (if $\delta_{g}=\delta_{g \, {\rm CO}}$).
{\it ii)} Comparing $\kappa^2$ terms in Eqs.\ (\ref{New-limit}) or (\ref{Deltag-2}) with Eq.\ (14) of  \cite{Schmidt:2008mb}, we note that the coefficient of $\Q$ is different  [in Eq.\ (\ref{New-limit}) or Eq. (\ref{Deltag-2}) it is $-2$ and in Eq.\ (14) of  \cite{Schmidt:2008mb} it is $-6$].  This discrepancy is related to the term $-2\Q\Delta (\M^{-1})^{(1)}\Delta V^{(1)}$that we find in the intermediate step of  Eq.\ (\ref{Deltag-2}).  {\it iii)}  There is no ${\p \Q/\p \ln \bar L}$   in Eq.\ (14) of   \cite{Schmidt:2008mb}.
{\it iv)} It is clear that in Eq.(\ref{Deltag-2}) we generalize the result obtained in Eq.\ (14) of  \cite{Schmidt:2008mb} adding all the relativistic contributions from velocities, Sachs-Wolfe, integrated SW and time-delay terms. {\it v)} If we set $\Q=\Q^{(1)}=0$ we find the same results that have been obtained in \cite{Bertacca:2014dra}, see Eqs. (\ref{Deltag1-Q=0}) and (\ref{Deltag2-Q=0}) in Appendix \ref{B}.

Let us conclude this section with the following comment on  the correct frame to define the local bias.  
Fluctuations of galaxy number density are related to the underlying matter density fluctuation $\delta_m$ on cosmological scales by a local bias.
In order to define this correctly,  we need to choose an appropriate frame where the baryon velocity perturbation vanishes. 
Then  the baryon rest frame coincides with the CDM rest frame and,  in $\Lambda$CDM, this rest frame is defined up to second order by the comoving-synchronous  gauge (S)  \cite{Matarrese:1997ay, Wands:2009ex, Bartolo:2010rw, Bartolo:2010ec, Bruni:2011ta, Bruni:2013qta}.  In this S-gauge, the galaxy and matter over-densities are gauge invariant \cite{Kodama:1985bj}.

By setting initial conditions correctly, the gauge mode in comoving-synchronous gauge can be removed (since it is a function only on spatial coordinates) and this gauge is equivalent to Lagrangian frame.  The correct frame to define the local bias is the Lagrangian frame. Indeed, this frame has the advantage that  the local (Lagrangian) bias  is related to the halo mass function through the peak-background split approach  \cite{Bouchet:1994xp, Catelan:1997qw, Wands:2009ex, Matsubara:2011ck}. 

The S-gauge is defined by the conditions  $g_{00}=-1$,    $g_{0i}=0$ and $v^{i}=0$. Then
 \begin{eqnarray} 
 \label{Comoving-ortnogonal_metric}
 \ud s^2 = a(\eta)^2\left\{-\ud\eta^2+\left[\delta_{ij} -2 \psi \delta_{ij} + \left(\p_i\p_j-\frac{1}{3} \delta_{ij}\nabla^2\right) \xi+\frac{1}{2}h_{ij {\rm S}}^{(2)}\right]\ud x^i\ud x^j\right\},
\end{eqnarray}
where $h^{(2)}_{ij {\rm S}} =- 2 \psi^{(2)} \delta_{ij} + F_{ij  {\rm S}}^{(2)}$, with $F^{(2)}_{ij   {\rm S}}= (\p_i\p_j- \delta_{ij}\nabla^2/3) \xi^{(2)}+\p_i \hat \xi^{(2)}_j+ \p_j \hat \xi^{(2)}_i+\hat h^{(2)}_{ij}$, $\p_i \hat \xi^{i(2)}=\p_i \hat h^{ij(2)}=0$. 
Here, for simplicity, we neglect vector and tensor perturbations at  first order, i.e.  $\hat \xi_j=  \hat h_{ij}=0$\;.

In order to obtain the galaxy fractional number overdensity  $\delta_{g {\rm S}}$, we transform the metric perturbations from the Poisson  to  comoving-synchronous  gauge \cite{Matarrese:1997ay, Bertacca:2014dra}
 \begin{eqnarray} 
 \label{dg1}
\delta_{g }&=&\delta_{g {\rm S}}- b_e \cH v+ 3 \cH v ,\\
 \label{dg2}
\delta_{g }^{(2)} &=&  \delta_{g  {\rm S}}^{(2)}- b_e \cH v^{(2)}+ 3 \cH v^{(2)} + \left( b_e \cH'-3 \cH' + \cH^2  \frac{\p b_e}{\p  \ln \bar a} + b_e^2 \cH^2  -6  b_e  \cH^2 + 9 \cH^2 \right) v^2 + \cH b_e  v  {v}' - 3 \cH   v  {v}' \nonumber\\
&&  -2\cH b_e  v \delta_{g  {\rm S}}   + 6 \cH  v \delta_{g  {\rm S}} - 2  v {\delta_{g{\rm S}}}' - \frac{1}{2} \p^i \xi \left(- b_e \cH \p_i v+ 3 \cH \p_i v + 2 \p_i \delta_{g  {\rm S}} \right) - \left(b_e-3\right) \cH  \nabla^{-2}\bigg( v \nabla^2 {v}' - {v}' \nabla^2 v  \nonumber \\
&&- 6 \p_i \Phi \p^i v - 6 \Phi \nabla^2 v  + \frac{1}{2} \p_i \xi \p^i \nabla^2 v + \frac{1}{2} \p_i v \p^i \nabla^2 \xi + \p_i \p_j \xi \p^i \p^j v\bigg).
\end{eqnarray} 
Note that  $v={\xi}'/2$.

Then the scale-independent  bias at first and at second order (down to mildly nonlinear scales) is given by \cite{Matsubara:2011ck,Bertacca:2015mca}
\be
\label{bias}
\delta_{g {\rm S}}^{(1)}+\frac{1}{2} \delta_{g  {\rm S}}^{(2)} = b_{1}^L \delta_{m  {\rm S}}^{(1)} +\frac{1}{2} b_{1}^L  \delta_{m  {\rm S}}^{(2)} +\frac{1}{2} b_{2}^L \big( \delta_{m  {\rm S}}^{(1)} \big)^2\;.
 \ee
Expressions \eqref{dg1}--\eqref{bias} can then be substituted into  \eqref{Deltag-2}, thus incorporating the bias correctly.

Finally,  in order to make a correct result, it is important to study the degrees of freedom of these equations have for given initial conditions. It will also be important for an accurate analysis of the `contamination' of primordial non-Gaussianity by relativistic projection effects. This is the subject of ongoing work  \cite{Bertacca:2014n}.


\section{Conclusions}
\label{Sec:Conclusions}

In this paper we have presented the observed galaxy counts to second order in redshift space on cosmological scales  for a $\Lambda$CDM model, including all general relativistic effects and as function of the magnification. The main result is given by Eq.\ (\ref{Deltag-2}).
 
We have found new terms and contributions that we cannot neglect: 

{\it Firstly}, comparing our result with \cite{Yoo:2014sfa}, we have new magnification terms  proportional to  \[\left({\p \Q}/{\p \ln \bar a}\right) \quad \quad {\rm and} \quad \quad \Q^{(1)}(\bar x^\alpha, \bar L)=- (\p b_1/\p \ln \bar L)\delta_{m \, {\rm CO}}^{(1)} \quad ({\rm if} ~ \delta_{g}=\delta_{g \, {\rm CO}}) .\]

{\it Secondly}, comparing $\kappa^2$ terms in Eqs.\ (\ref{New-limit}) or (\ref{Deltag-2}) with Eq.\ (14) of  \cite{Schmidt:2008mb}, we note that the coefficient of $\Q$ is different  [in Eq.\ (\ref{New-limit}) or Eq. (\ref{Deltag-2}) it is $-2$ and in Eq.\ (14) of  \cite{Schmidt:2008mb} it is $-6$].  This discrepancy is related to the term \[-2\Q\Delta (\M^{-1})^{(1)}\Delta V^{(1)}\] that we find in the intermediate step of  Eq.\ (\ref{Deltag-2}). Then, there is no ${\p \Q/\p \ln \bar L}$   in Eq.\ (14) of   \cite{Schmidt:2008mb}.

{\it Lastly} we generalize the result obtained in Eq.\ (14) of  \cite{Schmidt:2008mb} adding all the relativistic contributions from velocities, Sachs-Wolfe, integrated SW and time-delay terms.

The results presented in this work suggest that  we have to take into account the magnification corrections  when making measurements of non-Gaussianity. If we neglect these effects, we could potentially not estimate correctly the sensitivity of galaxy surveys to primordial non-Gaussianity \cite{Bertacca:2014n}. Finally, this allows for an investigation of whether general relativistic effects are measurable beyond the linear approximation in the mildly nonlinear regime in future surveys.

\[\]{\bf Acknowledgments:}\\
We thank Nicola Bartolo, Chris Clarkson, Roy Maartens, Sabino Matarrese, Prina Patel and Alvise Raccanelli for helpful discussions. DB  is  supported by the South African Square Kilometre Array Project. 
We thank Enea di Dio, Ruth Durrer, Kazuya Koyama, Giovanni Marozzi  for alerting us to the possibility of an error in our results.

\appendix

\section{Useful relations in the Poisson gauge}\label{A}

From Eq.\ (\ref{Poiss-metric}) the perturbation of  FRW metric metric $g_{\mu \nu}$ and $g^{\mu \nu}$ is
\begin{eqnarray}
\begin{array} {lll}
g_{00}=- a^2 \left(1+ 2 \Phi + \Phi^{(2)}\right),  & \quad& g^{00}=- a^{-2} \left[1-2  \Phi  - \Phi^{(2)} +4  {\Phi}^2\right], \\ \\
g_{0i}= a^2 \omega^{(2)}_i,  & \quad&  g^{0i}=a^{-2}\omega^{i (2)}, \\ \\
g_{ij}= a^2 \left(\delta_{ij} -2 \delta_{ij} \Phi  -\delta_{ij} \Psi^{(2)}+\hat h^{(2)}_{ij}/2 \right),  & \quad&   g^{ij}=a^{-2}  \left[ \delta^{ij}+2 \delta^{ij} \Phi  + \delta^{ij} \Psi^{(2)}-\hat h^{ij(2)}/2 +4\delta^{ij} (\Phi )^2\right], \\
  \end{array}  \nonumber \\
\end{eqnarray}

For four-velocity $u^\mu$, we find
\begin{eqnarray}
\label{Poiss-u0i}
u_0&=&-a\left[1+\Phi +\frac{1}{2}\Phi^{(2)}-\frac{1}{2} {\Phi}^2+\frac{1}{2}v_k v^{k }\right] ,\\ 
u_i&=&a\left[v_i +\frac{1}{2}\left(v_i^{(2)}+2\omega_i^{(2)}\right)- 2 \Phi v_i \right], \\
u^0&=&\frac{1}{a}\left[1-\Phi -\frac{1}{2}\Phi^{(2)}+\frac{3}{2} {\Phi}^2+\frac{1}{2}v_k v^{k }\right] ,\\
u^i&=&\frac{1}{a}\left(v^{i }+\frac{1}{2}v^{i(2)}\right)\;.
\end{eqnarray}

Given $T_m^{\mu \nu}=\rho_m u^\mu u^\nu$, i.e. the cold dark matter stress-energy tensor,  for  first and second-order perturbations we obtain
\begin{eqnarray}
&& \delta_m {'} + \p_i v^{i }- 3 \Phi  {'}=0\;, \nonumber \\  
&&v^{i }{'}+ \cH  v^{i }+ \p^i  \Phi  =0\;, \nonumber \\ 
&& \frac{1}{2} \delta_m^{(2)}{'} + \frac{1}{2}\p_i v^{i(2)}-\frac{3}{2} \Psi^{(2)}{'}+\frac{1}{4} \hat h^{i(2)}_i {'}-  \cH  v^{i } v _i + \left(\Phi  + \delta_m  \right)\p_i v^{i } +v^{i }  \p_i \delta_m    -3 \delta_m   \Phi {'} -3 v^{j } \p_j  \Phi - 6 \Phi  \Phi {'} =0\;, \nonumber \\
&& \left( \frac{1}{2} v^{i(2)} + \omega^{i(2)}\right){'}+ \cH  \left( \frac{1}{2} v^{i(2)}+ \omega^{i(2)} \right) + \frac{1}{2}  \p^i  \Phi^{(2)} 
+ v^{j } \p_j  v^{i }  -2 v^{i  } \Phi {'}  -   \Phi  \p^i  \Phi    +2 \p^i  \Phi    \Phi  =0 \;.
\label{ConsEq}
\end{eqnarray}

The geodesic equation for the comoving null geodesic vector $k^\mu (\chi)=(\ud x^\mu/\ud \chi)(\chi) $  is 
\begin{eqnarray} 
\frac{\ud k^\mu (\chi) }{ \ud \chi}+\hat \Gamma^\mu_{\alpha \beta}(x^\gamma)\, k^\alpha(\chi) \,  k^\beta(\chi) =0
\end{eqnarray}
where $\hat \Gamma^\mu_{\alpha \beta}$ are the Christoffel symbols defined using the comoving metric $\hat g_{\mu \nu}= g_{\mu \nu}/a^2$ or $\hat g^{\mu \nu}=a^2 g_{\mu \nu}$.
Expanding $k^\mu (\chi)$ and $\hat \Gamma^\mu_{\alpha \beta}(x^\gamma)$ up to second order in the observed coordinates, we have
\begin{eqnarray} 
\label{photon-gamma-pert}
 k^\mu (\chi) &=& k^\mu (\bar\chi)  + \delta\chi  \frac{\ud k^\mu }{ \ud \chi}(\bar \chi) +\frac{1}{2}  \delta\chi^2  \frac{\ud k^\mu }{ \ud \chi}(\bar \chi)\;, \nonumber \\
 \hat \Gamma^\mu_{\alpha \beta}( x^\gamma) &=&\hat \Gamma^\mu_{\alpha \beta}(\bar x^\gamma) + \Delta x^\nu \frac{\p }{\p \bar x^\nu} \hat \Gamma^\mu_{\alpha \beta}(\bar x^\gamma) + \frac{1}{2} \Delta x^\nu  \Delta x^\sigma \frac{\p^2 }{\p \bar x^\nu \p \bar x^\sigma} \hat \Gamma^\mu_{\alpha \beta}(\bar x^\gamma)\;. 
\end{eqnarray}
After some tedious calculations, we can write the geodesic equation for  $k^\mu (\bar \chi)$ in the following way\footnote{It is valid up to second order.}
\begin{eqnarray} 
\label{Eqphoton}
\frac{\ud k^\mu (\bar\chi) }{ \ud \bar \chi}+\left(\hat \Gamma^\mu_{\alpha \beta}(\bar x^\gamma)+ \delta x^{\sigma} \frac{\partial \hat \Gamma^\mu_{\alpha \beta}}{\partial \bar x^{\sigma}}(\bar x^\gamma)\right)k^\alpha(\bar\chi)  k^\beta(\bar\chi) =0 \;.
\end{eqnarray}
It is clear that we recover the result obtained in \cite{Pyne:1995bs}.

Now, at  zeroth order, we obtain Eq.\ (\ref{kmu-0}).
At first order, Eq.\ (\ref{Eqphoton}) yields 
\begin{eqnarray} 
\label{Poiss-Eq-dnu&de}
\frac{\ud}{\ud\bar \chi} \left(\delta \nu^{(1)}  - 2 \Phi \right) =2 \Phi {'}  \;,   \quad \quad \quad  \frac{\ud}{\ud\bar \chi} \left( \delta n^{i (1)} -2 \Phi  n^i \right) = -2 \p^i \Phi \;. 
\end{eqnarray}
At second order we have
\begin{eqnarray} 
\label{Poiss-dnu-2}
&&\frac{\ud}{\ud\bar \chi} \left(\delta \nu^{(2)} - 2 \Phi^{(2)} -2\omega_{\|}^{(2)}+4\Phi  \delta \nu^{(1)} \right) = \Phi^{(2)}{'} +2\omega_{\|}^{(2)}{'}+ \Psi^{(2)}{'} - \frac{1}{2}\hat h_{\|}^{(2)}{'}+ 4\delta n^{i }\p_i  \Phi +4\delta n _\| \Phi {'} \nonumber \\
 &&-4 \left(\frac{\ud}{\ud \bar \chi} \Phi{'} +   \Phi{''}     \right) T^{(1)}+4  \frac{\ud}{\ud \bar \chi}  \left(\frac{\ud}{\ud \bar \chi} \Phi+ \Phi{'} \right) \delta x_{\|}^{(1)}  + 4 \bigg[ \p_{\perp i}\left( \frac{\ud}{\ud \bar \chi} \Phi+ \Phi{'}  \right) - \frac{1}{\bar \chi} \p_{\perp i} \Phi \bigg] \delta x_{\perp}^{i (1)} 
\end{eqnarray}
and
\begin{eqnarray} 
\label{Poiss-de-2}
&&\frac{\ud}{\ud\bar \chi} \left( \delta n^{i(2)} - 2 \omega^{i(2)}-2\Psi^{(2)} n^i+ \hat h_{j}^{i(2)} n^j- 4 \delta n^{i (1)} \Phi  \right) =
- \p^i \Phi^{(2)} -2 \p^i\omega_{\|}^{(2)}+ \frac{2}{\bar \chi} \omega_{\perp}^{i(2)}- \p^i \Psi^{(2)}  + \frac{1}{2} \p^i\hat h_{\|}^{(2)}\nonumber \\
&&- \frac{1}{\bar \chi} \Perp^{ij} \hat h_{jk}^{(2)} n^k   + 4 \delta \nu^{(1)} \left(\p^i \Phi   +n^i \Phi  {'} \right) + 4 \delta n^{j (1)} \left(- n_j \p^i \Phi   +  n^i  \p_j \Phi    \right)  +4   {\left(  \p^i \Phi  -   n^i \frac{\ud}{\ud \bar \chi} \Phi \right)}'  T^{(1)}  \nonumber \\
 &&
 - 4 \frac{\ud}{\ud \bar \chi} \left( \p^i \Phi   -   n^i \frac{\ud}{\ud \bar \chi} \Phi \right) \delta x_{\|}^{(1)} - 4 \bigg[  \p_{\perp l} \left(  \p^i \Phi  -  n^i  \frac{\ud}{\ud \bar \chi}  \Phi \right)   + \frac{1}{\bar \chi} \Perp_l^j \left(  \delta_j^i\frac{\ud}{\ud \bar \chi}   \Phi  + n^i  \p_j  \Phi\right) \bigg]  \delta x_{\perp}^{l (1)} \;.
\end{eqnarray}

To solve Eqs.\  (\ref{Poiss-Eq-dnu&de}),  (\ref{Poiss-dnu-2})  and (\ref{Poiss-de-2}) we require the values of $\delta \nu^{(1)}$, $\delta \nu^{(2)}$, $\delta n^{i (1)}$ and  $\delta n^{i (2)}$ today. In this case we need all the components of the comoving tetrad  $E^{\hat \alpha}_\mu$ which is defined through the following  relations
\begin{eqnarray}
\label{LambdaE}
\hat g^{\mu \nu} E^{\hat \alpha}_\mu E^{\hat \beta}_\nu& = &\eta^{\hat  \alpha \hat \beta}\;, \quad  \quad  \eta_{\hat  \alpha \hat \beta} E^{\hat \alpha}_\mu E^{\hat \beta}_\nu = \hat g_{\mu \nu}\;,  \quad  \quad \hat g^{\mu \nu} E^{\hat \beta}_\nu = E^{\hat \beta \mu}\;,    \quad  \quad \eta_{\hat  \alpha \hat \beta}  E^{\hat \beta}_\nu  =  E_{\hat \beta \nu}  \;,
\end{eqnarray}
where $ \eta_{\hat  \alpha \hat \beta} $ the comoving Minkowski metric.
 If we choose $u^\mu$ as the time-like basis vector, then  $u_\mu=a \, E_{\hat{0} \mu}$  and $u^\mu= a^{-1}E_{\hat{0}}^\mu$. In the background $ E_{\hat{0}\mu}^{(0)}=(-1, {\bf 0})$, and, at first and second order, we get
\begin{eqnarray} \label{Poiss-E-1-2}
\begin{array} {lll}
 E_{\hat 0 0} = - \Phi \;, & \quad& E_{\hat 0 i} =  v_i \;,  \\  \\
E_{\hat a 0} = - v_{\hat a} \;, & \quad& E_{\hat a i} =- \delta_{\hat a i}\Phi \;,   \\  \\
\frac{1}{2}  E_{\hat 0 0}^{(2)} =  -\frac{1}{2}\Phi^{(2)}+\frac{1}{2} {\Phi}^2-\frac{1}{2}v_k v^{k }   \;,  \\  \\
\frac{1}{2}  E_{\hat 0 i}^{(2)}=  \frac{1}{2}\left(v_i^{(2)}+2 \omega_i^{(2)}\right)-2 \Phi v_i  \;, \\  \\
\frac{1}{2}  E_{\hat a 0}^{(2)}= -\frac{1}{2} v_{\hat a}^{(2)}\;, \\  \\  
\frac{1}{2}  E_{\hat a i}^{(2)}= -\frac{1}{2} \delta_{\hat a j} \Psi^{(2)}+\frac{1}{4}  \hat h_{\hat a j}^{(2)} +  \frac{1}{2} v_i  v_{\hat a} - \frac{1}{2}\delta_{\hat a j}   {\Phi}^2\;. \\  \\
\end{array} 
\end{eqnarray}

Assuming $a_o = 1$  for $\bar \chi=0$, we have $k_{\hat{0} o} =( E_{\hat{0} \mu}k^\mu)|_o=1\;,~k_{\hat{a}  o} = (E_{\hat{a} \mu} k^\mu)|_o=n_{\hat{a}} $. Then we find, at first order, 
\begin{eqnarray}
\label{Poiss-dnude-1o}
\delta\nu^{(1)}_o=\Phi _o+v _{\| o}\;, \quad \quad \quad \delta n^{\hat a  (1)}_o=-v^{\hat a  }_o + n^{\hat a} \Phi _{ \, o}  \;,
\end{eqnarray}
and, at second order,
\begin{eqnarray}
\label{Poiss-dnude-2o}
  \delta\nu^{(2)}_o&=&  \Phi^{(2)}_o+ v^{(2)}_{\| o}+ 2 \omega^{(2)}_{\| o}- 3  \left(\Phi _{ \, o}\right)^2 -4 v _{\| o} \Phi _o -  v_{k \, o}  v_o^{k  }  \nonumber \\
 \delta n^{\hat a (2)}_o&=&- v^{\hat a (2)}_o+ n^{\hat a} \Psi^{(2)}_{\, o} -\frac{1}{2}  n^i \hat h_{i \,o}^{\hat a (2)} +  v_o^{\hat a  }  v_{\| \, o} +  3 \, n^{\hat a}  \left(\Phi _{ \, o}\right)^2   \;,
\end{eqnarray}
where we define $v^{i  }_o \equiv (\p^i v )_o$, i.e. $\p^i v $ evaluated at the observer, and $v^{i (2)}_o \equiv (\p^i v^{(2)})_o+ \hat v^{i (2)}_o$.

From Eq.\ (\ref{Poiss-dnu&de}) and the constraint from Eq.\ (\ref{Poiss-dnude-1o}), we obtain at first order
\begin{eqnarray}
\label{Poiss-dnu&de}
\delta\nu^{(1)} &=&- \left (\Phi _o-v _{\| \, o}\right)+ 2 \Phi   + 2 \int_0^{\bar \chi} \ud \tilde \chi \, \Phi {'} = - \left (\Phi _o-v _{\| \, o}\right)+ 2 \Phi  - 2I^{(1)}  \;, \\
\delta n^{i (1)}&=& -v^{i  }_o- n^i \Phi _{\, o}  +2 n^i \Phi  -2 \int_0^{\bar \chi} \ud \tilde \chi\,  \tilde \p^i \Phi  =  n^i \delta n_\|^{ (1)}+\delta n_\perp^{i  (1)}\;,
\end{eqnarray}
where
\begin{eqnarray}
\label{Poiss-dnue-||perp-1}
\delta n_\|^{(1)}=\Phi _o-v _{\| \, o}+2I^{(1)} \;,   \quad \quad \quad \delta n_\perp^{i  (1)}=  -v^{i  }_{\perp \, o }+ 2S_{\perp}^{i (1)}  \;.
\end{eqnarray}
Let us point out the following useful relation $\delta n_\|^{(1)}+  \delta\nu^{(1)} =2 \Phi $\;.

At second order we find
\begin{eqnarray}
\label{Poiss-dnu2-2}
 &&\delta\nu^{(2)}=- \Phi^{(2)}_o+ v^{(2)}_{\| \, o}+ \left(\Phi _{ \, o}\right)^2-8\Phi _o v _{\| \, o}- v _{k\, o} v^{k  }_o  +4 \left(\Phi _o-v _{\| \, o}\right) \left(  \Phi   - I^{(1)} \right) - 8 v_{\| \, o} \int_0^{\bar \chi}   \frac{\ud \tilde{\chi}}{\tilde \chi} \Phi +2\Phi^{(2)}+ 2 \omega^{(2)}_{\| }-8  {\Phi}^2  \nonumber \\
&&+8 \, \Phi  I^{(1)}  - 2 I^{(2)} -8 \left(I^{(1)}\right)^2   -8 \Phi  \kappa^{(1)}   + 4\left( \frac{\ud}{\ud \bar \chi} \Phi + \Phi{'} \right) \delta x_{\|}^{(1)}  -4 \Phi{'}  T^{(1)}  +4  \p_{\perp i} \Phi ~\delta x_{\perp}^{i (1)}   \nonumber\\
          &&    +4 \int_0^{\bar \chi}  \ud \tilde{\chi} \bigg[  - \Phi{''}     T^{(1)}   - 2 \Phi{'}  \Phi   - 2  \Phi{'}    I^{(1)}   
             -  2 \Phi  \tilde \p_{\perp j}S_{\perp}^{j(1)}   + 2 \bigg( \frac{\ud}{\ud \tilde \chi} \Phi    -  \frac{1}{\tilde\chi} \Phi   \bigg) \kappa^{(1)} +   \tilde \p_{\perp i} \Phi{'}  ~   \delta x_{\perp}^{i (1)} \bigg] \;,
\end{eqnarray}
and, splitting $\delta n^{i(2)}= n^i \delta n_\|^{ (2)}+\delta n_\perp^{i (2)}$, we obtain
\begin{eqnarray}
\label{Poiss-de_||-2}  
&& \delta n_\|^{(2)}=\Phi^{(2)}_o- v^{(2)}_{\| \, o} +  {v _{\| \, o}}^2-  {\Phi _{\, o}}^2 + 4  \left (\Phi _o-v _{\| \, o}\right) I^{(1)} +4v_{\perp i \, o}  S_{\perp }^{i (1)}  -\Phi^{(2)}+ \Psi^{(2)} -\frac{1}{2}\hat h_{\|}^{(2)}  +2 I^{(2)} +4  {\Phi}^2  -8 \left(I^{(1)}\right)^2 \nonumber \\
 &&-4\delta_{ij} S_{\perp }^{i (1)} S_{\perp }^{j (1)}+ 8 \int_0^{\bar \chi} \ud \tilde \chi  \left( \Phi {'}  \Phi  \right)    - 4 \Phi{'} \delta x_{\|}^{(1)}  +4 \int_0^{\bar \chi}  \ud \tilde{\chi} \bigg[  \Phi{''}     T^{(1)}   +2  \Phi{'}  I^{(1)}     -  \tilde \p_{\perp i}  \Phi{'}  ~  \delta x_{\perp}^{i (1)} \bigg] 
\end{eqnarray}
and
\begin{eqnarray}
\label{Poiss-de_perp-2}  
 &&\delta n_\perp^{i(2)}= - 2 \omega^{i(2)}_{\perp \, o}-  v^{i(2)}_{\perp \, o} + \frac{1}{2} n^j \hat h_{ j k\, o}^{(2)} \Perp^{ki}+   v _{\| \, o}v^{i }_{\perp \, o} +4 \left(\Phi _o-v _{\| \, o}\right)  S_{\perp}^{i (1)} - 4 v^{i  }_{\perp \, o }   \Phi - 4 v_{\perp o}^{i}  \int_0^{\bar \chi} \frac{ \ud \tilde{\chi}}{\tilde \chi} \Phi  +2\omega^{i(2)}_{\perp} -   n^j \hat h_{jk}^{(2)} \Perp^{ki}\nonumber \\
&& +2 S_{\perp }^{i(2)} +8 \, \Phi  S_{\perp}^{i (1)}   -4 \p^i_{\perp } \Phi ~ \delta x_{\|}^{(1)}  -  \frac{4}{\bar \chi}  \Phi ~ \delta x_{\perp}^{i (1)}    +4 \int_0^{\bar \chi}  \ud \tilde{\chi} \bigg[ 2 \Phi \tilde\p^i_\perp \Phi   -2 \tilde\p^i_\perp \Phi   I^{(1)} +  \tilde \p_{\perp }^i    \Phi{'}        T^{(1)}    +   \frac{2}{\tilde \chi} \Phi~ S_{\perp}^{i(1)} \nonumber\\
&&    -   \bigg( \Perp^{im} \tilde \p_{\perp j}\tilde   \p_{\perp m}  \Phi +  \frac{1}{\tilde\chi} \Perp^i_j  \Phi{'}  +   \frac{1}{\tilde\chi^2} \Perp^i_j \Phi \bigg)  \delta x_{\perp}^{j (1)}   \bigg] \;,
\end{eqnarray}
where
\begin{eqnarray}
S_{\perp}^{i(2)} &=& -\frac{1}{2} \int_0^{\bar \chi} \ud \tilde \chi \left[ \tilde\p^i_\perp \left( \Phi^{(2)} +2  \omega^{(2)}_{\| } + \Psi^{(2)}- \frac{1}{2} \hat h^{(2)}_{\| }\right) + \frac{1}{\tilde \chi} \left(-2 \omega^{i (2)}_{\perp }+  n^k \hat h_{kj}^{(2)} \Perp^{ij}  \right)\right]\;.
\label{Poiss-varsigma}
\end{eqnarray}

Combining Eqs.\  (\ref{Poiss-dnu2-2}) and (\ref{Poiss-de_||-2})  we  obtain following useful relation

\begin{eqnarray} 
\label{Poiss-dnu+de_||-2}
&& \delta\nu^{(2)} +  \delta n_\|^{(2)}= -8\Phi _o v _{\| \, o}  -   v _{\perp k \, o} v^{k  }_{\perp \, o}+4 \left (\Phi _o-v _{\| \, o}\right)   \Phi  
  - 8  v_{\| \, o}  \int_0^{\bar \chi}   \frac{\ud \tilde{\chi}}{\tilde \chi} \Phi  + 4 \, v_{\perp i \, o}  \, S_{\perp }^{i (1) }  + \Phi^{(2)}+ 2 \omega^{(2)}_{\| } +  \Psi^{(2)}   \nonumber \\
&&-\frac{1}{2}\hat h_{\|}^{(2)}-4 {\Phi}^2+8\Phi  I^{(1)}     -4 S_{\perp }^{i (1)}S_{\perp }^{j (1)} \delta_{ij}  + 4 \frac{\ud \Phi}{\ud \bar \chi}  \delta x_{\|}^{(1)}  - 4 \Phi{'} T^{(1)}   +4  \p_{\perp i}\Phi   \delta x_{\perp}^{i (1)}   - 8 \Phi \kappa^{(1)}\nonumber \\
 &&   + 8 \int_0^{\bar \chi}  \ud \tilde{\chi} \bigg[     -  \Phi  \tilde \p_{\perp m}S_{\perp}^{m(1)}  +  \left( \frac{\ud \Phi}{\ud \tilde \chi}   -  \frac{1}{\tilde\chi} \Phi  \right) \kappa^{(1)}  \bigg]  \;.
\end{eqnarray}

From Eqs.\ (\ref{Poiss-dnu&de}) and  (\ref{Poiss-dnue-||perp-1}) we find
\begin{eqnarray}
\label{Poiss-dx0-1}
\delta x^{0 (1)}&=& -\bar \chi \left (\Phi _o-v _{\| \, o}\right)+ 2\int_0^{\bar \chi} \ud \tilde \chi \left[  \Phi  + \left(\bar \chi-\tilde \chi\right) \Phi {'} \right] =  -\bar \chi \left (\Phi _o-v _{\| \, o}\right) -T^{(1)} - 2\bar \chi I^{(1)} - 2\int_0^{\bar \chi} \ud \tilde \chi \tilde \chi  \Phi {'} \; \\
\label{Poiss-dx||-1}
\delta x_{\|}^{(1)} &=& \bar \chi \left(\Phi _o-v _{\| \, o}\right) -2\int_0^{\bar \chi} \ud \tilde \chi \left[  \left(\bar \chi-\tilde \chi\right)\Phi {'}  \right]= \bar \chi \left(\Phi _o-v _{\| \, o}\right) + 2\bar \chi I^{(1)} + 2\int_0^{\bar \chi} \ud \tilde \chi \tilde \chi  \Phi {'} \;,   \\
\label{Poiss-dxperp-1}
\delta x_{\perp}^{i  (1)}&=& - \bar \chi \, v^{i  }_{\perp \, o } - 2 \int_0^{\bar \chi} \ud \tilde \chi \left[ \left(\bar \chi-\tilde \chi\right) \tilde \p^i_\perp  \Phi \right] =- \bar \chi \, v^{i  }_{\perp \, o }+ 2  \bar \chi S_\perp^{i(1)} - \bar \chi \p_\perp^i T^{(1)}  \;,
\end{eqnarray}
to first order. 

At second order,
\begin{eqnarray}
\label{Poiss-dx0-2}
&& \delta x^{0(2)}=  \bar \chi \left(-\Phi^{(2)}_o+ v^{(2)}_{\| o} + \Phi _{o}^2 -8 \Phi_o v _{\| o}- v _{k\, o} v^{k  }_o \right) 
+ 4  \bar \chi \left(\Phi_o-v_{\| \, o}\right)  \left( \Phi   -   I^{(1)}   \right)   - 8 v_{\| \, o}  \left( \frac{1}{2}T^{(1)}+ \bar \chi  \int_0^{\bar \chi}   \frac{\ud \tilde{\chi}}{\tilde \chi} \Phi \right) \nonumber \\ 
&&  + 2 \bar \chi v^{i (1)}_{\perp \, o } \left(   \p_\perp^i T^{(1)} + 2\bar \chi \p_{\perp i}  I^{(1)} + 2\p_{\perp i} \int_0^{\bar \chi} \ud \tilde \chi ~  \tilde \chi  \Phi{'} \right)    + 8 \Phi \left(     \bar \chi  I^{(1)} + \int_0^{\bar \chi} \ud \tilde \chi ~  \tilde \chi  \Phi{'}  \right)   -4 \bar \chi  S_{\perp }^{i (1)}S_{\perp }^{j (1)} \delta_{ij} \nonumber\\
      && + 2 \int_0^{\bar \chi} \ud \tilde{\chi} \bigg[ \Phi^{(2)}+  \omega^{(2)}_{\| }-4 {\Phi}^2 -4 \left(I^{(1)}\right)^2+ 4\tilde \chi  \Phi{'}  I^{(1)} + 4 \Phi{'}  \int_0^{\tilde \chi} \ud \tilde{\tilde \chi} \tilde{\tilde \chi}  \Phi {'}  -  2 \Phi{'} T^{(1)} - 4 \Phi \kappa^{(1)}   + 2 S_{\perp }^{i (1)}S_{\perp }^{j (1)} \delta_{ij}  \nonumber\\
&&   - 2 \tilde \chi   \tilde \p_{\perp i} \Phi \p_\perp^i T^{(1)} \bigg]  +  \int_0^{\bar \chi} \ud \tilde \chi\left(\bar \chi-\tilde \chi\right) \bigg[\Phi^{(2)}{'} + 2 \omega^{(2)}_{\| }{'} + \Psi^{(2)}{'}- \frac{1}{2} \hat h^{(2)}_{\| }{'} -  8\Phi {'}  \Phi   -4  \Phi{''} T^{(1)}   - 8 \Phi{'}   I^{(1)}        - 8 \Phi    \tilde \p_{\perp m}S_{\perp}^{m(1)} \nonumber \\ 
    &&        + 8  \left( \frac{\ud}{\ud \tilde \chi} \Phi    -  \frac{1}{\tilde\chi} \Phi \right)  \kappa^{(1)}  + 4\tilde \chi  \tilde \p_{\perp i} \Phi{'}  \left( 2   S_\perp^{i(1)} - \p_\perp^i T^{(1)}\right)\bigg] \;,
\end{eqnarray}

\begin{eqnarray}
\label{Poiss-dx_||-2}  
&&\delta x_\|^{(2)} = \bar \chi \left( \Phi^{(2)}_o-  v^{(2)}_{\| o} + v _{\| o}^2-  \Phi _{o}^2 \right) 
+ 4\bar \chi  \left(\Phi _o-v _{\| \, o}\right)  I^{(1)}   +2  \bar \chi v_{\perp \, o}^{i } \bigg(  2S_\perp^{i(1)} - \p_\perp^i T^{(1)}      - 2\bar \chi \p_{\perp i}  I^{(1)} -  2 \p_{\perp i} \int_0^{\bar \chi} \ud \tilde \chi ~  \tilde \chi  \Phi{'} \bigg) \nonumber\\
&&+  \int_0^{\bar \chi} \ud \tilde \chi \bigg[- \Phi^{(2)} + \Psi^{(2)}-\frac{1}{2}\hat h_{\|}^{(2)} + 4  {\Phi}^2 +8 \left(I^{(1)}\right)^2-4\delta_{ij} S_{\perp }^{i (1)} S_{\perp }^{j (1)} -8 \tilde \chi  \Phi{'}   I^{(1)} - 8  \Phi{'}  \int_0^{\tilde \chi} \ud \tilde{\tilde \chi} \tilde{\tilde \chi}  \Phi {'}\bigg]  \nonumber \\
      &&  + \int_0^{\bar \chi}   \ud \tilde{\chi}   (\bar \chi - \tilde \chi) \bigg[  - \Phi^{(2)}{'} - 2 \omega^{(2)}_{\| }{'} - \Psi^{(2)}{'}  + \frac{1}{2} \hat h^{(2)}_{\| }{'} + 8  \Phi    \Phi {'}  + 4\Phi{''}   T^{(1)}  +8 \Phi{'} I^{(1)}    
   +4   \tilde \chi \tilde \p_{\perp i}  \Phi{'}  \left(   - 2   S_\perp^{i(1)} +   \p_\perp^i T^{(1)} \right) \bigg]\;,  \nonumber \\
 \end{eqnarray}

and

\begin{eqnarray}
\label{Poiss-dx_perp-2}  
&& \delta x_\perp^{i(2)}= \bar \chi \left(  -2 \omega^{i(2)}_{\perp  o}-  v^{i(2)}_{\perp o} + \frac{1}{2} n^j \hat h_{ j k\, o}^{(2)} \Perp^{ki}+   v _{\| \, o} v^{i }_{\perp o}    \right) + 4 \bar \chi \left (\Phi _o-v _{\|  o}\right)   S_\perp^{i(1)}  + 2 v_{\perp o}^j   \bigg[ -2 \Perp_j^i \left(  \bar \chi I^{(1)}   +  \int_0^{\bar \chi} \ud \tilde \chi \tilde \chi  \Phi {'}\right)
\nonumber\\
&& +  \bar \chi^2  \Perp^{im} \p_{\perp j}   \p_{\perp m}   \left( T^{(1)}+2 \bar \chi  \int_0^{\bar \chi}   \frac{\ud \tilde{\chi}}{\tilde \chi} \Phi \right)  \bigg]+  \int_0^{\bar \chi} \ud \tilde \chi \bigg( 2\omega^{i(2)}_{\perp} -  n^j \hat h_{jk}^{(2)} \Perp^{ki}  - 8\tilde \chi  \tilde \p^i_{\perp }  \Phi  ~ I^{(1)}  - 8  \tilde \p^i_{\perp }  \Phi \int_0^{\tilde \chi} \ud \tilde{\tilde \chi} \tilde{\tilde \chi}  \Phi {'}   + 4 \Phi \tilde \p_\perp^i T^{(1)} \bigg) \nonumber\\
&&+ \int_0^{\bar \chi} \ud \tilde \chi  \left(\bar \chi-\tilde \chi\right) \bigg\{-\bigg[\tilde \p^i_\perp \bigg( \Phi^{(2)} + 2 \omega^{(2)}_{\| }+ \Psi^{(2)}- \frac{1}{2} \hat h^{(2)}_{\| } \bigg)   + \frac{1}{\tilde \chi} \left(- 2 \omega^{i (2)}_{\perp }+  n^k \hat h_{kj}^{(2)} \Perp^{ij}  \right)\bigg]    +  8 \Phi \tilde\p^i_\perp \Phi   - 8  \tilde\p^i_\perp \Phi I^{(1)}   \nonumber \\
&&    - 4 \tilde \p_{\perp }^i    \Phi{'}  ~T^{(1)}    +  \frac{4}{\tilde \chi}  \Phi  \p^i_\perp T^{(1)} +  4\bigg(\tilde \chi \Perp^{im} \tilde \p_{\perp j}  \tilde  \p_{\perp m}   \Phi +  \Perp^i_j \Phi{'}  \bigg) \left(- 2   S_\perp^{j(1)} +  \p_\perp^j T^{(1)}\right)  \bigg\}\;.
\end{eqnarray}

Combining Eqs.\  (\ref{Poiss-dx0-2}) and (\ref{Poiss-dx_||-2})   we have
\begin{eqnarray} 
\label{Poiss-dx0+dx_||-2}
&& \delta x^{0 (2)} + \delta x_\|^{(2)}=
- \bar \chi \left( 8  \Phi _o v _{\| \, o}   + v _{\perp k \, o} v^{k  }_{\perp \, o} \right) +  4  \bar \chi\left(\Phi_o-v_{\| \, o}\right) \Phi     -4  v_{\| \, o}\left( T^{(1)}+ 2\bar \chi  \int_0^{\bar \chi}   \frac{\ud \tilde{\chi}}{\tilde \chi} \Phi \right)
 +4  \bar \chi v_{\perp \, o}^{i }  S_\perp^{i(1)}     - T^{(2)}\nonumber \\
&&   + 8 \Phi \left(     \bar \chi  I^{(1)} + \int_0^{\bar \chi} \ud \tilde \chi ~  \tilde \chi  \Phi{'}  \right)-4 \bar \chi  S_{\perp }^{i (1)}S_{\perp }^{j (1)} \delta_{ij} + 4 \int_0^{\bar \chi} \ud \tilde \chi \left(-{\Phi}^2  -  \Phi{'} T^{(1)}  -2\Phi \kappa^{(1)}     - \tilde \chi \tilde \p_{\perp i}\Phi  \p_\perp^i T^{(1)} \right) \nonumber \\
 && +8 \int_0^{\bar \chi}  \ud \tilde{\chi} ~ (\bar \chi - \tilde \chi) \bigg[   -  \Phi   \tilde \p_{\perp m}S_{\perp}^{m(1)}     
 +  \left( \frac{\ud}{\ud \tilde \chi} \Phi   -  \frac{1}{\tilde\chi} \Phi  \right) \kappa^{(1)}  \bigg]\;.
\end{eqnarray}

\section{Final result with $Q=0$ or $Q=1$}
\label{B}

In this Appendix we show two particular cases of Eqs.\ (\ref{Deltag-1}) and (\ref{Deltag-2}).

\subsection{$Q=0$, $Q^{(1)}=0$}

At first order  (this particular solution has been computed previously in \cite{Bertacca:2014dra})
\begin{eqnarray}
\label{Deltag1-Q=0}
\Delta_g^{(1)} &=&\delta_g^{(1)} +\left( b_e  - \frac{\cH'}{\cH^2}  -  \frac{2}{\bar \chi \cH} \right)  \Delta \ln a^{(1)}  - \Phi - \frac{1}{\cH}  \p_\|^2 v  + \frac{1}{\cH} \Phi {'} - \frac{2}{\bar \chi}  T^{(1)}  - 2 \kappa^{(1)} \;,  
\end{eqnarray}

and at second order (see also \cite{Bertacca:2014dra})
\begin{eqnarray}
\label{Deltag2-Q=0}
&&\Delta_g^{(2)} = \delta_g^{(2)}  +\left( b_e -   \frac{\cH'}{\cH^2} - \frac{2}{\bar \chi \cH} \right) \, \Delta \ln a^{(2)}  -2  \Psi^{(2)} -\frac{1}{2}\hat h_{\|}^{(2)}     - \frac{2}{\bar \chi} T^{(2)}- 2 \kappa^{(2)} + \Phi^{(2)}+  \frac{1}{ \cH} \Psi^{(2)}{'}-  \frac{1}{2 \cH} \hat h^{(2)}_{\| }{'} \nonumber \\  
&&  -\frac{1}{ \cH }\p_\|^2 v^{(2)} -\frac{1}{ \cH }   \p_\|\hat v^{(2)}_\|  - 2\Phi \delta_g^{(1)} - \frac{2}{\cH} \delta_g^{(1)} \p_\|^2 v  +\frac{2}{\cH}\delta_g^{(1)}  \Phi {'}+ 2\frac{\cH'}{\cH^3}\Phi \Phi {'} - 5 {\Phi}^2 + \left( \p_\| v  \right)^2  - \frac{2}{\cH}\left(1+\frac{\cH'}{\cH^2}\right)\Phi \p_\|^2 v\nonumber \\ 
&&   + \frac{2}{\cH^2}\left( \Phi {'}  \right)^2  + \frac{2}{\cH^2}\left(\p_\|^2 v  \right)^2  + \frac{2}{\cH^2} \p_\| v  \p_\|^2 \Phi  +\frac{4}{ \cH }\p_\| v  \p_\| \Phi   - \frac{2}{\cH^2} \Phi \p_\|^3 v   -\frac{2}{\cH}\Phi \p_\| \Phi    + \frac{2}{\cH^2}\Phi \frac{\ud \Phi {'} }{\ud \bar \chi} - \frac{2}{\cH^2} \p_\| v \frac{\ud \Phi {'} }{\ud \bar \chi} \nonumber \\
 &&  +\frac{2}{\cH} \left(1+\frac{\cH'}{\cH^2} \right) \p_\| v \p_\|^2 v   - \frac{2}{\cH^2}\Phi  \p_\|^2 \Phi  +\frac{2}{\cH} \left(1 -\frac{\cH'}{\cH^2} \right)  \p_\| v  \Phi {'}  -  \frac{4}{\cH^2}\p_\|^2 v   \Phi {'}  +\frac{2}{\cH}\p_{\perp i} v \p^i_\perp \Phi    -\frac{4}{ \cH } \p_{\perp i} v   \p_{\perp}^i \p_\|  v \nonumber \\
 &&   +\left(-1  +\frac{4}{ \bar \chi \cH } \right) \p_{\perp i} v    \p_{\perp}^i v  +  \frac{2}{\cH^2}\p_\| v \p_\|^3v  + \Bigg[ \bigg(  -2 b_e    + 2  \frac{\cH'}{\cH^2}   - \frac{4}{\bar \chi \cH} \bigg) \Phi  +2 \left( b_e  -  \frac{\cH'}{\cH^2}  - \frac{2}{\bar \chi \cH} \right) \delta_g^{(1)}   \nonumber \\
  &&    - \frac{2}{\cH}  \frac{\ud  \delta_g^{(1)} }{\ud \bar \chi}+  \frac{2}{\cH}  \left( -   b_e   +  \frac{\cH' }{\cH^2}    +  \frac{2}{\bar \chi \cH} \right)  \p_\|^2 v   + \frac{2}{\cH}\left(  -2  +  b_e  - \frac{\cH' }{\cH^2}  -   \frac{2}{\bar \chi \cH}\right)  \Phi {'}     + 4\bigg(  - \left(b_e -\frac{\p \Q}{\p \ln \bar a} \right) +  \frac{\cH'}{\cH^2} + \frac{1}{\bar \chi \cH}   \bigg)  \nonumber \\    
&&   \times \left(\frac{1}{\bar \chi}T^{(1)}+\kappa^{(1)} \right)  \Bigg]\,\Delta \ln a^{(1)}+ \Bigg[-b_e+b_e^2+ \frac{\p b_e}{\p \ln \bar a}  + \left(1-2 b_e \right) \frac{\cH' }{\cH^2} -\frac{\cH'' }{\cH^3} +3\left( \frac{\cH' }{\cH^2} \right)^2 + \frac{6}{\bar \chi} \frac{\cH' }{\cH^3}    +  \frac{2}{\bar \chi^2 \cH^2}  \nonumber \\
&& + \frac{2}{\bar \chi \cH}     \Bigg] \left(\Delta \ln a^{(1)}\right)^2  + 4  \bigg[  +\frac{1}{\cH} \left(1 -\frac{\cH' }{\cH^2}\right)\Phi {'} + \frac{1}{\cH}   \p_\| \Phi  +  \frac{1}{\cH} \left(1 +  \frac{\cH' }{\cH^2} \right)\p_\|^2 v    +  \frac{1}{\cH^2} \p_\|^2 \Phi  + \frac{1}{\cH^2} \p_\|^3 v   - \frac{1}{\cH^2}\frac{\ud \Phi {'} }{\ud \bar \chi}  \bigg] I^{(1)}   \nonumber \\
&&+ \bigg( - \frac{4}{\bar \chi}  \delta_g^{(1)} - 2   \p_{\|}\delta_g^{(1)}   - \frac{4}{\bar \chi \cH}   \Phi {'} - \frac{4}{\bar \chi}  \Phi +2 \p_{\|}\Phi  + \frac{4}{\bar \chi \cH} \p_\|^2 v   +\frac{2}{\cH} \p_\|^3 v  -\frac{2}{\cH} \p_\| \Phi {'}    \bigg) T^{(1)}  +  \frac{2}{\bar \chi^2} \left(  T^{(1)} \right)^2 + \frac{4}{\bar \chi} T^{(1)}   \kappa^{(1)}   \nonumber \\ 
 &&+4 \bigg( - \Phi  + \frac{1}{\cH}  \p_\|^2 v   -\frac{1}{\cH} \Phi {'}  -   \delta_g^{(1)} \bigg)\kappa^{(1)} +  \vartheta_{ij}^{(1) }\vartheta^{ij(1)}  + 2 \left(\kappa^{(1)}\right)^2   -2  \big|\gamma^{(1)}\big|^2  + 4 \bigg[  \frac{\bar \chi}{\cH}\bigg(  \p_{\perp i}\Phi {'}  -  \p_{\perp i} \p_\|^2 v  \bigg)   +   \bar \chi  \p_{\perp i} \delta_g^{(1)}  \nonumber \\
 &&   +  \bar \chi  \p_{\perp i} \Phi  -2\bar \chi \p_{\perp i} \Phi    +  \frac{1}{\cH}  \p_{\perp i}  \Delta \ln a^{(1)} \bigg] S_{\perp}^{i (1)}  -4 S_{\perp }^{i (1)}S_{\perp }^{j (1)} \delta_{ij}  +2 \bigg(  \frac{2}{\bar \chi \cH} \p_{\perp i} v   -  \frac{\bar \chi}{\cH} \p_{\perp i} \Phi {'}     + \frac{\bar \chi}{\cH} \p_{\perp i} \p_\|^2 v   -\frac{2}{\cH}\p_{\perp i}\p_\| v \nonumber \\
 &&      -   \bar \chi  \p_{\perp i} \delta_g^{(1)}     - \bar \chi  \p_{\perp i} \Phi       +  2 \bar \chi \p_{\perp i} \Phi     \bigg)  \p_\perp^i T^{(1)}     + 8\Bigg\{  \int_0^{\bar \chi}  \ud \tilde{\chi}\bigg[       -  \Phi  \tilde \p_{\perp m}S_{\perp}^{m(1)}  +  \left( \frac{\ud \Phi}{\ud \tilde \chi}   -  \frac{1}{\tilde\chi} \Phi  \right) \kappa^{(1)}  \bigg]   \nonumber \\
 && - \frac{1}{\bar \chi}\int_0^{\bar \chi}  \ud \tilde{\chi} \bigg(    { \Phi}^2  +  \Phi{'} T^{(1)} +2 \Phi \kappa^{(1)}   + \tilde \chi \tilde \p_{\perp i}\Phi  \p_\perp^i T^{(1)}   \bigg) +  \frac{1}{\bar \chi} \int_0^{\bar \chi}  \ud \tilde{\chi} ~ (\bar \chi - \tilde \chi) \bigg[  -  2 \Phi   \tilde \p_{\perp m}S_{\perp}^{m(1)}     
 + 2  \left( \frac{\ud  \Phi}{\ud \tilde \chi}   -  \frac{1}{\tilde\chi} \Phi  \right) \kappa^{(1)}  \nonumber \bigg] \Bigg\}   \nonumber \\
           && - \left( 24\Phi _o v _{\| \, o}  +   v _{\perp k \, o} v^{k  }_{\perp \, o} \right)    - 8 v_{\| \, o}  \left(  \frac{1}{\bar \chi} T^{(1)}+ 3 \int_0^{\bar \chi}   \frac{\ud \tilde{\chi}}{\tilde \chi} \Phi \right)    + 2\left(\Phi _o-v _{\| \, o}\right) \bigg[\left(\frac{\cH' }{\cH^3}+ \frac{1}{\cH}\right) \p_\|^2 v   +\left(-\frac{\cH' }{\cH^3}+ \frac{1}{\cH}\right)  \Phi {'}     \nonumber \\
           &&   + \frac{1}{\cH^2} \p_\|^2 \Phi  + \frac{1}{\cH^2}\p_\|^3 v + \frac{1}{\cH}  \p_\| \Phi - \frac{1}{\cH^2}\frac{\ud \Phi {'} }{\ud \bar \chi} \bigg]   +v^{i  }_{\perp \, o }\bigg[  -2   \bar \chi  \p_{\perp i} \Phi  + 2 \frac{\bar \chi}{\cH} \p_{\perp i}  \left(-\Phi {'} +\p_\|^2v  \right)    - 2 \bar \chi    \p_{\perp i} \delta_g^{(1)}   \bigg]  \nonumber \\
 &&     +  v_{\perp  i\, o} \bigg(   4  S^{i }_{\perp}  - \frac{2}{\cH}  \p^i_{\perp}  \Delta \ln a^{(1)} + 4 \bar \chi  \p^i_{\perp} \Phi  \bigg) \;.
   \end{eqnarray}

\subsection{$Q=1$, $Q^{(1)}=0$}

At first order
 \begin{eqnarray}
\label{Deltag1-Q=1}
\Delta_g^{(1)} &=&\delta_g^{(1)} +\left( b_e  - \frac{\cH'}{\cH^2} - 2  \right)  \Delta \ln a^{(1)}  +\Phi - \frac{1}{\cH}  \p_\|^2 v  + \frac{1}{\cH} \Phi {'} \;,   
\end{eqnarray}
and, at second order, 
\begin{eqnarray}
\label{Deltag2-Q=1}
&&\Delta_g^{(2)} = \delta_g^{(2)}  +\left( b_e-2  -   \frac{\cH'}{\cH^2}  \right) \, \Delta \ln a^{(2)}  + \Phi^{(2)}+  \frac{1}{ \cH} \Psi^{(2)}{'}-  \frac{1}{2 \cH} \hat h^{(2)}_{\| }{'} -\frac{1}{ \cH }\p_\|^2 v^{(2)} -\frac{1}{ \cH }   \p_\|\hat v^{(2)}_\| +2\Phi \delta_g^{(1)} - \frac{2}{\cH} \delta_g^{(1)} \p_\|^2 v  \nonumber \\ 
&&  +\frac{2}{\cH}\delta_g^{(1)}  \Phi {'} + \frac{2}{\cH}\left(2+\frac{\cH'}{\cH^2}\right)\Phi \Phi {'} +3 {\Phi}^2 
+\left( \p_\| v  \right)^2   - \frac{2}{\cH}\left(3  +\frac{\cH'}{\cH^2}\right)\Phi \p_\|^2 v + \frac{2}{\cH^2}\left( \Phi {'}  \right)^2 \nonumber \\
 && + \frac{2}{\cH^2}\left(\p_\|^2 v  \right)^2  + \frac{2}{\cH^2} \p_\| v  \p_\|^2 \Phi  +\frac{4}{ \cH }\p_\| v  \p_\| \Phi   - \frac{2}{\cH^2} \Phi \p_\|^3 v   -\frac{2}{\cH}\Phi \p_\| \Phi    + \frac{2}{\cH^2}\Phi \frac{\ud \Phi {'} }{\ud \bar \chi}- \frac{2}{\cH^2} \p_\| v \frac{\ud \Phi {'} }{\ud \bar \chi}  +\frac{2}{\cH} \left(1 +\frac{\cH'}{\cH^2} \right) \p_\| v \p_\|^2 v \nonumber \\
 && - \frac{2}{\cH^2}\Phi  \p_\|^2 \Phi  + \frac{2}{\cH} \left(1-\frac{\cH'}{\cH^2}\right) \p_\| v  \Phi {'}  -  \frac{4}{\cH^2}\p_\|^2 v   \Phi {'}  +\frac{2}{\cH}\p_{\perp i} v \p^i_\perp \Phi   -\frac{4}{ \cH } \p_{\perp i} v   \p_{\perp}^i \p_\|  v   +\left(-1  +\frac{4}{ \bar \chi \cH } \right) \p_{\perp i} v    \p_{\perp}^i v +  \frac{2}{\cH^2}\p_\| v \p_\|^3v  \nonumber \\
  && 
  + \Bigg[ 2\bigg(   - 6   +  b_e - \frac{\cH'}{\cH^2}  + \frac{4}{\bar \chi \cH} \bigg) \Phi    +2 \left( b_e - 2  -  \frac{\cH'}{\cH^2}    \right) \delta_g^{(1)}  
 - \frac{2}{\cH}  \frac{\ud  \delta_g^{(1)} }{\ud \bar \chi}+  \frac{2}{\cH}  \left( -   b_e   + 2     +  \frac{\cH' }{\cH^2}   
  \right)  \p_\|^2 v  \nonumber \\    
 &&   + \frac{2}{\cH}\left(  -2  +  b_e  - \frac{\cH' }{\cH^2}  \right)  \Phi {'}    - \frac{4}{\cH}  \p_\|  \Phi  + 8\bigg(  - 1   + \frac{1}{\bar \chi \cH}  \bigg) \left(\frac{1}{\bar \chi}T^{(1)}+\kappa^{(1)} \right)  \Bigg]\,\Delta \ln a^{(1)} + \Bigg[- 5 b_e+b_e^2+ \frac{\p b_e}{\p \ln \bar a}+10 \nonumber \\
&& + \left(5 -2 b_e \right) \frac{\cH' }{\cH^2} -\frac{\cH'' }{\cH^3} +3\left( \frac{\cH' }{\cH^2} \right)^2 
  +  \frac{4}{\bar \chi^2 \cH^2}  - \frac{8}{\bar \chi \cH}  \Bigg] \left(\Delta \ln a^{(1)}\right)^2  + 4  \bigg[  \frac{1}{\cH} \left(1 -\frac{\cH' }{\cH^2}\right)\Phi {'} + \frac{1}{\cH}   \p_\| \Phi   +  \frac{1}{\cH} \left(1 +  \frac{\cH' }{\cH^2} \right)\p_\|^2 v  \nonumber \\
 &&      +  \frac{1}{\cH^2} \p_\|^2 \Phi  + \frac{1}{\cH^2} \p_\|^3 v   - \frac{1}{\cH^2}\frac{\ud \Phi {'} }{\ud \bar \chi}  \bigg] I^{(1)}  
 +2 \bigg( -    \p_{\|}\delta_g^{(1)} + \frac{4}{\bar \chi}   \Phi -  \p_{\|}\Phi   +\frac{1}{\cH} \p_\|^3 v  -\frac{1}{\cH} \p_\| \Phi {'}    \bigg) T^{(1)} \nonumber \\ 
 &&  +4 \left[  \frac{1}{\bar \chi^2} \left(  T^{(1)} \right)^2 + \frac{2}{\bar \chi} T^{(1)}   \kappa^{(1)}  \right] 
 +8 \Phi \kappa^{(1)}   + 4 \left(\kappa^{(1)}\right)^2     + 4 \bigg[  \frac{\bar \chi}{\cH}\bigg(  \p_{\perp i}\Phi {'}  -  \p_{\perp i} \p_\|^2 v  \bigg)   +   \bar \chi  \p_{\perp i} \delta_g^{(1)}   +  \bar \chi  \p_{\perp i} \Phi  \bigg] S_{\perp}^{i (1)}\nonumber \\
 && +2 \bigg[  \frac{2}{\bar \chi \cH} \p_{\perp i} v   -  \frac{\bar \chi}{\cH} \p_{\perp i} \Phi {'} + \frac{\bar \chi}{\cH} \p_{\perp i} \p_\|^2 v   -\frac{2}{\cH}\p_{\perp i}\p_\| v  \   -   \bar \chi  \p_{\perp i} \delta_g^{(1)}     - \bar \chi  \p_{\perp i} \Phi        \bigg]  \p_\perp^i T^{(1)}       + 2\left(\Phi _o-v _{\| \, o}\right) \bigg[\left(\frac{\cH' }{\cH^3}+ \frac{1}{\cH}\right) \p_\|^2 v   \nonumber \\
           &&  +\left(-\frac{\cH' }{\cH^3}+ \frac{1}{\cH}\right)  \Phi {'}      + \frac{1}{\cH^2} \p_\|^2 \Phi  + \frac{1}{\cH^2}\p_\|^3 v + \frac{1}{\cH}  \p_\| \Phi - \frac{1}{\cH^2}\frac{\ud \Phi {'} }{\ud \bar \chi} \bigg]   +v^{i  }_{\perp \, o }\bigg[  -2   \bar \chi  \p_{\perp i} \Phi  + 2 \frac{\bar \chi}{\cH} \p_{\perp i}  \left(-\Phi {'} +\p_\|^2v  \right)    - 2 \bar \chi    \p_{\perp i} \delta_g^{(1)}   \bigg]  \nonumber \\
 &&     \nonumber \\
   \end{eqnarray}

\end{document}